\documentclass{aa}  
 \usepackage[varg]{txfonts}

\usepackage{orcidlink}

\usepackage{natbib}
\bibpunct{(}{)}{;}{a}{}{,}
\usepackage{graphicx}
\usepackage{subcaption}

\usepackage{siunitx}
\usepackage{txfonts}
\usepackage{url}
\usepackage{hyperref}
\hypersetup{
     colorlinks=true, 
     linkcolor=red,  
     filecolor=magenta,   
     citecolor=blue,   
     urlcolor=blue
     }
\begin{document}

   \title{Hubble Space Telescope proper motions of Large Magellanic
Cloud star clusters}

   \subtitle{II. Kinematic structure of young and intermediate-age clusters}

      \author{F. Niederhofer\inst{1}\orcidlink{0000-0002-4341-9819}
          \and
          L. Cullinane\inst{1}\orcidlink{0000-0001-8536-0547}
          \and
          D. Massari\inst{2}\orcidlink{0000-0001-8892-4301}
          \and 
          N. Bastian\inst{3,4}\orcidlink{0000-0001-5679-4215}
          \and
          A. Bellini\inst{5}\orcidlink{0000-0003-3858-637X}
          \and
          F. Aguado-Agelet\inst{6}\fnmsep\inst{7}\orcidlink{0000-0003-2713-1943}
          \and
          S. Cassisi\inst{8}\fnmsep\inst{9}\orcidlink{0000-0001-5870-3735}
          \and
          D. Erkal\inst{10}\orcidlink{0000-0002-8448-5505}
          \and 
          M. Libralato\inst{11}\orcidlink{0000-0001-9673-7397}
          \and 
          N. Kacharov\inst{1}\orcidlink{0000-0002-6072-6669}
          \and
          I. Cabrera-Ziri\inst{12}\orcidlink{0000-0001-9478-5731}
          \and
          E. Ceccarelli\inst{2,13}
          \and 
          M.-R. L. Cioni\inst{1}\orcidlink{0000-0002-6797-696X}
          \and
          F. Dresbach\inst{14}\orcidlink{0000-0003-0808-8038}
          \and
          M. Häberle\inst{15}\orcidlink{0000-0002-5844-4443}
          \and 
          S. Martocchia\inst{16}\orcidlink{0000-0001-7110-6775}
          \and
          S. Saracino\inst{17}\orcidlink{0000-0003-4746-6003}
      }

   \institute{Leibniz-Institut für Astrophysik Potsdam, An der Sternwarte 16, D-14482 Potsdam, Germany\\
        \email{fniederhofer@aip.de}
        \and INAF – Osservatorio di Astrofisica e Scienza dello Spazio di Bologna, Via Gobetti 93/3, 40129 Bologna, Italy\
         \and Donostia International Physics Center (DIPC), Paseo Manuel de Lardizabal, 4, 20018 Donostia-San Sebastián, Guipuzkoa, Spain
        \and IKERBASQUE, Basque Foundation for Science, 48013 Bilbao, Spain
        \and Space Telescope Science Institute, 3700 San Martin Drive, Baltimore, MD 21218, USA\
        \and atlanTTic, Universidade de Vigo, Escola de Enxeñaría de Telecomunicación, 36310 Vigo, Spain\
        \and Universidad de La Laguna, Avda. Astrofísico Fco. Sánchez, 38205 La Laguna, Tenerife, Spain\
        \and INAF – Osservatorio Astronomico d' Abruzzo, Via M. Maggini, 64100 Teramo, Italy\
        \and INFN – Sezione di Pisa, Universitá di Pisa, Largo Pontecorvo 3, 56127 Pisa, Italy\
        \and Department of Physics, University of Surrey, Guildford GU2 7XH, UK\
        \and INAF - Osservatorio Astronomico di Padova, Vicolo dell'Osservatorio 5, Padova I-35122, Italy\
        \and Vyoma GmbH, Karl-Theodor-Straße 55, 80803 Munich, Germany\
        \and Dipartimento di Fisica e Astronomia, Università degli Studi di Bologna, via Gobetti 93/2, I-40129 Bologna, Italy\
        \and Lennard-Jones Laboratories, School of Chemical and Physical Sciences, Keele University, Keele ST5 5BG, UK
        \and ESO, European Southern Observatory, Karl-Schwarzschild-Str. 2, D-85748 Garching bei München, Germany\
        \and Aix Marseille Universit{\'e}, CNRS, CNES, LAM, Marseille, France
        \and INAF – Osservatorio Astrofisico di Arcetri, Largo E. Fermi 5, 50125 Firenze, Italy
}

   \date{Received date /
Accepted date}

% \abstract{}{}{}{}{} 
% 5 {} token are mandatory
 
  \abstract{In this paper, we explore the kinematic properties of a sample of 19 young ($<$1~Gyr) and intermediate-age (1--2.5~Gyr) massive star clusters within the Large Magellanic Cloud (LMC). We analyse the proper motions of the clusters, which have been measured based on multi-epoch \textit{Hubble} Space Telescope \textit{(HST)} observations. Additionally, we infer from the \textit{HST} data homogeneous and robust estimates for the distances, ages and metallicities of the clusters. This collection of information, in combination with literature line-of-sight velocities, allows us to investigate the full 3D dynamics of our sample of clusters within the frame of the LMC in a self-consistent way. While most young clusters orbit the LMC close to the stellar disc plane, NGC~1850 ($\sim$100~Myr old) depicts a peculiar case. Depending on the exact distance from the disc, it follows either a highly inclined, retrograde orbit or an eccentric orbit along the bar structure. The orbits of young clusters that formed North of the LMC centre show signs that might be connected to the resettling motion of the LMC bar structure. Based on the dynamic properties in combination with the positions of the clusters in the age-metallicity space, we find no clear-cut evidence for clusters in our sample that could have been stripped from the Small Magellanic Cloud (SMC) onto the LMC. We finally compare the kinematics of the intermediate-age clusters with a suite of simple numerical simulations of the Magellanic system to interpret the cluster motions. A possible interaction history of the LMC with the SMC, where the SMC had two past crossings of the LMC disc plane (about 300 and 900~Myr ago), in combination with the recent SMC pericentre passage, can qualitatively explain the observed kinematic structure of the clusters analysed in this work.
   }


   \maketitle
%
%-------------------------------------------------------------------

\section{Introduction}\label{sec:intro}

On the cosmic stage, we can witness from the front row the dance of two galaxies, the Large and the Small Magellanic Clouds (LMC and SMC). Both galaxies are dynamically interacting with each other and are in the early phases of a minor merger event. Thanks to their close proximity \citep[LMC$\sim$50~kpc, SMC$\sim$60~kpc;][]{Pietrzynski19, Graczyk20}, they provide an ideal opportunity to study in unparalleled detail the dynamic effects that are at play during the merging of galaxies. The two Clouds show periods of enhanced star formation \citep[e.g.][]{Mazzi21,Massana22}, leading to the formation of massive star clusters \citep[e.g.][]{Forbes18} that span the full cosmic age range \citep[e.g.][]{Horta21}. 
Studying the properties of these cluster populations will provide us with invaluable information about the formation and evolution of the Clouds, as well as the origin of their system of star clusters. Exploring high-resolution spectra of old ($>$10~Gyr) LMC clusters, \citet{Mucciarelli21} discovered a peculiar chemical composition of the old LMC cluster NGC~2005, suggesting this cluster has been accreted onto the LMC from a low-mass galaxy. 

Studies of the full three-dimensional (3D) kinematics of the clusters within the two galaxies, that complement the information of the ages and chemical compositions of the clusters, have so far been impeded by the availability of sufficiently precise proper motion (PM) measurements in crowded environments, such as extragalactic star clusters. Luckily, this situation has changed during the last couple of years, thanks to high-precision, multi-epoch space-based data; e.g.\ from the \textit{Gaia} mission or the \textit{Hubble} Space Telescope (\textit{HST}). In a first attempt, \citet{Piatti19} combined PM data from the \textit{Gaia} data release 2 \citep[DR2;][]{Gaia18} catalogue with spectroscopic line-of-sight (LOS) measurements to analyse the 3D kinematic structure of the old star-cluster population of the LMC. They claim the existence of two spatially and dynamically distinct cluster populations, with some clusters showing disc-like properties, and others belonging to the halo. This result contradicts the findings based on LOS velocity data alone \citep[e.g.][]{Freeman83, Grocholski06,Sharma10}, which suggest that all old clusters reside within a disc-like structure. \citet{Bennet22} used a combination of \textit{Gaia} DR3 \citep{Gaia23} and \textit{HST} data to measure the bulk motion of 31 star clusters of various ages within the LMC. Their results showed no evidence for a halo population with all studied clusters following disc-like kinematics.

\citet{Massari21} showed that it is now feasible to measure precise stellar PMs within star clusters at the distance of the Clouds using multi-epoch HST data: based on archival data with long temporal baselines, they determined the PMs towards the star cluster NGC~419 in the SMC. Using a similar method, in a recent work \citep{Niederhofer24}, we presented HST-based astro-photometric catalogues of a sample of 26 star clusters within the LMC.
Analysing the PM data of the young ($\sim$100~Myr old) LMC cluster NGC~1850, we were able to distinguish the kinematics of the various stellar populations within the \textit{HST} field and showed that NGC~1850 is not dynamically related to a close-by association of very young stars. In a following study \citep{Niederhofer25}, we employed these catalogues to analyse for the first time the chemo-dynamics of a sample of seven old LMC star clusters in a self-consistent way. We found conclusive evidence that the cluster NGC~1841 has been accreted by the
LMC from a smaller galaxy, based on its position in age-metallicity space and its peculiar motion within the LMC.

In this next paper of the series, we use the astro-photometric catalogues from \citet{Niederhofer24} to study the kinematic structure of a sample of 19 young ($<$1~Gyr) and intermediate-age (1--2.5~Gyr) star clusters within the LMC. Employing the isochrone fitting framework from the Cluster Ages to Reconstruct the Milky Way Assembly (CARMA) project\footnote{All cluster ages derived within the CARMA project are listed and continuously updated here: \href{https://www.oas.inaf.it/en/research/m2-en/carma-en/}{https://www.oas.inaf.it/en/research/m2-en/carma-en/}} \citep{Massari23}, we self-consistently derive the ages, metallicities and distances of the clusters. 
This work continues the successful method of combining dynamical information with accurate determinations of the cluster properties (age, metallicity) to investigate the evolution of the LMC based on its star cluster system.

The paper is organised as follows. In Section~\ref{sec:data}, we introduce
the used data sets and the compilation of the catalogues for the isochrone fitting. In Section~\ref{sec:isoc_fit}, we describe the isochrone fitting method and present the results for our cluster sample. We present the kinematic structure of the clusters within the LMC in Section~\ref{sec:kinematic_structure}. In Section~\ref{sec:dyn_models}, we compare the kinematics of the intermediate-age clusters with a suite of dynamical models, and conclude the paper in Section~\ref{sec:conclusions}.

%--------------------------------------------------------------------

\section{Data\label{sec:data}}

\subsection{Photometry and PM measurements}

The astro-photometric catalogues presented by \citet{Niederhofer24}\footnote{The catalogues are publicly available as a High Level Science Product
at MAST under: \href{https://archive.stsci.edu/hlsp/hamsters}{https://archive.stsci.edu/hlsp/hamsters}} are based on multi-epoch \textit{HST} observations using the Ultraviolet-Visible (UVIS) channel of the Wide Field Camera 3 (WFC3) and the Wide-Field Channel (WFC) of the Advanced Camera for Surveys (ACS)\footnote{A detailed list of all observations is presented in the appendix of \citet{Niederhofer24} and in the Mikulski Archive for Space Telescopes (MAST) under the following DOI: \href{https://doi.org/10.17909/7d5e-s940}{10.17909/7d5e-s940}.}. 
To increase the sample of clusters, we supplemented these data with two additional clusters, namely NGC~1751 and NGC~1818, which have available archival long time-baseline data, suitable for PM determinations.
Observation logs for NGC~1751 and NGC~1818 can be found in Tables~\ref{tab:ngc1751obs} and \ref{tab:ngc1818obs}. 

The photometric and astrometric reduction of the data sets is described in detail in \citet{Niederhofer24}. The photometric measurements follow the well-established state-of-the-art methods for \textit{HST} data \citep[see][]{Bellini17b, Bellini18}. 
In brief, the photometric measurements have been performed on the individual un-resampled \texttt{\_flc} images, which have been corrected for imperfect charge transfer efficiency (CTE). We measure positions and fluxes of neighbour-subtracted sources using a first- and second-pass photometric run, based on the \texttt{Fortran} tools \texttt{hst1pass} \citep{Anderson22} and \texttt{KS2} \citep[see][for details]{Sabbi16, Bellini17b}. We corrected the measured stellar positions for geometric distortions, applying the precise distortion solutions for WFC3/UVIS and ACS/WFC \citep{Anderson10, Bellini09, Bellini11} and transformed the corrected stellar positions to a common frame, registered on the \textit{Gaia} DR3 astrometric frame \citep{Gaia23}. The instrumental magnitudes have been calibrated to the Vega-system, as described in \citet{Bellini17b}.

We calculated stellar PMs relative to the bulk motions of the clusters in an iterative way, following the methods developed by \citet{Bellini14, Bellini18, Libralato18b, Libralato22}. 
We transformed the observed stellar positions within all exposures to a common reference frame using general six-parameter transformations, and fitted these transformed positions as a function of time with a least-squares straight line. The slopes of these lines directly correspond to the PMs of the stars. For the transformations, we initially used a sample of well-measured likely cluster members selected on a colour-magnitude diagram (CMD) and refined this selection within each iteration, excluding stars not in agreement with the motion of the cluster. After the last iteration, we applied a posteriori corrections to the measured PMs to account for spatially variable systematic effects, \citep[e.g. uncorrected CTE and distortion effects, see][]{Bellini14}. 
To calibrate these relative PMs to absolute motions, we cross-identified sources within our catalogues with the \textit{Gaia} DR3 catalogue. For each cluster, we determined as the PM zero-point the 2.5$\sigma$-clipped median PM difference between the \textit{HST} and \textit{Gaia} PMs. We did this for each PM component separately. The absolute motions of the clusters are then simply given by the negative values of the zero-points. 
Table~\ref{tab:clusters_params} lists the resulting absolute PMs of the star clusters.

\begin{table*}\footnotesize
\centering
\caption{Positions and velocities of our sample of young and intermediate-age LMC star clusters. \label{tab:clusters_params}}
\begin{tabular} {l c c c c r@{\,}c@{\,}l r@{\,}c@{\,}l r@{\,}c@{\,}l r}
\hline\hline
\noalign{\smallskip}
Cluster ID
      & RA$_0$
         & $\Delta$RA$_0$
            & Dec$_0$
               & $\Delta$Dec$_0$
                   &  \multicolumn{3}{c}{$\mu_{\alpha}$cos($\delta$)} 
                       & \multicolumn{3}{c}{$\mu_{\delta}$} 
                          & \multicolumn{3}{c}{LOS Velocity}
                               & Ref
                    \\
&   [h:m:s]
    & [arcsec]
      & [\degr:\arcmin:\arcsec]
         & [arcsec]
           & \multicolumn{3}{c}{[mas\,yr$^{-1}$]}
             & \multicolumn{3}{c}{[mas\,yr$^{-1}$]}
              & \multicolumn{3}{c}{[km\,s$^{-1}$]}
                 &
      \\
\noalign{\smallskip}
\hline
\noalign{\smallskip}
NGC 1651  &  04:37:32.29  &  0.46  &  $-$70:35:10.5  &  0.43 &  1.956 & $\pm$ & 0.034 &  $-$0.305 & $\pm$ & 0.038  &   228.2 & $\pm$ & 2.3   & (1)  \\
NGC 1718  &  04:52:26.02  &  0.37  &  $-$67:03:05.1  &  0.31 &  1.851 & $\pm$ & 0.040 &  $-$0.462 & $\pm$ & 0.036  &   278.4 & $\pm$ & 2.2   & (1)  \\
NGC 1751  &  04:54:12.42  &  1.10  &  $-$69:48:27.8  &  0.60 &  1.943 & $\pm$ & 0.033 &  $-$0.132 & $\pm$ & 0.036  &   241.3 & $\pm$ & 0.2   & (2)  \\
NGC 1783  &  04:59:08.94  &  0.38  &  $-$65:59:14.8  &  0.34 &  1.648 & $\pm$ & 0.036 &  $-$0.025 & $\pm$ & 0.031  &   279.6 & $\pm$ & 0.2   & (2)  \\
NGC 1805  &  05:02:21.66  &  0.34  &  $-$66:06:42.4  &  0.31 &  1.593 & $\pm$ & 0.033 &     0.100 & $\pm$ & 0.031  &   301.7 & $\pm$ & 3.8   & (3)  \\
NGC 1806  &  05:02:12.18  &  0.31  &  $-$67:59:10.0  &  0.43 &  1.823 & $\pm$ & 0.035 &  $-$0.040 & $\pm$ & 0.034  &   229.7 & $\pm$ & 0.3   & (2)  \\
NGC 1818  &  05:04:13.64  &  0.40  &  $-$66:26:02.9  &  1.20 &  1.537 & $\pm$ & 0.044 &     0.046 & $\pm$ & 0.041  &   311.1 & $\pm$ & 3.9   & (4)  \\
NGC 1831  &  05:06:16.19  &  0.38  &  $-$64:55:08.8  &  0.43 &  1.714 & $\pm$ & 0.035 &  $-$0.014 & $\pm$ & 0.033  &   276.8 & $\pm$ & 0.2   & (2)  \\
NGC 1846  &  05:07:33.90  &  0.47  &  $-$67:27:43.2  &  0.51 &  1.745 & $\pm$ & 0.051 &     0.176 & $\pm$ & 0.054  &   239.2 & $\pm$ & 0.2   & (2)  \\
NGC 1850  &  05:08:45.33  &  0.66  &  $-$68:45:40.5  &  0.38 &  2.011 & $\pm$ & 0.028 &     0.119 & $\pm$ & 0.030  &   248.8 & $\pm$ & 0.3   & (2)  \\
NGC 1856  &  05:09:30.17  &  0.28  &  $-$69:07:43.6  &  0.24 &  1.795 & $\pm$ & 0.035 &     0.136 & $\pm$ & 0.035  &   270.9 & $\pm$ & 1.5   & (5)  \\
NGC 1866  &  05:13:38.63  &  0.38  &  $-$65:27:52.7  &  0.42 &  1.558 & $\pm$ & 0.030 &     0.177 & $\pm$ & 0.039  &   298.5 & $\pm$ & 0.4   & (6)  \\
NGC 1868  &  05:14:36.02  &  0.28  &  $-$63:57:15.0  &  0.25 &  1.725 & $\pm$ & 0.046 &  $-$0.038 & $\pm$ & 0.053  &   287.0 & $\pm$ & 1.0   & (5)  \\
NGC 1978  &  05:28:45.13  &  0.18  &  $-$66:14:11.9  &  0.20 &  1.813 & $\pm$ & 0.033 &     0.448 & $\pm$ & 0.041  &   293.1 & $\pm$ & 0.3   & (2)  \\
NGC 2108  &  05:43:57.00  &  0.48  &  $-$69:10:52.0  &  0.46 &  1.610 & $\pm$ & 0.033 &     0.728 & $\pm$ & 0.034  &   248.0 & $\pm$ & 3.4   & (3)  \\
NGC 2173  &  05:57:58.50  &  0.48  &  $-$72:58:42.6  &  0.55 &  1.978 & $\pm$ & 0.037 &     0.851 & $\pm$ & 0.038  &   237.4 & $\pm$ & 0.7   & (1)  \\
NGC 2203  &  06:04:42.50  &  0.50  &  $-$75:26:15.0  &  0.48 &  1.955 & $\pm$ & 0.033 &     0.841 & $\pm$ & 0.038  &   252.8 & $\pm$ & 0.2   & (2)  \\
NGC 2209  &  06:08:36.36  &  0.80  &  $-$73:50:14.6  &  0.81 &  1.852 & $\pm$ & 0.054 &     0.936 & $\pm$ & 0.053  &   251.2 & $\pm$ & 0.2   & (2)  \\
NGC 2213  &  06:10:42.19  &  0.35  &  $-$71:31:45.6  &  0.36 &  1.829 & $\pm$ & 0.046 &     0.988 & $\pm$ & 0.042  &   242.7 & $\pm$ & 1.2   & (1)  \\
\noalign{\smallskip}
\hline
\end{tabular}
\tablefoot{References: (1): \citet{Grocholski06}; (2): \citet{Song21}; (3): this work; (4): \citet{Marino18}; (5): \citet{Usher19}; (6): \citet{Mucciarelli11}.
}
\end{table*}

\subsection{Selection criteria\label{sec:selection}}

We applied several selection criteria in order to keep only well-measured likely cluster members in the input catalogues for the isochrone fitting routine. To select stars with good photometric as well as astrometric measurements, we applied the same quality cuts as in \citet{Niederhofer24}. 
These selections are based on the following diagnostic parameters provided by \texttt{KS2}: the quality-of-fit parameter (which indicates how well a source is fitted by the PSF model), the shape parameter \texttt{RADXS} (which describes how extended a source is with respect to the PSF model), the photometric RMS error, the isolation parameter of a source (defined as the fraction of flux within the PSF fitting aperture that comes from neighbouring sources, before neighbour subtraction), and the fraction of good measurements of a source with respect to the total number of detections. The astrometric selection criteria are based on the quality of the PM measurements and include the reduced $\chi^2$ of the PM fit, the fraction of data points of a source actually used for the determination of its PM, and the PM uncertainty.

We subsequently selected from the sample of well-measured stars those sources that likely belong to a given cluster, using the measured relative PMs of the stars. To this aim, we constructed a diagram of the relative 1D motions as a function of the $m_{\rm F814W}$ magnitude (as an example, see Fig.~\ref{fig:ngc1856_pm_sel} for the selection of cluster stars of NGC~1856). Since we are considering here the relative motions, stars belonging to the cluster have PMs close to zero, whereas field stars have larger motions, due to the larger velocity dispersion of the field stars and any offset in velocity between the cluster and field stars. From this diagram, we selected (by hand) the stars that follow the bulk of the cluster stars. 
We opted for this procedure over a fixed cut in PM to account for the varying PM uncertainties as a function of magnitude. A manual selection further allows for higher flexibility. Panel (c) in Fig.~\ref{fig:ngc1856_pm_sel} illustrates that for NGC~1866, which is one of the clusters most severely affected by field-star contamination, our PM selection yields a much cleaner CMD.
We note that usually the PM distribution of the field stars substantially overlaps with the PMs of the cluster stars, which results in the inclusion of field interlopers in the PM-selected catalogue at the 2 to 5 per cent level for the different clusters.
The risk of including field stars is higher at the magnitude level of the lower main-sequence, given the larger PM uncertainties there and the resulting larger overlap between cluster and field stars, but these faint stars are generally not used for isochrone fitting anyway (see Section~\ref{sec:isoc_fit}).
To refine our selection, we considered for our final catalogue only stars that are within 1--2 times (depending on the size of the cluster and the number of stars within the cluster/field) the effective radius of the clusters \citep[as determined by][]{Niederhofer24}.

\subsection{Differential reddening correction\label{sec:dr_corr}}
After we defined our input lists, we corrected the photometry of each cluster for the effects of differential reddening, following the steps described by \citet{Milone23a}. This method, which we briefly recapitulate below, takes into account simultaneously the photometric information from all available filters. As a first step, we determined for each star in the catalogue of a given cluster the temperature dependent extinction coefficients, A$_{\lambda}$, for each filter. To calculate A$_{\lambda}$, we created a list of extinction coefficients across a wide range of stellar effective temperatures. We assumed a standard \citet*{Cardelli89} reddening law with R$_{\rm V}=3.1$. Then, we overlaid a Bag of Stellar Tracks and Isochrones (BaSTI) model \citep{Hidalgo18, Pietrinferni21} over the CMD of the cluster that resembles the cluster sequences, from an initial run of the isochrone-fitting algorithm (see Section~\ref{sec:isoc_fit}). We estimated the effective temperature of each star as the temperature of the closest isochrone point in the CMD and linearly interpolated the grid of extinction coefficients to determine the A$_{\lambda}$ values for each star.

In the next step, we selected for each cluster a sample of reference stars for the determination of the local reddening. We chose main sequence stars below the turn-off and omitted stars along the binary sequence. Then, we constructed for each cluster different CMDs: $m_{\lambda} - m_{\rm F814W}$ vs $m_{\rm F814W}$, where $\lambda$ corresponds to all broadband filters with available observations (except for F814W) for a given cluster. Within each CMD, we then determined the fiducial line of the reference stars and
measured their distances from this line along the direction of the reddening vector. We did this for all CMDs and compared these distances with expectations from a grid of $E(B-V)$ values ranging from $-$0.3~mag to 0.3~mag (as conservative estimates) in increments of 0.001~mag (we also used negative values for $E(B-V)$, since the correction is relative to the mean reddening of the cluster). For each $E(B-V)$ value, we calculated the corresponding $\chi^2$ and chose the $E(B-V)$ value that provides the minimum $\chi^2$ as the best-fitting reddening for a given star. For clusters with observations in two filters, we constructed the CMD resulting from these two filters and estimated the reddening values of the reference stars in this single CMD from the offsets of these stars from the fiducial line in the direction of the reddening vector. 

In the last step, we determined the local reddening for each star in the photometric catalogues as the 2.5$\sigma$-clipped median value of the closest 75 reference stars, excluding the target star itself. To illustrate the effect of the differential reddening correction, Fig.~\ref{fig:ngc1751_dr} shows, using NGC~1751 as an example, a comparison between the un-corrected and corrected CMD of the cluster and the corresponding reddening map across the cluster field.

\section{Isochrone fitting\label{sec:isoc_fit}}

To derive homogeneous estimates of the ages, distances and metallicities of the clusters in our sample, we took advantage of the isochrone-fitting methodology developed for the CARMA project \citep[see][for a detailed description of the methods and functionalities]{Massari23}. This algorithm simultaneously determines the best-fitting values for age, [M/H], distance and $E(B-V)$ and assigns each value a robust uncertainty through a Bayesian statistical framework. Several studies within the CARMA project have already successfully applied the fitting code \citep[]{Aguado-Agelet25, Ceccarelli25, Niederhofer25}, and we briefly summarise the main steps of the procedure below. As theoretical models, we adopt isochrones from the newest release of the BaSTI stellar evolution library \citep{Hidalgo18, Pietrinferni21}. We built two separate grids of isochrones for the fitting of the young and intermediate-age clusters, respectively, to optimally sample the age ranges covered by these clusters. The first grid spans the age range from 15~Myr to 500~Myr (in steps of 5~Myr) that we used for fitting the young clusters, and the second grid covers ages going from 500~Myr to 4~Gyr (in steps of 50~Myr) that we used for the fitting of the intermediate-age clusters. For each age, we created models with metallicities [M/H] from $-$2.0~dex to 0.0~dex in increments of 0.01~dex. We emphasise here that we use solar-scaled models and our final results are in terms of the global metallicity [M/H], instead of the iron abundance [Fe/H]. This way, we avoid making any assumptions about the $\alpha$-element abundances of the individual clusters. It has been demonstrated \citep[e.g.][]{Salaris93, Cassisi04}, that in optical filter bands, the photometric difference between solar-scaled and $\alpha$-enhanced models at the same global metallicity is negligible. Thus, this approach is well-justified. To incorporate the effect of interstellar reddening into the fitting routine, we created a grid of temperature-dependent extinction coefficients in the various ACS/WFC and WFC3/UVIS filters, as described in Section~\ref{sec:dr_corr}.

From the input catalogues that we constructed as described in Section~\ref{sec:selection}, the isochrone fitting code selected stars within each CMD that follow the cluster sequence, i.e.\ excluding stars on the binary sequence, blue straggler stars and left-over field interlopers.
For this, the routine constructs the mean-ridge line of the stars in the CMD, and selects all stars within 1.5$-$2.0 times (depending on the photometric quality of the cluster) the local colour uncertainty from the ridge line. 
We also applied lower and upper cuts in magnitude to the cleaned catalogue of stars in order to prevent stars on the main sequence dominate the fitting. We optimised these limits individually for each cluster.

We performed the isochrone fitting using Gaussian priors for the distance modulus, reddening and metallicity. For the first two parameters, we adopted the values provided by \citet{Milone23a}. For the metallicity (initially assuming solar-scaled [$\alpha$/Fe] mixture), we used the spectroscopic measurements, along with the associated uncertainties from \citet{Grocholski06, Mucciarelli06, Mucciarelli08, Mucciarelli11, Marino18, Asa'd22} and \citet{Song21}. These values provide the most robust and reliable prior information for the fitting. Some clusters in our sample, however, do not have any spectroscopically determined metallicity. For these clusters, we use estimates derived photometrically by \citet{Milone23a}, with a broad dispersion of 0.1~dex, as a conservative estimate. We did not impose any prior on the ages of the clusters, but let the code freely explore the parameter space. 

For each cluster, we ran the code on all possible colour-magnitude combinations that can be constructed using the available filters that exist for that cluster. For any given CMD, we determined the best-fit solution and its uncertainties as the 50th, 16th and the 84th percentiles of the posterior distributions of the four free parameters, respectively. 
As an example, we show in Fig.~\ref{fig:ngc1978_isofit} the result of the isochrone fitting routine for NGC~1978. Shown are two different CMDs with the best-fitting isochrones (stars actually used for the fitting are highlighted in green in both CMDs), as well as the corresponding corner plots of the posterior probability distribution. Representative CMDs with the best-fitting isochrone models for the remaining clusters are presented in Fig.~\ref{fig:cmds}.
The final results in terms of age, metallicity, distance modulus and reddening for each cluster are calculated as the average values from the fits to the individual CMDs, and the associated overall uncertainties are derived accounting for the upper and lower limits of the inferred quantities from all fits.

\begin{table}\small
\centering
\caption{Results of the isochrone fits. \label{tab:isochrone_fits}}
\begin{tabular}{lccccc}%{@{}l@{ }c@{ }c@{ }c@{ }}c@{ }r@{ }}
\hline\hline
\noalign{\smallskip}
Cluster ID & [M/H] & E(B$-$V) & (m$-$M) & Age & Flag\\
& [dex] & [mag] & [mag] & [Gyr] & \\
\noalign{\smallskip}
\hline
\noalign{\smallskip}
NGC 1651   &  $-0.46^{+0.03}_{-0.03}$  &  $0.11^{+0.01}_{-0.01}$ &   $18.45^{+0.01}_{-0.01}$  &   $1.99^{+0.01}_{-0.01}$ &  g \\
\noalign{\smallskip}
NGC 1718   &  $-0.33^{+0.01}_{-0.01}$  &  $0.15^{+0.01}_{-0.01}$ &   $18.48^{+0.02}_{-0.02}$  &   $1.83^{+0.03}_{-0.03}$ &  g \\
\noalign{\smallskip}
NGC 1751   &  $-0.33^{+0.02}_{-0.02}$  &  $0.11^{+0.01}_{-0.01}$ &   $18.46^{+0.01}_{-0.02}$  &   $1.41^{+0.04}_{-0.04}$ &  g \\
\noalign{\smallskip}
NGC 1783   &  $-0.29^{+0.05}_{-0.04}$  &  $0.00^{+0.00}_{-0.00}$ &   $18.45^{+0.01}_{-0.01}$  &   $1.69^{+0.05}_{-0.06}$ &  g \\
\noalign{\smallskip}
NGC 1805   &  $-0.31^{+0.01}_{-0.01}$  &  $0.07^{+0.01}_{-0.01}$ &   $18.37^{+0.02}_{-0.04}$  &   $0.04^{+0.01}_{-0.01}$ &  u \\
\noalign{\smallskip}
NGC 1806   &  $-0.29^{+0.02}_{-0.03}$  &  $0.01^{+0.01}_{-0.01}$ &   $18.47^{+0.02}_{-0.03}$  &   $1.67^{+0.48}_{-0.06}$ &  g \\
\noalign{\smallskip}
NGC 1818   &  $-0.28^{+0.09}_{-0.13}$  &  $0.07^{+0.01}_{-0.01}$ &   $18.43^{+0.02}_{-0.04}$  &   $0.06^{+0.01}_{-0.01}$ &  g \\
\noalign{\smallskip}
NGC 1831   &  $-0.34^{+0.10}_{-0.10}$  &  $0.04^{+0.05}_{-0.05}$ &   $18.41^{+0.10}_{-0.10}$  &   $0.70^{+0.20}_{-0.20}$ &  e \\
\noalign{\smallskip}
NGC 1846   &  $-0.24^{+0.01}_{-0.01}$  &  $0.00^{+0.00}_{-0.00}$ &   $18.48^{+0.02}_{-0.02}$  &   $1.59^{+0.01}_{-0.01}$ &  g \\
\noalign{\smallskip}
NGC 1850   &  $-0.18^{+0.10}_{-0.11}$  &  $0.13^{+0.01}_{-0.01}$ &   $18.40^{+0.01}_{-0.01}$  &   $0.09^{+0.01}_{-0.01}$ &  g \\
\noalign{\smallskip}
NGC 1856   &  $-0.17^{+0.05}_{-0.08}$  &  $0.17^{+0.01}_{-0.01}$ &   $18.44^{+0.06}_{-0.10}$  &   $0.26^{+0.03}_{-0.02}$ &  g \\
\noalign{\smallskip}
NGC 1866   &  $-0.27^{+0.05}_{-0.03}$  &  $0.07^{+0.02}_{-0.03}$ &   $18.35^{+0.05}_{-0.04}$  &   $0.19^{+0.02}_{-0.01}$ &  g \\
\noalign{\smallskip}
NGC 1868   &  $-0.44^{+0.04}_{-0.04}$  &  $0.07^{+0.01}_{-0.01}$ &   $18.37^{+0.05}_{-0.05}$  &   $1.06^{+0.04}_{-0.05}$ &  u \\
\noalign{\smallskip}
NGC 1978   &  $-0.39^{+0.04}_{-0.02}$  &  $0.05^{+0.02}_{-0.01}$ &   $18.44^{+0.07}_{-0.02}$  &   $2.40^{+0.10}_{-0.31}$ &  g \\
\noalign{\smallskip}
NGC 2108   &  $-0.34^{+0.10}_{-0.10}$  &  $0.14^{+0.05}_{-0.05}$ &   $18.44^{+0.10}_{-0.10}$  &   $1.00^{+0.20}_{-0.20}$ &  e \\
\noalign{\smallskip}
NGC 2173   &  $-0.22^{+0.03}_{-0.02}$  &  $0.05^{+0.01}_{-0.01}$ &   $18.41^{+0.02}_{-0.03}$  &   $1.56^{+0.01}_{-0.01}$ &  g \\
\noalign{\smallskip}
NGC 2203   &  $-0.36^{+0.02}_{-0.01}$  &  $0.07^{+0.01}_{-0.01}$ &   $18.39^{+0.01}_{-0.01}$  &   $1.69^{+0.03}_{-0.03}$ &  g \\
\noalign{\smallskip}
NGC 2209   &  $-0.20^{+0.10}_{-0.10}$  &  $0.08^{+0.05}_{-0.05}$ &   $18.38^{+0.10}_{-0.10}$  &   $1.15^{+0.20}_{-0.20}$ &  e \\
\noalign{\smallskip}
NGC 2213   &  $-0.24^{+0.04}_{-0.04}$  &  $0.04^{+0.01}_{-0.01}$ &   $18.44^{+0.01}_{-0.01}$  &   $1.57^{+0.01}_{-0.01}$ &  g \\
\noalign{\smallskip}
\hline

\end{tabular}

\tablefoot{Flags: g: good fit; u: fit including the F336W filter; e: fit by eye.
}

\end{table}

The final results of the isochrone fitting for the 19 clusters in our sample are presented in Table~\ref{tab:isochrone_fits}. 
We note that for some clusters, the algorithm provided solutions in certain filter combinations that clearly miss the cluster sequence, especially the red giant branch. This often happened for CMDs combining the F438W and F555W filters. In these cases, the code often confuses the red clump with the red giant branch due to the small difference in colour. We excluded these fits from the calculation of the final results. We also did not use observations in the F336W filter for the fitting, since for UV wavelengths, the above-mentioned equivalency between solar-scaled and $\alpha$-enhanced models at the same global metallicity is not accurate anymore. There are, however, two clusters -- namely NGC~1805 and NGC~1868 -- for which only observations in the filters F336W and F814W exists. The isochrone fits for these two clusters (flagged with the label \textit{u} in Table~\ref{tab:isochrone_fits}) therefore might be biased and not as accurate. For three more clusters -- namely NGC~1831, NGC~2108 and NGC~2209 -- the routine was not able to find a robust solution. We thus decided to fit the CMDs of these clusters (flagged with the label \textit{e} in Table~\ref{tab:isochrone_fits}) by eye. We assigned conservative uncertainties for the estimated parameters of 0.1~dex for [M/H], 0.05~mag for $E(B-V)$, 0.1~mag for (m$-$M) and 0.20~Gyr for the age. We will include these five clusters in our further analysis but treat them as special cases with caution. Since the distance to these clusters is less certain, and might be biased, the reliability of their positions within the LMC, and thus the projection of their 3D velocity vectors in the reference frame of the LMC, is also reduced.  

%--------------------------------------------------------------------

\section{The kinematic structure within the LMC \label{sec:kinematic_structure}}

The distances to the clusters that we determined through the isochrone-fitting routine allow us -- together with the on-sky positions -- to obtain the 3D positions of the clusters. To collect the full 6D phase-space information (3D position and 3D velocities), we combine the PMs of the clusters with spectroscopic measurements of their LOS velocities from the literature (see Table~\ref{tab:clusters_params}). All but two clusters (namely NGC~1805 and NGC~2108) in our sample have existing LOS velocity measurements. For these two clusters, we used measurements from the \textit{Gaia} DR3 catalogue \citep{Gaia23} to derive estimates for their velocities. For NGC~1805, we found two stars within the effective radius of the cluster with measured LOS velocities that would also be consistent with being evolved cluster stars based on their photometry. These stars have velocities of 301.8~km\,s$^{-1}$ and 301.6~km\,s$^{-1}$ with uncertainties of 4.5~km\,s$^{-1}$ and 3.0~km\,s$^{-1}$, respectively. The weighted mean of these stars is 301.7~km\,s$^{-1}$ with a mean uncertainty of 3.8~km\,s$^{-1}$ \footnote{There is an old estimate for the LOS velocity of NGC~1805 from \citet{Fehrenbach74} who measured 390.0$\pm$10~km\,s$^{-1}$. This value, however, would result in a velocity of NGC~1805 close to the local escape velocity of the LMC \citep[as determined by][]{Boubert17}.}. For NGC~2108, we only found one such star (248.0$\pm$3.4~km\,s$^{-1}$). We will use these values as our best guesses for the clusters' LOS velocities during our further analysis. We note that both clusters are also in our separate sample with less accurate isochrone-fitting results. 

Since LOS velocities for the clusters are taken from different literature studies, they provide a heterogeneous set of measurements. To test whether this inhomogeneity would affect our final results in a significant way, we performed the following consistency check. For the clusters in our sample, we performed the analysis presented in this work again, assuming different literature determinations for their LOS velocities. Some clusters have already multiple measurements among the studies listed in Table~\ref{tab:clusters_params}. We complemented them with determinations from further studies \citep[][and the database of Structural Parameters of Local Group Star Clusters\footnote{\href{https://people.smp.uq.edu.au/HolgerBaumgardt/globular/lgclusters/parameter.html}{Structural
Parameters of Local Group Star Clusters}}]{Mucciarelli08, Mucciarelli14, Sakari17, Kamann23}. We found that different assumptions for the literature LOS velocities leave the overall kinematic structure of the clusters unaffected, making our results largely insensitive to the specific choice of the adopted LOS velocities. Moreover, the varying precisions of the different LOS measurements only marginally affect the final uncertainties of the velocity components within the LMC, since the latter are the combination of several measurement uncertainties.

\subsection{Velocity and coordinate transformation \label{sec:trafo}}

To analyse the dynamics of the clusters within the LMC, we first need to de-project their measured positions and velocities as seen from Earth into the reference frame of the galaxy (i.e. a frame where the LMC is at rest). For this, we follow the formalism developed by \citet{vanderMarel01b} and \citet{vanderMarel02} \citep[see also appendix~A of][for a detailed overview of the formalism]{Jimenez23}. Solving the transformation equations gives us the positions and velocities of the clusters within a right-handed orthogonal coordinate system, centred on the dynamical centre of the LMC. The system is oriented such that the X--Y plane is aligned with the disc of the LMC, and the X axis is along the line of nodes (the line where the LMC-disc plane intersects the sky plane). The transformations require a knowledge of the position, orientation and velocity of the LMC relative to an observer. We performed the de-projection adopting the following parameters: the coordinates of the LMC dynamical centre ($\alpha_0$, $\delta_0$)= (79.945\degr, $-$69.306\degr), the LMC centre-of-mass PM ($\mu_{\alpha}$cos($\delta$)$_0$, $\mu_{\delta,0}$)= (1.867, 0.314)~mas\,yr$^{-1}$, the inclination angle $i=33.5\degr$, and the position angle of the line-of-nodes $\Theta=129.8\degr$. These values are taken from \citet{Niederhofer22}. For the distance to the LMC, we assumed 49.59~kpc \citep{Pietrzynski19} and for the LOS velocity we adopt a value of 262.2~km\,s$^{-1}$ \citep{vanderMarel02}. We propagated the measurement uncertainties of the clusters trough the transformation equations and finally transformed the positions and velocities of the clusters within the LMC reference frame from the Cartesian coordinate system into cylindrical coordinates. These values are presented in Table~\ref{tab:clusters_lmc_params}. The exact values of these quantities are defined by the specific choice of the adopted parameters of the LMC. However, adopting other parameters from the literature does not significantly affect our results.

\begin{table*}\small
\centering
\caption{Positions and velocities of the young and intermediate-age star clusters projected into the reference frame of the LMC. The cluster NGC~1850 is the only cluster with a retrograde orbit (negative V$_{\phi}$) in our sample. \label{tab:clusters_lmc_params}}
\begin{tabular} {l r@{\,}c@{\,}l c  r@{\,}c@{\,}l r@{\,}c@{\,}l r@{\,}c@{\,}l r@{\,}c@{\,}l}
\hline\hline
\noalign{\smallskip}
Cluster ID
      &  \multicolumn{3}{c}{R} 
            & $\phi$
        & \multicolumn{3}{c}{Z} 
            & \multicolumn{3}{c}{V$_{\phi}$}
                & \multicolumn{3}{c}{V$_{R}$} 
                    & \multicolumn{3}{c}{V$_{Z}$} 
                    \\
&   \multicolumn{3}{c}{[kpc]}
    & [deg]
      & \multicolumn{3}{c}{[kpc]}
      & \multicolumn{3}{c}{[km\,s$^{-1}$]}
      & \multicolumn{3}{c}{[km\,s$^{-1}$]}
      & \multicolumn{3}{c}{[km\,s$^{-1}$]}
      \\
\noalign{\smallskip}
\hline
\noalign{\smallskip}
NGC 1651  &   2.55 & $\pm$ & 0.11  &   124.00 &     2.15 & $\pm$ &  0.18  &    64.2 & $\pm$ &   8.1  &     3.7 &  $\pm$ &   7.5  &    33.3  & $\pm$ &   5.4  \\ 
NGC 1718  &   2.94 & $\pm$ & 0.03  &   178.00 &  $-$0.02 & $\pm$ &  0.38  &   106.0 & $\pm$ &   7.8  & $-$70.1 &  $\pm$ &   9.6  &    22.2  & $\pm$ &   5.5  \\ 
NGC 1751  &   1.61 & $\pm$ & 0.09  &   133.69 &     1.17 & $\pm$ &  0.18  &    53.9 & $\pm$ &   7.5  &  $-$5.9 &  $\pm$ &   7.4  &    26.0  & $\pm$ &   4.7  \\ 
NGC 1783  &   3.35 & $\pm$ & 0.04  &   200.24 &  $-$0.01 & $\pm$ &  0.19  &    78.3 & $\pm$ &   6.8  &     4.2 &  $\pm$ &   7.8  &    12.5  & $\pm$ &   4.3  \\ 
NGC 1805  &   3.53 & $\pm$ & 0.15  &   218.39 &     1.41 & $\pm$ &  0.37  &   105.1 & $\pm$ &   7.0  &     0.4 &  $\pm$ &   7.2  &  $-$2.3  & $\pm$ &   5.2  \\ 
NGC 1806  &   1.79 & $\pm$ & 0.02  &   179.80 &     0.13 & $\pm$ &  0.38  &    28.6 & $\pm$ &   7.1  & $-$16.3 &  $\pm$ &   8.6  &    43.9  & $\pm$ &   4.8  \\ 
NGC 1818  &   2.90 & $\pm$ & 0.12  &   208.95 &     0.33 & $\pm$ &  0.38  &   112.8 & $\pm$ &   9.0  &     1.9 &  $\pm$ &   9.8  &  $-$2.9  & $\pm$ &   6.4  \\ 
NGC 1831  &   4.14 & $\pm$ & 0.71  &   218.15 &     0.19 & $\pm$ &  1.89  &    62.3 & $\pm$ &  13.3  & $-$23.4 &  $\pm$ &  14.1  &    15.5  & $\pm$ &   8.1  \\ 
NGC 1846  &   1.83 & $\pm$ & 0.06  &   192.53 &  $-$0.41 & $\pm$ &  0.38  &    17.0 & $\pm$ &  10.8  &    20.8 &  $\pm$ &  12.5  &    28.1  & $\pm$ &   6.9  \\ 
NGC 1850  &   1.18 & $\pm$ & 0.08  &   218.89 &     1.48 & $\pm$ &  0.18  & $-$11.1 & $\pm$ &   6.1  & $-$22.7 &  $\pm$ &   6.3  &    10.8  & $\pm$ &   3.8  \\ 
NGC 1856  &   0.69 & $\pm$ & 0.10  &   187.14 &     0.85 & $\pm$ &  1.12  &    40.0 & $\pm$ &   9.7  &     2.2 &  $\pm$ &  11.3  &     8.0  & $\pm$ &   6.7  \\ 
NGC 1866  &   4.07 & $\pm$ & 0.45  &   233.84 &     1.22 & $\pm$ &  0.92  &   101.7 & $\pm$ &   9.0  & $-$19.4 &  $\pm$ &   8.7  &     6.6  & $\pm$ &   5.2  \\ 
NGC 1868  &   5.07 & $\pm$ & 0.42  &   231.16 &     0.39 & $\pm$ &  0.94  &    56.1 & $\pm$ &  12.2  & $-$50.8 &  $\pm$ &  11.7  &    16.5  & $\pm$ &   6.4  \\ 
NGC 1978  &   2.75 & $\pm$ & 0.74  &   247.00 &  $-$0.69 & $\pm$ &  1.35  &    24.4 & $\pm$ &  12.0  &    13.3 &  $\pm$ &  12.0  & $-$24.7  & $\pm$ &   7.2  \\ 
NGC 2108  &   1.99 & $\pm$ & 0.85  &   314.47 &     0.00 & $\pm$ &  1.90  &    70.3 & $\pm$ &  14.0  & $-$27.0 &  $\pm$ &  14.0  &    14.8  & $\pm$ &   9.7  \\ 
NGC 2173  &   3.88 & $\pm$ & 0.04  &   359.53 &     1.88 & $\pm$ &  0.36  &    68.4 & $\pm$ &   7.7  &    17.9 &  $\pm$ &   8.9  &     3.3  & $\pm$ &   5.3  \\ 
NGC 2203  &   5.27 & $\pm$ & 0.03  &    10.17 &     3.27 & $\pm$ &  0.18  &    42.5 & $\pm$ &   6.8  &    23.9 &  $\pm$ &   7.8  &     2.2  & $\pm$ &   5.2  \\ 
NGC 2209  &   4.73 & $\pm$ & 0.22  &   357.60 &     2.66 & $\pm$ &  1.78  &    47.7 & $\pm$ &  18.0  &     4.2 &  $\pm$ &  17.2  &     2.5  & $\pm$ &  11.4  \\ 
NGC 2213  &   4.15 & $\pm$ & 0.03  &   347.66 &     0.51 & $\pm$ &  0.19  &    77.9 & $\pm$ &   8.6  &    14.0 &  $\pm$ &  10.4  &     3.2  & $\pm$ &   5.6  \\  
\noalign{\smallskip}
\hline

\end{tabular}

\tablefoot{$\phi$ denotes the position angle within the LMC disc, measured anti-clockwise from the positive X-axis. A positive tangential velocity V$_{\phi}$ follows the clockwise rotation pattern of the LMC.
}
\end{table*}

\subsection{Kinematics within the LMC\label{sec:kinematics}}

\begin{figure*}
\begin{tabular}{cc}
\includegraphics[width=0.95\columnwidth]{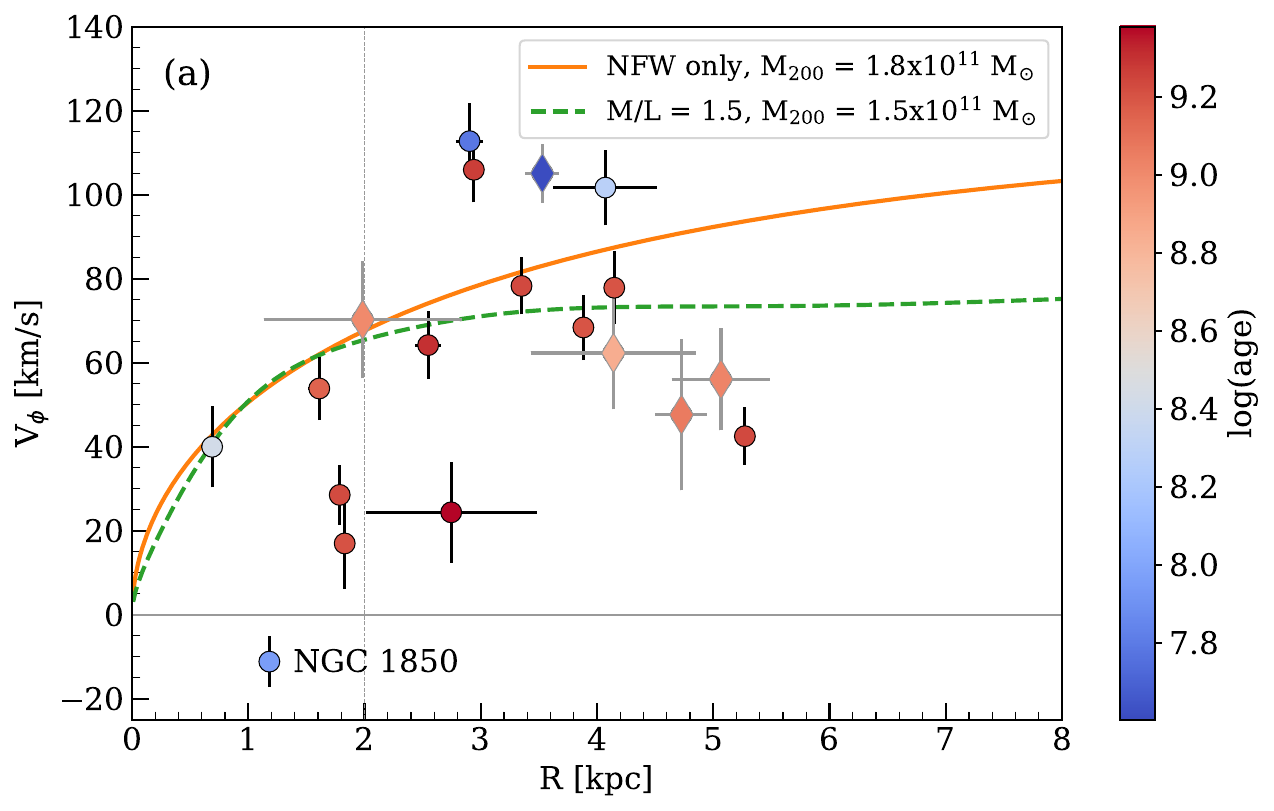} &
\includegraphics[width=0.95\columnwidth]{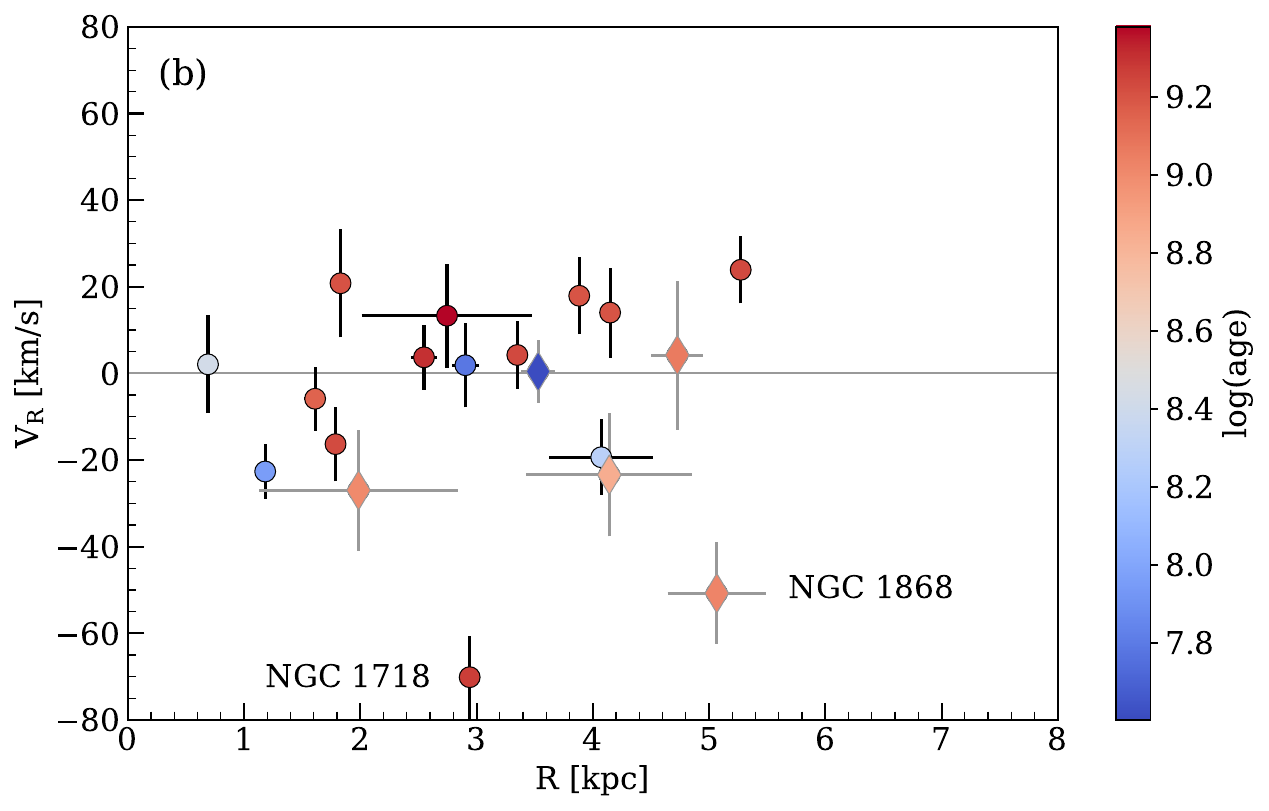} \\
\includegraphics[width=0.95\columnwidth]{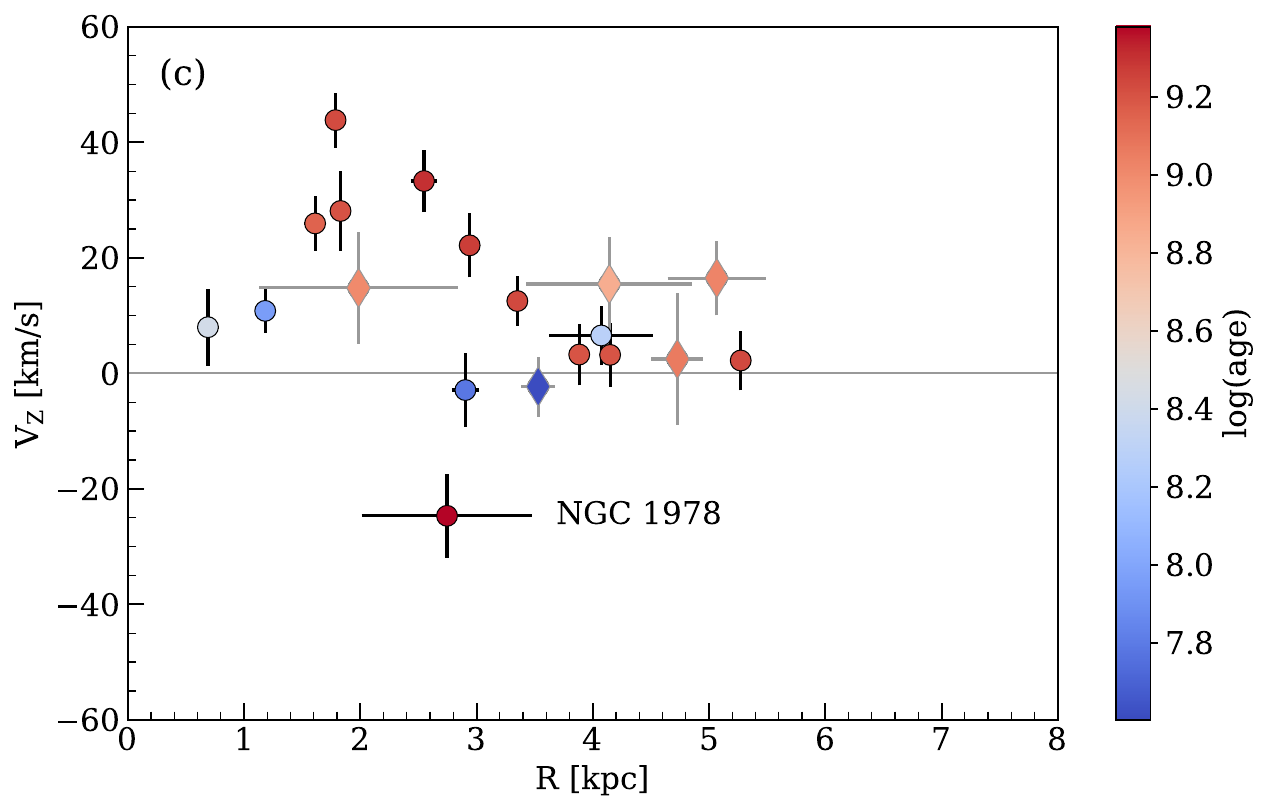} &
\includegraphics[width=0.95\columnwidth]{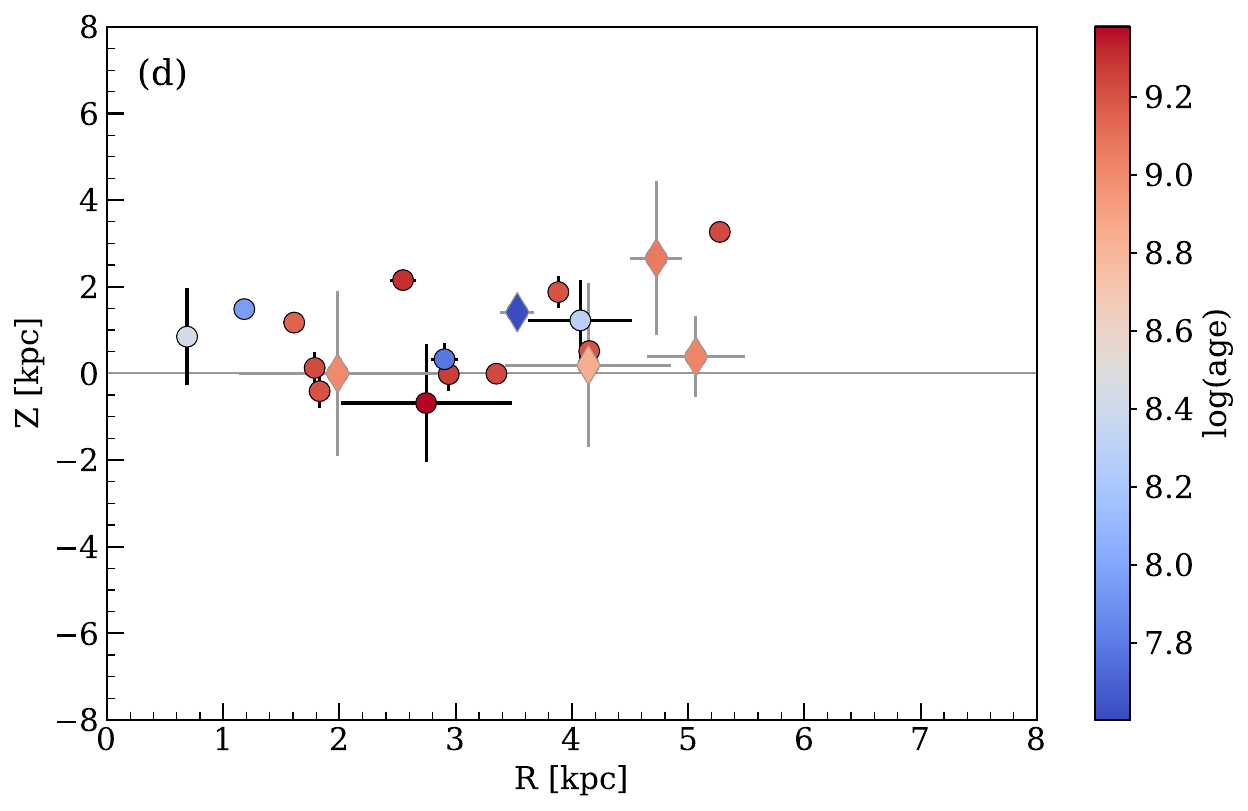} \\
\end{tabular}
\caption{Velocities and positions of the young and intermediate-age clusters within the LMC as a function of the radial distance from the galaxy's centre, R. In each panel, the clusters with less robust isochrone fits are denoted with diamond symbols and grey error bars. Panel (a) presents the tangential velocity V$_{\phi}$. Theoretical circular velocities resulting from a pure NFW profile (orange solid line) and a model composed of a dark matter halo and a stellar disc and bar (green dashed line) are also shown. The size of the stellar bar is indicated by the vertical dashed line. Panel (b) shows the radial velocity V$_{R}$; the out-of-plane velocity V$_{Z}$ is presented in panel (c); and panel (d) shows the vertical distance from the plane Z. In all panels, the clusters are colour-coded by the logarithm of their ages (in years). 
\label{fig:clusters_kinematics}
}
\end{figure*}

We start our analysis of the kinematic structure of the clusters by looking at the three velocity components (V$_{\phi}$, V$_{R}$, V$_{Z}$) as well as the vertical distance from the galaxy plane, Z, as a function of the cylindrical galactocentric radius R. This is illustrated in the four panels of Fig.~\ref{fig:clusters_kinematics}. Panel (a) shows the tangential velocity component (V$_{\phi}$) of the clusters, along with two model rotation curves of the LMC, as determined by \citet{Kacharov24}. The first model (orange line) follows a pure dark-matter halo with a spherical \citet*[][NFW]{Navarro97} mass profile and results from fitting axisymmetric Jeans dynamical models to stars in the \textit{Gaia} DR3 catalogue with measured 3D velocities.
This model suggests a virial mass of the LMC of M$_{200} = 1.81\times10^{11}$~M$_{\sun}$ \citep[see also][]{Kallivayalil13,Erkal19,Shipp21,Vasiliev21b}. The second model (green dashed line) follows a mass distribution that includes also the contribution from the stellar component of the galaxy, which is described as a triaxial bar and an axisymmetric disc. Here, \citet{Kacharov24} employed a Schwarzschild orbit superposition method, and fitted the LOS velocities of LMC stars. The model shown here assumes a mass-to-light ratio M/L of
1.5~M$_{\sun}$/L$_{\sun}$ and a total luminosity of the LMC of $1.3\times10^9$~L$_{\sun}$. While several clusters follow the predicted rotation curves from the models, the tangential velocities of most of the clusters lie below these curves. Only three young (NGC~1805, NGC~1818 and NGC~1866) and one intermediate-age cluster (NGC~1718) rotate faster than what is expected from the models. This deviation of the young clusters might be related to the resettling motion of the bar structure (see Section~\ref{sec:young_orbits}), whereas the orbit of NGC~1718 might have been influenced by a past encounter of the LMC with the SMC (see Section~\ref{sec:dyn_models}).
Beyond R$\sim$4~kpc, the rotational velocity of the clusters seem to decrease, a behaviour not predicted by the models. However, three of the four clusters (NGC~1831, NGC~1868, NGC~2203 and NGC~2209) that show this decline belong to the sample with the less-reliable isochrone fits, thus this feature might be spurious. The fact that shifts in the distances to the these clusters of more than 3~kpc would be required to make them consistent with the models, however, points towards a real feature. Moreover, NGC~2203, the cluster with the largest deviation from the model velocity curve among the four clusters, has a reliably determined distance, supporting a decrease in rotational velocity.
Also, \citet{Bennet22} found some hints of a declining rotation curve for clusters with R$\gtrsim$4~kpc. 
Within the region of the bar (indicated as a dashed vertical line in panel (a)), some clusters show very small V$_{\phi}$ values, suggesting that their orbits have been influenced by the stellar bar feature. The young cluster NGC~1850 represents an interesting case, since it is the only cluster in our sample with a negative tangential velocity, i.e.\ it is on a retrograde orbit. We will analyse the orbits of the young clusters in more detail in Section~\ref{sec:young_orbits} below. 

Looking at the plot of the radial velocity (V$_{R}$) as a function of R (displayed in panel (b) of Fig.~\ref{fig:clusters_kinematics}), shows that more than half of the clusters in our sample have radial velocities that are not consistent with zero, suggesting they are on non-circular, more elongated orbits. While for most clusters V$_{R}$ is less than $\sim$25~km\,s$^{-1}$, there are two outliers with very negative radial velocities, NGC~1718 (V$_{R}=-70.1$~km\,s$^{-1}$) and NGC~1868 (V$_{R}=-50.8$~km\,s$^{-1}$, however the isochrone fits were less robust for this cluster). This distribution in radial velocities is similar to what has been determined by \citet{Bennet22}, although we find an overall smaller scatter in the velocities. 

The out-of-plane velocity component (V$_{Z}$) and the vertical distance (Z) of the clusters as a function of R are illustrated in panels (c) and (d) of Fig.~\ref{fig:clusters_kinematics}. Our measurements suggest that most clusters in our sample have absolute vertical velocities smaller than $\sim$20~km\,s$^{-1}$. The most remarkable feature in this plot, however, is the group of intermediate-age clusters at R$\sim$2--3~kpc with large vertical velocities, up to $\sim$44~km\,s$^{-1}$ \citep[this feature is also evident in figure~3 of the study from][]{Bennet22}. This could indicate that the motion of these clusters have been disturbed in the vertical direction by a past encounter with the SMC. Interestingly, these clusters seem to be located very close to the LMC disc plane (see panel (d)). In panel (c), NGC~1978 stands out as an outlier, as it is the only cluster in our sample that has a significant negative out-of-plane velocity (V$_{Z}\sim-$25~km\,s$^{-1}$). We will discuss the impact of past LMC--SMC interactions on the motions of the clusters in detail in Section~\ref{sec:dyn_models}.

\subsection{Orbits of young clusters\label{sec:young_orbits}}

\begin{figure}
\centering
\includegraphics[width=0.9\columnwidth]{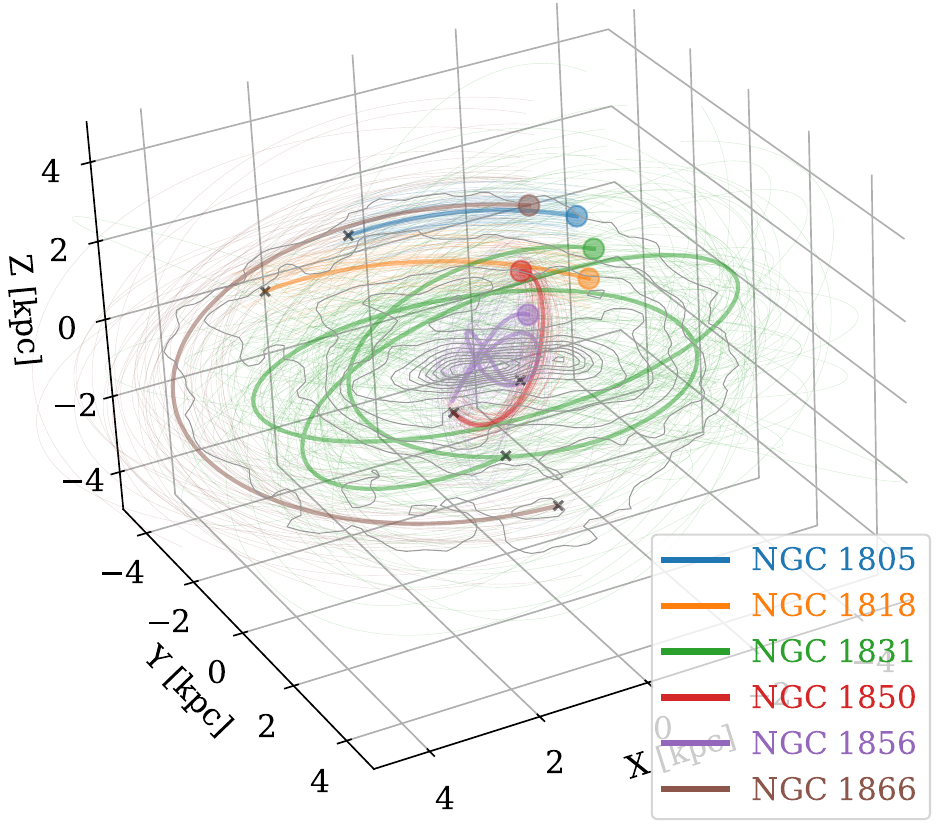} 
\caption{Reconstructed orbits of the six young ($<$1~Gyr) clusters within the LMC. The orbits have been integrated backwards for the inferred lifetime of the clusters. The coloured circles represent their current positions within the galaxy and the black crosses indicate the birth places of the clusters. For orientation, the shown contours follow the density of the LMC field star population. The plot is oriented such that North is approximately to the top and East approximately to the left of the image. 
\label{fig:young_clusters_orbits}
}
\end{figure}

A more illustrative way to look at the dynamics of the clusters is to reconstruct and study their orbits within the galaxy. To this end, we used the observed 6D phase-space information of the clusters and modelled their orbits within the LMC. As the potential of the LMC in which we place the clusters, we assumed the two LMC models by \citet{Kacharov24} shown in Fig.~\ref{fig:clusters_kinematics} (see Section~\ref{sec:kinematics}). 
As in \citet{Niederhofer25}, for clusters within the inner 2~kpc of the LMC (where the bar plays a significant role in influencing the dynamics of the clusters), we used the multi-component model, consisting of a spherical NFW dark matter halo, and a stellar component (disc and triaxial bar). For clusters located at larger radii, we used the pure NFW mass profile, which depicts the best representation of the outer potential of the LMC \citep[see also][]{Watkins24}. 

We will only perform the orbital analysis for the young ($<$1~Gyr) clusters within our sample, and will use the measured kinematics of the intermediate-age clusters to constrain the LMC--SMC interaction history (see Section~\ref{sec:dyn_models} below). We used the python package \texttt{galpy}\footnote{\url{http://github.com/jobovy/galpy}} \citep{Bovy15} for the reconstruction of the orbits and integrated the orbits backward for the inferred lifetimes of the young clusters (see Table~\ref{tab:clusters_params}). To assess the uncertainties in the resulting orbital parameters of the clusters, we created for each of the young clusters 500 realisations of the initial conditions by randomly drawing from Gaussian distributions centred around the measurements for their positions and velocities, and with standard deviations corresponding to the measurement uncertainties. Table~\ref{tab:orbit_params} provides an overview of the reconstructed orbital parameters of the young clusters.

The resulting orbits of the young clusters in the frame of the LMC, along with 50 of these random realisations, are illustrated in Fig.~\ref{fig:young_clusters_orbits}. The current positions of the clusters are indicated by coloured circles while the black crosses mark the inferred birth locations of the clusters. As expected, most of the young clusters rotate within, or very close to, the disc of the LMC.

NGC~1856 (purple lines in Fig.~\ref{fig:young_clusters_orbits}) is the innermost cluster in our sample. Since it is located close to the bar, its orbit is largely affected by this central structure. The cluster follows a box-like orbit \citep{Binney82} that is oriented approximately parallel to the bar. NGC~1856 seems to be confined within the inner regions of the LMC, with an apocentre distance of 1.36$\pm$0.59~kpc and a maximum vertical height above the disc of 1.16$\pm$0.74~kpc.

NGC~1805, NGC~1818, NGC~1831 and NGC~1866 (blue, orange, green and brown lines, respectively, in Fig.~\ref{fig:young_clusters_orbits}) orbit the LMC within the plane of the disc. Of the four clusters, NGC~1818 follows an orbit that is closest aligned with the disc of the galaxy (inclination angle of 9$\pm$5\degr relative to the plane of the LMC disc). The orbits of the other clusters show moderate inclination angles of 22$\pm$5\degr (NGC~1805), 15$\pm$13\degr (NGC~1831) and 19$\pm$10\degr (NGC~1866). While tracing back the orbits of the two youngest clusters in our sample, NGC~1805 (40~Myr) and NGC~1818 (60~Myr), we noted that the inferred birth places of both clusters are at larger galactocentric distances than their current positions. At present, NGC~1805 and NGC~1818 are located at R=3.53~kpc and R=2.90~kpc respectively, whereas they likely formed at R$_0$=4.35$\pm$0.31~kpc and R$_0$=4.85$\pm$0.55~kpc respectively. \citet{Piatti19b} determined the metallicities of a sample of young clusters within the LMC and found that for some of them (like NGC~1711 and NGC~1847, which are not within our sample of clusters), the metallicity does not match the one of the surrounding field population. The authors concluded that these clusters might have migrated from the places they have formed. Based on our results, it seems plausible that young clusters can be scattered from their birth places and radially migrate to different galactocentric radii. Indeed, \citet{Vijayasree25} recently studied the PM field of the LMC disc using data from the VISTA Survey of the Magellanic Cloud system \citep[]{Cioni11} and found evidence for streaming motions of stars North of the bar structure towards the centre of the LMC. This motion seems to be reflected in the orbits of NGC~1805 and NGC~1818, as well. Recently, \citet{Rathore25} investigated the response of the LMC bar to the recent SMC pericentre passage. They suggest that the bar is currently resettling within the potential of the LMC. The observed motions in the northern parts of the LMC are consistent with the predictions of the simulations from \citet{Rathore25}, and can thus be interpreted as a direct consequence of the bar response.

\subsubsection{NGC 1850}

An interesting exception is NGC~1850 (red-coloured line in Fig.~\ref{fig:young_clusters_orbits}), which follows an orbit that is considerably different from the ones of the other young clusters. NGC~1850 rotates almost perpendicular to the LMC disc, with an inclination angle of 75.33$\pm$6.16\degr relative to the disc plane. Additionally, the orbital plane is oriented such that the cluster is on a slight retrograde orbit. NGC~1850 has completed approximately half an orbit around the LMC since its birth. It has just passed apocentre, and is currently located 1.48$\pm$0.18~kpc above the disc plane. The birth location of NGC~1850 resulting from rewinding its orbit for 100~Myr lies very close to the LMC disc plane (Z$_0$=--0.24$\pm$0.41~kpc). Further, the inferred metallicity of NGC~1850 is typical for LMC clusters at the age of 100~Myr (see Fig.~\ref{fig:amr}). Thus, it seems unlikely that NGC~1850 formed out of gas that was accreted from the SMC onto the LMC. 

Since the velocity and position of NGC~1850 results in a rather unexpected orbit that cannot easily be explained by the latest LMC--SMC interaction, we now assess the robustness of this result and explore how sensitive the reconstructed orbit is to the assumed parameters of the cluster. 
Assuming literature PM \citep[from][]{Milone23a,Milone23b} and LOS velocity measurements \citep[from][and the database of Structural
Parameters of Local Group Star Clusters]{Kamann23}, we found that these different assumptions for the motion of NGC~1850 do not have a significant effect on its reconstructed orbit. The resulting orbits all follow very similar trajectories, with inclination angles with respect to the LMC disc between 75\degr and 85\degr, eccentricities between 0.26 and 0.33, vertical distances at the time of the birth of the cluster, Z$_0$, between --0.23~kpc and --0.31~kpc, as well as consistent apocentric distances of r$_{\mathrm{apo}}$ = 1.91. We then tested the dependence on the inferred distance to the cluster. Among literature studies, there exists a broad range of distance determinations for NGC~1850, indicating a large overall uncertainty. For our tests, we adopted the following measurements for the distance modulus: 18.35~mag \citep{Bastian16, Milone18}, 18.38~mag \citep{Milone23a} and 18.45~mag \citep{Correnti17, Yang18}. The range of these measurements correspond to physical distances between 46.77~kpc and 48.98~kpc. 
Distance moduli smaller than the one inferred by us (18.40~mag) have no significant effects on the overall shape and orientation of the final orbit. Adopting a distance modulus of 18.38~mag would result in a consistent (with respect to the one resulting from our determined distance modulus)  inclination angle of 78$\pm$5.75\degr, but a larger maximum vertical distance from the LMC disc of 1.84$\pm$0.17~kpc, and a larger apocentric distance of 2.29$\pm$0.15~kpc. A value of 18.35~mag would put NGC~1850 on an even more extreme orbit with a maximum vertical distance from the disc plane of 2.39$\pm$0.20~kpc and an apocentric distance of 2.89$\pm$0.17~kpc. However, adopting 18.45~mag for the distance modulus changes significantly the resulting orbit. This distance would place NGC~1850 closer to the disc plane (Z=0.55~kpc), and put it on a radial orbit along the bar structure (similar to the orbit of NGC~1856), oscillating through the disc plane. These tests suggest that the shape of the orbit of NGC~1850 is very sensitive to the assumed distance from the LMC disc plane and small uncertainties lead to relatively large differences in the final orbit. 
Although our isochrone fitting method provides homogeneous and precise distances for the clusters, the adopted distance and orientation for the LMC disc have not been estimated with the same methodology. Thus, the relative distance between the LMC and NGC~1850 might suffer for additional systematics. Moreover, young clusters have less populated and irregular upper CMDs, making their distance estimates possibly more uncertain. 
We note that also the shape of the orbit of NGC~1856 depends on the assumed distance. If adopting a distance modulus of 18.32~mag, as determined by \citet{Milone23a}, which would place the cluster about 3.6~kpc above the disc, would result in a wide orbit that is highly inclined with respect to the LMC disc. 

For an alternative scenario, we followed the simplistic assumption that all young clusters are located right in the disc plane at their respective positions, a likely scenario for young clusters. We reconstructed again the orbits for this configuration and inspected them. The orbits for all young clusters, except for NGC~1850, are very similar to their original orbits, but now moving very close to the disc plane. We note that the shapes of the orbits were not affected significantly, thus our above discussion would still be valid in this scenario. The most significant change in the orbits is only for NGC~1850, which is now on an extreme radial orbit moving along the LMC bar (see Fig.~\ref{fig:young_cluster_orbits_disc}). 

In summary, we found the orbit of NGC~1850 is very sensitive to the assumed distance to the LMC disc. If we take our inferred value of the distance modulus for NGC~1850 and the literature values for the distance and orientation of the LMC disc as face value, our results suggest a peculiar orbit for the cluster, orbiting the galaxy with a large inclination angle. Alternatively, if we assume that all young clusters reside in the disc (resulting in a distance modulus for NGC~1850 of 18.48~mag), then NGC~1850 follows the LMC bar on a highly eccentric orbit.

%--------------------------------------------------------------------

\subsection{Searching for accreted clusters in the LMC}

\begin{figure}
\centering
\includegraphics[width=1.0\columnwidth]{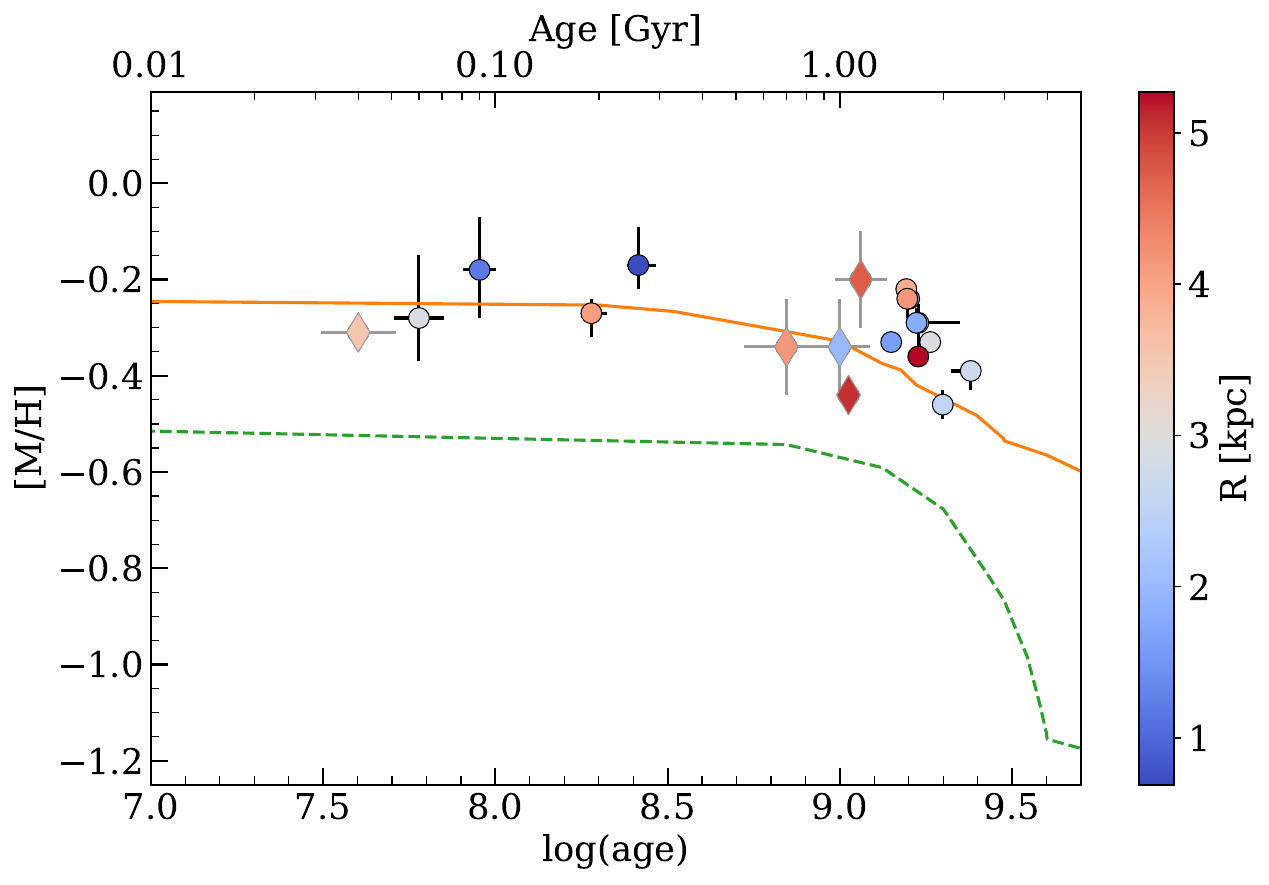} 
\caption{Age-metallicity relation of the young and intermediate-age clusters studied in this work. The clusters are colour-coded according to their cylindrical galactocentric distance R. Clusters with less robust isochrone fits are illustrated with diamond symbols and grey error bars. Also shown are analytical models of the chemical evolution of the LMC (orange solid line) and SMC (green dashed line), assuming a bursting star formation rate \citep{Pagel98}.
\label{fig:amr}
}
\end{figure}

In this section, we use the collection of kinematic data, combined with the inferred ages and metallicities of our sample of clusters, to search for any evidence of clusters that might have been stripped from the SMC and accreted by the LMC during the last close encounter of the two galaxies. \citet{Olsen11} studied the LOS dynamics of the LMC disc using a sample of 6\,000 giant stars, and identified a population of about 5\% of the stars as outliers that are apparently on retrograde orbits with respect to the direction of the bulk rotation of the LMC disc. This kinematic signature can be interpreted as actual counter-rotating stars, or stars that reside within a rotating disc that is inclined compared to the main LMC disc plane. Regardless of interpretation, as these stars also show lower metallicities than the average LMC disc population, \citet{Olsen11} concluded that these outliers have been accreted from the SMC onto the LMC. 

Based on LOS velocity measurements, the intermediate-age clusters NGC~1806 and NGC~1846 would belong to the outlier population of apparently counter-rotating stars. However, thanks to the full 6D phase space information collected in this study, we are able to assess their 3D kinematics within the LMC and establish the nature of these objects. We found that both clusters have positive tangential velocities, meaning that they follow the main rotation pattern of the LMC and are not on retrograde orbits (see Table~\ref{tab:clusters_lmc_params} and Fig.~\ref{fig:clusters_kinematics}). We note, however, that the rotational velocity of the two clusters, V$_{\phi}$ = 28.6$\pm$7.1~km\,s$^{-1}$ (NGC~1806) and V$_{\phi}$ = 17.0$\pm$10.8~km\,s$^{-1}$ (NGC~1846), is much smaller than predicted by the theoretical circular velocity curves at the location of the clusters. Both NGC~1806 and NGC~1846 have moderate velocities in the radial direction (V$_{R}$ = $-16.3\pm$8.6~km\,s$^{-1}$ for NGC~1846 and V$_{R}$ = $-20.8\pm$8.6~km\,s$^{-1}$ for NGC~1846) and are located close to the disc of the LMC. However, the two clusters show considerable vertical out-of-plane velocities (V$_{Z}$ = $-43.9\pm$4.8~km\,s$^{-1}$ for NGC~1846 and V$_{Z}$ = $-28.1\pm$6.9~km\,s$^{-1}$ for NGC~1846). These kinematic signatures of NGC~1806 and NGC~1846 do not provide any clear evidence for the clusters being accreted from the SMC. Rather, the orbits of the two clusters might have been disturbed in the past by the encounters with the SMC (see Section~\ref{sec:dyn_models}). Further, based on the kinematics, the other clusters in our sample do not show any peculiarities that can be seen as clear signs of an ex-situ origin.

Since the SMC is on average more metal-poor than the LMC \citep[e.g.][and references therein]{Choudhury20, Choudhury21, Grady21, Omkumar25}, star clusters that originate from the smaller galaxy can also reveal themselves based on their chemical composition. By comparing predicted age-metallicity relations of massive star clusters from the E-MOSAICS simulation with observed properties of LMC and SMC star clusters, \citet{Horta21} showed that clusters follow the properties of their host galaxies. Thus, star clusters formed within galaxies of different masses and star-formation histories follow distinct age-metallicity relations.  
We thus now focus on the ages and metallicities of the clusters that we derived in Section~\ref{sec:isoc_fit}. In Fig.~\ref{fig:amr}, we present the positions of the 19 studied clusters in the age-metallicity space. Also shown as an orange solid line is an analytical model of the chemical evolution of the LMC, assuming a bursting star formation rate \citep{Pagel98}. It is immediately evident that the overall observed age-metallicity relation of the clusters closely follows the analytical model of the LMC. The intermediate-age clusters reproduce well the increase in metallicity at $\sim$2~Gyr (log(age)$\sim$9.3). This increase is likely caused by a burst in the star formation rate at these times, as recovered by \citet{Mazzi21}. Further, Fig.~\ref{fig:amr} also displays the model from \citet{Pagel98} of the SMC as a green dashed line. This relation follows an enrichment history distinct from the one of the LMC, with a period of rapid enrichment about 2--3~Gyr ago and an overall lower metallicity than the LMC. As can be seen from the figure, all of the clusters studied in this work are located significantly above the SMC sequence, suggesting they are all genuine LMC member clusters. 
Analysis of Strömgren photometry of 110 LMC star clusters by \citet{Narloch22} confirms that the bursty star formation of \citet{Pagel98} is appropriate for LMC clusters, and that clusters from the LMC and SMC follow distinct chemical enrichment histories.

Taken everything together, 
we found no clear-cut evidence that a cluster within our sample has has been accreted from the SMC during a previous interaction event.

%--------------------------------------------------------------------

\section{Comparison with dynamical models\label{sec:dyn_models}}

\begin{figure*}
\centering
\includegraphics[width=1.5\columnwidth]{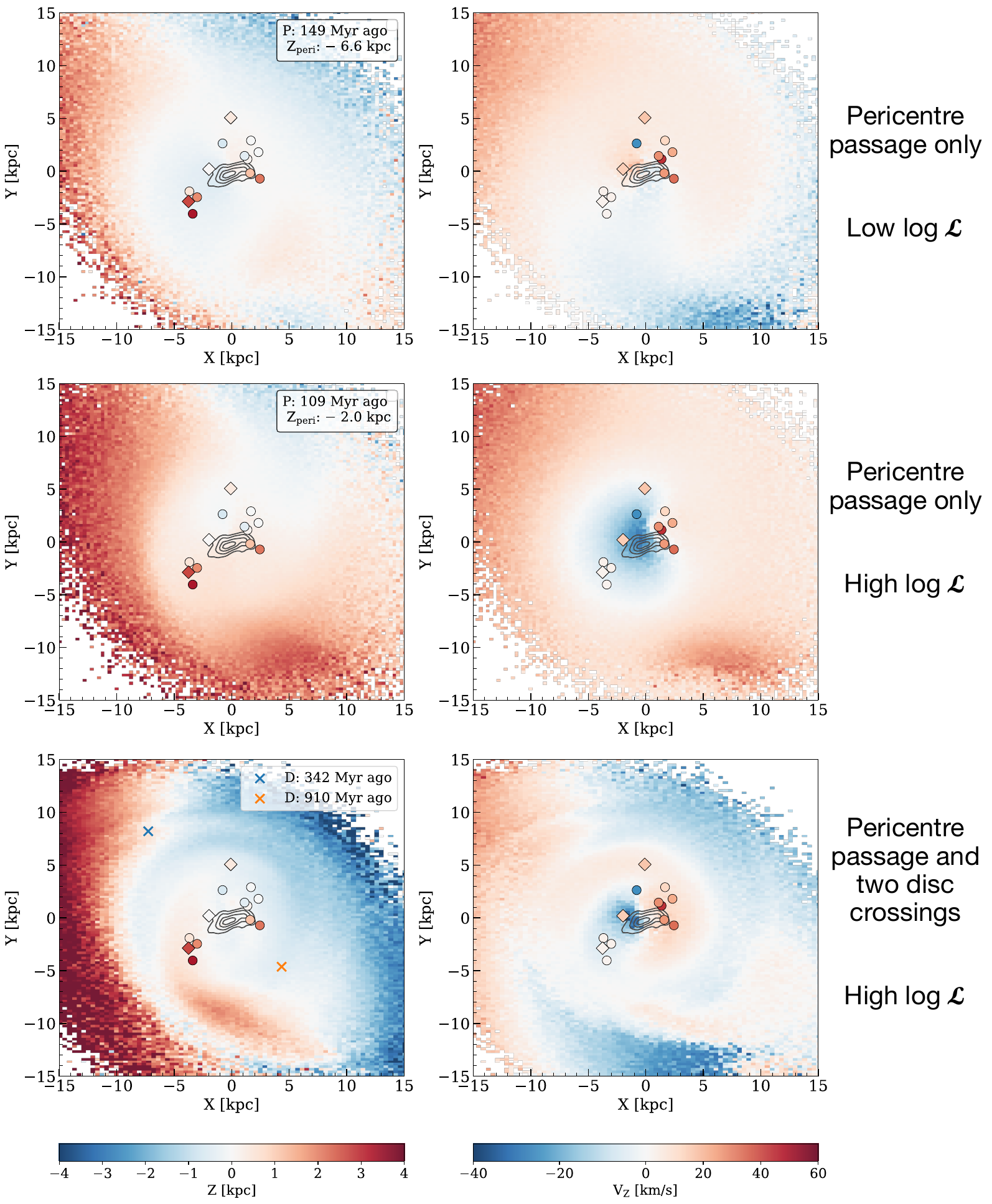} 
\caption{Predictions for three individual model realizations with different LMC--SMC orbital histories. The left column shows for each model the vertical out-of-plane distance (Z) while the right column presents the vertical out-of-plane velocity (V$_{Z}$). The rows, from top to bottom, show the following model realisations for which the LMC experiences different interaction events with the SMC: only an SMC pericentre passage about 150~Myr ago with an impact parameter Z$_{\mathrm{peri}}=-$6.6~kpc; only an SMC pericentre passage, but now more recently, about 110~Myr ago with a smaller impact parameter of only Z$_{\mathrm{peri}}=-$2~kpc; two events of the SMC crossing the disc plane of the LMC at $\sim$340~Myr and $\sim$910~Myr ago, in addition to the SMC pericentre passage of the SMC $\sim$150~Myr ago. In the bottom-left panel, the locations of the disc crossings are indicated with a blue and orange cross. The intermediate-age clusters in our sample are indicated by coloured circles, where the colour represents the measured out-of-plane distance Z (left panels), and the out-of-plane velocity V$_{Z}$ (right panels), respectively. Clusters with less robust isochrone fits are illustrated with diamond symbols. The black contours shown in all panels follow the stellar density of the LMC bar structure. 
\label{fig:model_compar}
}
\end{figure*}

\subsection{Model set-up}

In the final section of this study, we attempt to interpret the derived kinematic and structural properties of the sample of intermediate-age clusters within the LMC in terms of the interaction history of the LMC with the SMC. To do so, we compare our results to a suite of simple dynamical simulations that model the combined effects of the SMC and the Milky Way on the LMC for a set of different orbital histories. These models were initially developed to interpret the kinematic substructures in the outskirts of the LMC and were presented in detail by \citet{Cullinane22a, Cullinane22b}. We provide here a brief description of the basic model parameters and also discuss important limitations and caveats of the simulations. 
The LMC is modelled as a system of test particles orbiting a particle sourcing a rigid potential that consists of two components: a dark matter halo that follows a \citet{Hernquist90} profile with a total mass of $1.5\times10^{11}$~M$_{\sun}$ and a scale radius of 20~kpc, as well as an exponential stellar disc with a mass of $2\times10^{9}$~M$_{\sun}$, a scale radius of 1.5~kpc, and a scale height of 0.4~kpc. 
The SMC is modelled as a particle sourcing a pure Hernquist dark matter halo with a mass of $2.5\times10^{9}$~M$_{\sun}$ and a scale radius of 0.043~kpc, with no test particles placed within the potential. The model for the Milky Way potential is set up similar to the \texttt{MWPotential2014} \citep{Bovy15} and consists of three components; namely a stellar disc, bulge, and dark matter halo. All three galaxies in the simulation were treated as particles sourcing the associated potentials, which allows us to account for the reflex motion of the Milky Way in response to the LMC. Note that the model accounts for the effects of dynamical friction between the motions of the LMC and the Milky Way, but not between the LMC and the SMC. 

The simulations start by initialising the LMC and the SMC with their present-day positions and velocities. Then, their orbits were rewound back for 1~Gyr in the presence of each other and the Milky Way. The disc of the LMC was aligned such that its orientation matches the values as derived by \citet{Choi18}, and that orientation was held fixed during the simulation due to the rigid nature of the assumed potential. This sets the initial conditions for the models. The LMC disc is then initialised with $\sim2.5\times10^6$ tracer particles using the \texttt{AGAMA} software package \citep{Vasiliev19} to account for the velocity dispersion of the LMC disc and the thickness of the disc. Finally, the system is evolved forward to the present time. 

For the present study, we use the 'base-case' suite of models. For this set of models, 100 individual realizations of the initial conditions were run, sampling from literature values and their associated uncertainties for the present-day distances and velocities of the LMC and SMC (see Table~\ref{tab:model_params}).
These realisations result in a variety of different possible orbital histories and interaction events for the Clouds. It is worth noting that these different runs result all in consistent orbits of the Clouds around our Galaxy, with the two galaxies just past their first percentric passage around the Milky Way. However, the orbit of the SMC around the LMC can vary significantly between the realisations, especially for times later than 250~Myr ago. In all models, the SMC had a recent close pericentre passage around the LMC about 150~Myr ago. 
This is in agreement with the results from other studies \citep[e.g.][]{Zivick18, Choi22} that predict a close encounter of the SMC with the LMC about 150~Myr ago.
These passages occurred significantly below the disc of the LMC, with Z$_{\rm peri}=-6.8\pm2.6$~kpc. About half of the realisations show an SMC disc crossing about 400~Myr ago at a broad range of in-disc distances ($\sim$20--40~kpc). Additionally, another 10 per cent of the models show another disc-crossing event at $\sim$900~Myr ago, although this crossing would occur within a larger number of realisations if the models were rewound beyond 1~Gyr.

\subsection{Caveats}

This suite of models allows to explore a large variety of possible orbital histories of the Clouds; however, due to the simplicity of the simulations, the models have some important limitations. In particular, the use of rigid potentials can have significant effects on the evolution of the system. For one, the dark matter halos of the galaxies do not deform as a response to the gravitational interactions, potentially affecting the orbits of the galaxies. Additionally, the test particles in the LMC are not able to directly affect each other (i.e.\ there is no self-gravity), thus the potential of the disc also does not change in shape and orientation. This will limit the effect of perturbations induced in the LMC disc as responses to interactions. 

As discussed by \citet{Cullinane22a}, these limitations will have the most significant effects for close collisions between the Clouds, where the SMC crosses the LMC disc plane at small galactocentric radii. Such interactions can introduce additional asymmetries in the disc, such as offsets between the dynamical centre of the disc and the bar, or density waves and spiral arms in the disc. These simple models will not be able to fully reproduce such effects. 

Finally, it is important to note that the models were originally designed to study dynamic features within the outer regions of the LMC. To save computation time, the models therefore only include test particles with apocentres $>$7~kpc. This means that the density of particles within the inner regions, where most of our studied clusters are at present, is significantly reduced. Thus, the models might not be able to fully capture features induced by the different interaction events within the central parts of the LMC and the results should not be over-interpreted. Nevertheless, the models provide valuable explorations of the possible effects the various LMC--SMC interactions can have on the dynamical and structural properties of the clusters within the LMC. Given these limitations associated with the simple models, we will provide here only qualitative comparisons with our observations.

\subsection{Comparison with observations}

The models provide, for each realisation, the following mock observables of each test particle at the present time: the RA and Dec position on the sky, the distance and velocity along the LOS, as well as the PMs in RA and Dec direction. We used the formalism described in Section~\ref{sec:trafo} to transform these model observations into the reference frame of the LMC for a direct comparison with the properties of the clusters. Since the models assume for the geometry and orientation of the LMC disc the values as determined by \citet{Choi18}, we also use these parameters to de-project the positions and velocities of the model particles. 

For the comparison of the models with our observations, we concentrate on two quantities: the vertical distance from the LMC disc plane, Z, and the vertical out-of-plane velocity, $V_Z$. Due to the absence of tracer particles with apocentres less than 7~kpc, we refrain from using the tangential velocity, V$_{\phi}$, and radial velocity, V$_R$, for our comparisons, since these quantities will not provide useful constraints. To select from the 100 realisations those simulations that best reproduce our observations we follow a similar approach to the one used by \citet{Navarrete23}, employing a log-likelihood (log~$\mathcal{L}$) method. For each cluster, we selected the 75 spatially closest tracer particles (as a compromise between statistics and spatial resolution) in a given realisation and determined their median Z and $V_Z$ values. We then calculated the combined log~$\mathcal{L}$ for measuring these observables for all clusters given the present model.
We performed this step for all 100 realisations. 

To illustrate the qualitative effects of the different interaction histories on the LMC and their relations to the measured properties of our sample of star clusters, we show in Fig.~\ref{fig:model_compar} spatially binned maps for three different model realisations. For each of these realisations we show the median vertical out-of-plane distance Z (left panels) and median out-of-plane velocity V$_Z$ (right panels). Also shown as coloured circles are the positions of the studied clusters, colour-coded according to the corresponding quantities, using the same scale as for the maps. The top row of Fig~\ref{fig:model_compar} shows one of the model instances with a low log~$\mathcal{L}$, i.e.\ it poorly resembles the observations. In this realisation, the LMC only experiences the most recent SMC pericentric passage that occurred about 150~Myr ago, with the closest approach at Z$_{\rm peri}=-6.6$~kpc below the plane of the LMC. While this realisation is able to qualitatively reproduce the overall observed vertical velocity structure of the clusters (although we measure larger velocities than predicted by the models), there are discrepancies in the distances perpendicular to the disc plane. The clusters located closest to the centre have vertical displacements close to zero, as suggested by the models, however, the group of clusters located at X$\sim$--3~kpc and Y$\sim$--3~kpc are displaced consistently above the disc, while the model predicts a perturbation towards negative Z values. 

The middle and bottom panels in Fig.~\ref{fig:model_compar} show the two model instances with the largest log~$\mathcal{L}$ values, i.e.\ they best resemble the observed cluster properties. Similar to the model in the top panels, the realisation shown in the middle panels only experiences the most recent encounter with the SMC. In this case, however, the passage of the SMC happened more recently -- at about 110~Myr ago, at a much closer distance below the disc (Z$_{\rm peri}$ of only $-$2.0~kpc). This close interaction had more severe effects on the kinematics of the disc, as is evident from the maps. There is a reasonably overall good agreement between this model instance and the observations. The simulation is able to reproduce the displacement towards positive Z values of the clusters South-East of the LMC centre. Similarly, the predicted vertical velocity structure broadly follows the observations, with the clusters to the South-East having $V_Z$ close to zero and the clusters towards the West of the LMC centre moving with positive vertical velocities. This model is also able to reproduces the observed negative $V_Z$ of NGC~1978 (located at X$\sim$--1~kpc and Y$\sim$3~kpc). Based on these qualitative comparisons, the observed properties of the clusters would support a scenario of a recent close LMC--SMC interaction. However, this possibility is not supported by other studies. Specifically, \citet{Choi22} studied the kinematic pattern of the LMC disc based on \textit{Gaia} DR3 PMs and compared it to numerical simulations, with the level of disc heating they measured largely excluding an interaction of the two galaxies with small impact parameters $\sim$110~Myr ago. 

The bottom two panels in Fig.~\ref{fig:model_compar} show a realisation in which the LMC experiences two disc crossings of the SMC, about 340 and 910~Myr ago, besides the recent SMC pericentre passage $\sim150$~Myr ago. The predicted locations where the SMC crossed the disc are indicated in the bottom-left panel as blue and orange crosses, respectively. The orbital history of this model also reproduces reasonably well the observations. In terms of the vertical distance from the disc plane, the clusters near the centre are very close to the plane, and the displacement toward positive Z-values for the group of four clusters towards the South-East is reproduced by the models. Only the two clusters at the West end of the bar structure (NGC~1651 and NGC~1751), located above the disc, are not predicted by the model. The overall vertical velocity structure of the clusters is also in good agreement with the predictions from the model. The vertical velocities of the four clusters in the South-East are almost zero, while the clusters towards the West of the centre move with positive in-plane velocities. This model realisation, as the previous one, predicts a negative vertical velocity feature near the centre of the galaxy, which would be in agreement with the motion of NGC~1978 (X$\sim$--1~kpc and Y$\sim$3~kpc); however, the cluster seems to be somewhat displaced with respect to this feature. In contrast, NGC~2108 (located at X,Y$\sim$--2~kpc,$\sim$0~kpc) for which the model suggests a negative velocity shows a positive vertical motion. It is worth noting here that the kinematics of NGC~2108 and the two clusters at the West end of the bar could potentially be impacted by the bar, especially in a scenario where the bar was rotating in the past \citep{Jimenez25}. This influence by the bar is not captured in the models. We verified that these clusters did not impact our selection of the best-fitting models by performing the log~$\mathcal{L}$ again excluding these clusters.

Overall, our qualitative comparisons with this suite of simple models favours a scenario in which the LMC experiences two disc crossings of the SMC within the last Gyr, besides the most recent pericentric passage of the SMC about 150~Myr ago (the model with a recent close SMC pericentre passage is excluded by the measured level of disc heating in the LMC). This result is in agreement with the findings by \citet{Cullinane22b} and \citet{Navarrete23}. These two studies analysed the kinematics of different substructures within the outskirts of the LMC and used the same set of simple simulations to interpret their results. Both studies found that the observed kinematic structures in the outer regions of the LMC can be explained by at least one additional disc crossing of the SMC. \citet{Navarrete23} additionally expanded the models to cover the age range up to 2~Gyr ago and found the best agreement between models and observations for a scenario with three disc crossings where the last one happened at 1.9~Gyr ago. Although the results obtained for our sample of clusters and the findings for the outer LMC regions lead to the same conclusions, 
we emphasise again that due to the simplistic nature of the used models, the simulations are not able to fully capture all effects induced by the interactions of the galaxies. Thus, more detailed models, especially for the inner regions, would be required to definitively constrain the complex interaction history of the Clouds.

%--------------------------------------------------------------------

\section{Summary and Conclusions}\label{sec:conclusions}

In this work, we studied the kinematic structure of a sample of 19 young and intermediate-age LMC star clusters. We derived homogeneous estimates for their ages, distances and metallicities, and determined their full 3D motions, resulting from multi-epoch \textit{HST} PMs and literature LOS velocities. This allowed us to reconstruct the positions and velocity components of the clusters in the frame of the LMC. 

As expected, young clusters ($<$1~Gyr) orbit close to the disc plane of the galaxy. A remarkable exception is NGC~1850. Our derived distance to the cluster would put it on a highly inclined, slightly retrograde orbit. Additional tests, however, suggest that resulting orbit is very sensitive to the exact distance of cluster to the LMC disc plane. When assuming a distance that would put the cluster within the disc plane, NGC~1850 would follow a highly eccentric orbit along the LMC bar structure. The orbits of young clusters that formed North of the LMC centre show high tangential velocities and elongated orbits, suggesting they formed at larger galactocentric distances. These features could be a consequence of the resettling motion of the LMC bar. 

From the combined information coming from the orbital properties of the clusters and their location in the age-metallicity space, we found no evidence for clusters in our sample that might have been accreted onto the LMC from the SMC during one of their last encounters, or have been formed out of stripped, low-metallicity material from the SMC. 

In the last part of the paper, we compared the inferred positions and velocities of the intermediate-age clusters with a suite of simple dynamical models, to interpret our results. A scenario in which the SMC crossed the LMC disc plane twice (about 300 and 900~Myr ago), in combination with a recent SMC pericentre passage (about 150~Myr ago), can qualitatively explain the observed kinematic structure of the clusters. This is in accordance with the results for the kinematics of substructures in the outer parts of the LMC. Self-gravitating models that also focus on the inner regions of the LMC, along with a larger sample of clusters with measured velocities and distances, will be required to 
definitively confirm this scenario.

%--------------------------------------------------------------------

\begin{acknowledgements}  
We thank the anonymous referee for constructive comments
and suggestions that improved the quality of our paper. This research was funded by DLR grant 50 OR 2216.
DM, and SC acknowledge financial support from PRIN-MIUR-22: CHRONOS: adjusting the clock(s) to unveil the CHRONO-chemo-dynamical Structure of the Galaxy” (PI: S. Cassisi). 
SS acknowledges funding from the European Union under the grant ERC-2022-AdG, {\em "StarDance: the non-canonical evolution of stars in clusters"}, Grant Agreement 101093572, PI: E. Pancino.
Support for this work was provided by NASA through grants for program GO-16478 from the Space Telescope Science Institute (STScI), which is operated by the Association of Universities for Research in Astronomy (AURA), Inc., under NASA contract NAS5-26555. This work is based on observations made with the NASA/ESA Hubble Space Telescope, obtained from the Data Archive at the Space Telescope Science Institute. This work has made use of data from the European Space Agency (ESA) mission \textit{Gaia} (\url{https://www.cosmos.esa.int/gaia}), processed by the Gaia Data Processing and Analysis Consortium (DPAC, \url{https://www.cosmos.esa.int/web/gaia/dpac/consortium}). Funding for the DPAC has been provided by national institutions, in particular the institutions participating in the Gaia Multilateral Agreement.
This research made use of \texttt{astropy},\footnote{\href{http://www.astropy.org}{http://www.astropy.org}} a community-developed core \texttt{python} package for Astronomy \citep{Astropy22}, \texttt{iphython} \citep{Perez07}, \texttt{Jupyter Notebook} \citep{Kluyver16}, \texttt{matplotlib} \citep{Hunter07}, \texttt{numpy} \citep{Harris2020} and \texttt{scipy} \citep{Virtanen20}.
\end{acknowledgements}

%\empty

%\newpage

\bibliographystyle{aa}
\bibliography{references}

@ARTICLE{Aguado-Agelet25,
       author = {{Aguado-Agelet}, Fernando and {Massari}, Davide and {Monelli}, Matteo and {Cassisi}, Santi and {Gallart}, Carme and {Ceccarelli}, Edoardo and {Gonz{\'a}lez-Koda}, Yllari Kay and {Ruiz-Lara}, Tom{\'a}s and {Pancino}, Elena and {Saracino}, Sara and {Salaris}, Maurizio},
        title = "{Cluster Ages to Reconstruct the Milky Way Assembly (CARMA). II. The age-metallicity relation of Gaia-Sausage-Enceladus globular clusters}",
      journal = {arXiv e-prints},
         year = 2025,
        month = feb,
          eid = {arXiv:2502.20436},
        pages = {arXiv:2502.20436},
          doi = {10.48550/arXiv.2502.20436},
archivePrefix = {arXiv},
       eprint = {2502.20436},
 primaryClass = {astro-ph.GA},
       adsurl = {https://ui.adsabs.harvard.edu/abs/2025arXiv250220436A}
}

@ARTICLE{Anderson10,
       author = {{Anderson}, Jay and {Bedin}, Luigi R.},
        title = "{An Empirical Pixel-Based Correction for Imperfect CTE. I. HST{\textquoteright}s Advanced Camera for Surveys}",
      journal = {\pasp},
         year = 2010,
        month = sep,
       volume = {122},
       number = {895},
        pages = {1035},
          doi = {10.1086/656399},
archivePrefix = {arXiv},
       eprint = {1007.3987},
 primaryClass = {astro-ph.IM},
       adsurl = {https://ui.adsabs.harvard.edu/abs/2010PASP..122.1035A}
}

@MISC{Anderson22,
       author = {{Anderson}, Jay},
        title = "{One-Pass HST Photometry with hst1pass}",
 howpublished = {Instrument Science Report WFC3 2022-5, 55 pages},
         year = 2022,
        month = jul,
        pages = {5},
       adsurl = {https://ui.adsabs.harvard.edu/abs/2022wfc..rept....5A}
}

@ARTICLE{Astropy22,
       author = {{Astropy Collaboration} and {Price-Whelan}, Adrian M. and {Lim}, Pey Lian and {Earl}, Nicholas and {Starkman}, Nathaniel and {Bradley}, Larry and {Shupe}, David L. and {Patil}, Aarya A. and {Corrales}, Lia and {Brasseur}, C.~E. and {N{\"o}the}, Maximilian and {Donath}, Axel and {Tollerud}, Erik and {Morris}, Brett M. and {Ginsburg}, Adam and {Vaher}, Eero and {Weaver}, Benjamin A. and {Tocknell}, James and {Jamieson}, William and {van Kerkwijk}, Marten H. and {Robitaille}, Thomas P. and {Merry}, Bruce and {Bachetti}, Matteo and {G{\"u}nther}, H. Moritz and {Aldcroft}, Thomas L. and {Alvarado-Montes}, Jaime A. and {Archibald}, Anne M. and {B{\'o}di}, Attila and {Bapat}, Shreyas and {Barentsen}, Geert and {Baz{\'a}n}, Juanjo and {Biswas}, Manish and {Boquien}, M{\'e}d{\'e}ric and {Burke}, D.~J. and {Cara}, Daria and {Cara}, Mihai and {Conroy}, Kyle E. and {Conseil}, Simon and {Craig}, Matthew W. and {Cross}, Robert M. and {Cruz}, Kelle L. and {D'Eugenio}, Francesco and {Dencheva}, Nadia and {Devillepoix}, Hadrien A.~R. and {Dietrich}, J{\"o}rg P. and {Eigenbrot}, Arthur Davis and {Erben}, Thomas and {Ferreira}, Leonardo and {Foreman-Mackey}, Daniel and {Fox}, Ryan and {Freij}, Nabil and {Garg}, Suyog and {Geda}, Robel and {Glattly}, Lauren and {Gondhalekar}, Yash and {Gordon}, Karl D. and {Grant}, David and {Greenfield}, Perry and {Groener}, Austen M. and {Guest}, Steve and {Gurovich}, Sebastian and {Handberg}, Rasmus and {Hart}, Akeem and {Hatfield-Dodds}, Zac and {Homeier}, Derek and {Hosseinzadeh}, Griffin and {Jenness}, Tim and {Jones}, Craig K. and {Joseph}, Prajwel and {Kalmbach}, J. Bryce and {Karamehmetoglu}, Emir and {Ka{\l}uszy{\'n}ski}, Miko{\l}aj and {Kelley}, Michael S.~P. and {Kern}, Nicholas and {Kerzendorf}, Wolfgang E. and {Koch}, Eric W. and {Kulumani}, Shankar and {Lee}, Antony and {Ly}, Chun and {Ma}, Zhiyuan and {MacBride}, Conor and {Maljaars}, Jakob M. and {Muna}, Demitri and {Murphy}, N.~A. and {Norman}, Henrik and {O'Steen}, Richard and {Oman}, Kyle A. and {Pacifici}, Camilla and {Pascual}, Sergio and {Pascual-Granado}, J. and {Patil}, Rohit R. and {Perren}, Gabriel I. and {Pickering}, Timothy E. and {Rastogi}, Tanuj and {Roulston}, Benjamin R. and {Ryan}, Daniel F. and {Rykoff}, Eli S. and {Sabater}, Jose and {Sakurikar}, Parikshit and {Salgado}, Jes{\'u}s and {Sanghi}, Aniket and {Saunders}, Nicholas and {Savchenko}, Volodymyr and {Schwardt}, Ludwig and {Seifert-Eckert}, Michael and {Shih}, Albert Y. and {Jain}, Anany Shrey and {Shukla}, Gyanendra and {Sick}, Jonathan and {Simpson}, Chris and {Singanamalla}, Sudheesh and {Singer}, Leo P. and {Singhal}, Jaladh and {Sinha}, Manodeep and {Sip{\H{o}}cz}, Brigitta M. and {Spitler}, Lee R. and {Stansby}, David and {Streicher}, Ole and {{\v{S}}umak}, Jani and {Swinbank}, John D. and {Taranu}, Dan S. and {Tewary}, Nikita and {Tremblay}, Grant R. and {de Val-Borro}, Miguel and {Van Kooten}, Samuel J. and {Vasovi{\'c}}, Zlatan and {Verma}, Shresth and {de Miranda Cardoso}, Jos{\'e} Vin{\'\i}cius and {Williams}, Peter K.~G. and {Wilson}, Tom J. and {Winkel}, Benjamin and {Wood-Vasey}, W.~M. and {Xue}, Rui and {Yoachim}, Peter and {Zhang}, Chen and {Zonca}, Andrea and {Astropy Project Contributors}},
        title = "{The Astropy Project: Sustaining and Growing a Community-oriented Open-source Project and the Latest Major Release (v5.0) of the Core Package}",
      journal = {\apj},
         year = 2022,
        month = aug,
       volume = {935},
       number = {2},
          eid = {167},
        pages = {167},
          doi = {10.3847/1538-4357/ac7c74},
archivePrefix = {arXiv},
       eprint = {2206.14220},
 primaryClass = {astro-ph.IM},
       adsurl = {https://ui.adsabs.harvard.edu/abs/2022ApJ...935..167A}
}

@ARTICLE{Bastian16,
       author = {{Bastian}, N. and {Niederhofer}, F. and {Kozhurina-Platais}, V. and {Salaris}, M. and {Larsen}, S. and {Cabrera-Ziri}, I. and {Cordero}, M. and {Ekstr{\"o}m}, S. and {Geisler}, D. and {Georgy}, C. and {Hilker}, M. and {Kacharov}, N. and {Li}, C. and {Mackey}, D. and {Mucciarelli}, A. and {Platais}, I.},
        title = "{A young cluster with an extended main-sequence turnoff: confirmation of a prediction of the stellar rotation scenario}",
      journal = {\mnras},
         year = 2016,
        month = jul,
       volume = {460},
       number = {1},
        pages = {L20-L24},
          doi = {10.1093/mnrasl/slw067},
archivePrefix = {arXiv},
       eprint = {1604.01046},
 primaryClass = {astro-ph.GA},
       adsurl = {https://ui.adsabs.harvard.edu/abs/2016MNRAS.460L..20B}
}

@ARTICLE{Bellini09,
       author = {{Bellini}, A. and {Bedin}, L.~R.},
        title = "{Astrometry and Photometry with HST WFC3. I. Geometric Distortion Corrections of F225W, F275W, F336W Bands of the UVIS Channel}",
      journal = {\pasp},
         year = 2009,
        month = dec,
       volume = {121},
       number = {886},
        pages = {1419},
          doi = {10.1086/649061},
archivePrefix = {arXiv},
       eprint = {0910.3250},
 primaryClass = {astro-ph.IM},
       adsurl = {https://ui.adsabs.harvard.edu/abs/2009PASP..121.1419B}
}

@ARTICLE{Bellini11,
       author = {{Bellini}, A. and {Anderson}, J. and {Bedin}, L.~R.},
        title = "{Astrometry and Photometry with HST WFC3. II. Improved Geometric-Distortion Corrections for 10 Filters of the UVIS Channel}",
      journal = {\pasp},
         year = 2011,
        month = may,
       volume = {123},
       number = {903},
        pages = {622},
          doi = {10.1086/659878},
archivePrefix = {arXiv},
       eprint = {1102.5218},
 primaryClass = {astro-ph.IM},
       adsurl = {https://ui.adsabs.harvard.edu/abs/2011PASP..123..622B}
}

@ARTICLE{Bellini14,
       author = {{Bellini}, A. and {Anderson}, J. and {van der Marel}, R.~P. and {Watkins}, L.~L. and {King}, I.~R. and {Bianchini}, P. and {Chanam{\'e}}, J. and {Chandar}, R. and {Cool}, A.~M. and {Ferraro}, F.~R. and {Ford}, H. and {Massari}, D.},
        title = "{Hubble Space Telescope Proper Motion (HSTPROMO) Catalogs of Galactic Globular Clusters. I. Sample Selection, Data Reduction, and NGC 7078 Results}",
      journal = {\apj},
         year = 2014,
        month = dec,
       volume = {797},
       number = {2},
          eid = {115},
        pages = {115},
          doi = {10.1088/0004-637X/797/2/115},
archivePrefix = {arXiv},
       eprint = {1410.5820},
 primaryClass = {astro-ph.SR},
       adsurl = {https://ui.adsabs.harvard.edu/abs/2014ApJ...797..115B}
}

@ARTICLE{Bellini17b,
       author = {{Bellini}, A. and {Bianchini}, P. and {Varri}, A.~L. and {Anderson}, J. and {Piotto}, G. and {van der Marel}, R.~P. and {Vesperini}, E. and {Watkins}, L.~L.},
        title = "{Hubble Space Telescope Proper Motion (HSTPROMO) Catalogs of Galactic Globular Clusters. V. The Rapid Rotation of 47 Tuc Traced and Modeled in Three Dimensions}",
      journal = {\apj},
         year = 2017,
        month = aug,
       volume = {844},
       number = {2},
          eid = {167},
        pages = {167},
          doi = {10.3847/1538-4357/aa7c5f},
archivePrefix = {arXiv},
       eprint = {1706.08974},
 primaryClass = {astro-ph.GA},
       adsurl = {https://ui.adsabs.harvard.edu/abs/2017ApJ...844..167B}
}

@ARTICLE{Bellini18,
       author = {{Bellini}, Andrea and {Libralato}, Mattia and {Bedin}, Luigi R. and {Milone}, Antonino P. and {van der Marel}, Roeland P. and {Anderson}, Jay and {Apai}, D{\'a}niel and {Burgasser}, Adam J. and {Marino}, Anna F. and {Rees}, Jon M.},
        title = "{The HST Large Programme on {\ensuremath{\omega}} Centauri. II. Internal Kinematics}",
      journal = {\apj},
         year = 2018,
        month = jan,
       volume = {853},
       number = {1},
          eid = {86},
        pages = {86},
          doi = {10.3847/1538-4357/aaa3ec},
archivePrefix = {arXiv},
       eprint = {1801.01504},
 primaryClass = {astro-ph.GA},
       adsurl = {https://ui.adsabs.harvard.edu/abs/2018ApJ...853...86B}
}

@ARTICLE{Bennet22,
       author = {{Bennet}, Paul and {Alfaro-Cuello}, Mayte and {Pino}, Andr{\'e}s del and {Watkins}, Laura L. and {van der Marel}, Roeland P. and {Sohn}, Sangmo Tony},
        title = "{Kinematic Structure of the Large Magellanic Cloud Globular Cluster System from Gaia eDR3 and Hubble Space Telescope Proper Motions}",
      journal = {\apj},
         year = 2022,
        month = aug,
       volume = {935},
       number = {2},
          eid = {149},
        pages = {149},
          doi = {10.3847/1538-4357/ac81c9},
archivePrefix = {arXiv},
       eprint = {2207.13100},
 primaryClass = {astro-ph.GA},
       adsurl = {https://ui.adsabs.harvard.edu/abs/2022ApJ...935..149B}
}

@ARTICLE{Binney82,
       author = {{Binney}, J.},
        title = "{Regular and irregular orbits in galactic bars}",
      journal = {\mnras},
         year = 1982,
        month = oct,
       volume = {201},
        pages = {1-14},
          doi = {10.1093/mnras/201.1.1},
       adsurl = {https://ui.adsabs.harvard.edu/abs/1982MNRAS.201....1B}
}

@ARTICLE{Boubert17,
       author = {{Boubert}, D. and {Erkal}, D. and {Evans}, N.~W. and {Izzard}, R.~G.},
        title = "{Hypervelocity runaways from the Large Magellanic Cloud}",
      journal = {\mnras},
         year = 2017,
        month = aug,
       volume = {469},
       number = {2},
        pages = {2151-2162},
          doi = {10.1093/mnras/stx848},
archivePrefix = {arXiv},
       eprint = {1704.01373},
 primaryClass = {astro-ph.GA},
       adsurl = {https://ui.adsabs.harvard.edu/abs/2017MNRAS.469.2151B}
}

@ARTICLE{Bovy15,
       author = {{Bovy}, Jo},
        title = "{galpy: A python Library for Galactic Dynamics}",
      journal = {\apjs},
         year = 2015,
        month = feb,
       volume = {216},
       number = {2},
          eid = {29},
        pages = {29},
          doi = {10.1088/0067-0049/216/2/29},
archivePrefix = {arXiv},
       eprint = {1412.3451},
 primaryClass = {astro-ph.GA},
       adsurl = {https://ui.adsabs.harvard.edu/abs/2015ApJS..216...29B}
}

@ARTICLE{Cardelli89,
       author = {{Cardelli}, Jason A. and {Clayton}, Geoffrey C. and {Mathis}, John S.},
        title = "{The Relationship between Infrared, Optical, and Ultraviolet Extinction}",
      journal = {\apj},
         year = 1989,
        month = oct,
       volume = {345},
        pages = {245},
          doi = {10.1086/167900},
       adsurl = {https://ui.adsabs.harvard.edu/abs/1989ApJ...345..245C}
}

@ARTICLE{Cassisi04,
       author = {{Cassisi}, Santi and {Salaris}, Maurizio and {Castelli}, Fiorella and {Pietrinferni}, Adriano},
        title = "{Color Transformations and Bolometric Corrections for Galactic Halo Stars: {\ensuremath{\alpha}}-Enhanced versus Scaled-Solar Results}",
      journal = {\apj},
         year = 2004,
        month = nov,
       volume = {616},
       number = {1},
        pages = {498-505},
          doi = {10.1086/424907},
archivePrefix = {arXiv},
       eprint = {astro-ph/0408111},
 primaryClass = {astro-ph},
       adsurl = {https://ui.adsabs.harvard.edu/abs/2004ApJ...616..498C}
}

@ARTICLE{Ceccarelli25,
       author = {{Ceccarelli}, E. and {Massari}, D. and {Aguado-Agelet}, F. and {Mucciarelli}, A. and {Cassisi}, S. and {Monelli}, M. and {Pancino}, E. and {Salaris}, M. and {Saracino}, S.},
        title = "{Cluster Ages to Reconstruct the Milky Way Assembly (CARMA). III. NGC 288 as the first Splashed globular cluster}",
      journal = {arXiv e-prints},
         year = 2025,
        month = mar,
          eid = {arXiv:2503.02939},
        pages = {arXiv:2503.02939},
          doi = {10.48550/arXiv.2503.02939},
archivePrefix = {arXiv},
       eprint = {2503.02939},
 primaryClass = {astro-ph.GA},
       adsurl = {https://ui.adsabs.harvard.edu/abs/2025arXiv250302939C}
}

@ARTICLE{Choi18,
       author = {{Choi}, Yumi and {Nidever}, David L. and {Olsen}, Knut and {Blum}, Robert D. and {Besla}, Gurtina and {Zaritsky}, Dennis and {van der Marel}, Roeland P. and {Bell}, Eric F. and {Gallart}, Carme and {Cioni}, Maria-Rosa L. and {Johnson}, L. Clifton and {Vivas}, A. Katherina and {Saha}, Abhijit and {de Boer}, Thomas J.~L. and {No{\"e}l}, Noelia E.~D. and {Monachesi}, Antonela and {Massana}, Pol and {Conn}, Blair C. and {Martinez-Delgado}, David and {Mu{\~n}oz}, Ricardo R. and {Stringfellow}, Guy S.},
        title = "{SMASHing the LMC: A Tidally Induced Warp in the Outer LMC and a Large-scale Reddening Map}",
      journal = {\apj},
         year = 2018,
        month = oct,
       volume = {866},
       number = {2},
          eid = {90},
        pages = {90},
          doi = {10.3847/1538-4357/aae083},
archivePrefix = {arXiv},
       eprint = {1804.07765},
 primaryClass = {astro-ph.GA},
       adsurl = {https://ui.adsabs.harvard.edu/abs/2018ApJ...866...90C}
}

@ARTICLE{Choi22,
       author = {{Choi}, Yumi and {Olsen}, Knut A.~G. and {Besla}, Gurtina and {van der Marel}, Roeland P. and {Zivick}, Paul and {Kallivayalil}, Nitya and {Nidever}, David L.},
        title = "{The Recent LMC-SMC Collision: Timing and Impact Parameter Constraints from Comparison of Gaia LMC Disk Kinematics and N-body Simulations}",
      journal = {\apj},
         year = 2022,
        month = mar,
       volume = {927},
       number = {2},
          eid = {153},
        pages = {153},
          doi = {10.3847/1538-4357/ac4e90},
archivePrefix = {arXiv},
       eprint = {2201.04648},
 primaryClass = {astro-ph.GA},
       adsurl = {https://ui.adsabs.harvard.edu/abs/2022ApJ...927..153C}
}

@ARTICLE{Choudhury20,
       author = {{Choudhury}, Samyaday and {de Grijs}, Richard and {Rubele}, Stefano and {Bekki}, Kenji and {Cioni}, Maria-Rosa L. and {Ivanov}, Valentin D. and {van Loon}, Jacco Th and {Niederhofer}, Florian and {Oliveira}, Joana M. and {Ripepi}, Vincenzo},
        title = "{The VMC survey - XXXIX. Mapping metallicity trends in the Small Magellanic Cloud using near-infrared passbands}",
      journal = {\mnras},
         year = 2020,
        month = sep,
       volume = {497},
       number = {3},
        pages = {3746-3760},
          doi = {10.1093/mnras/staa2140},
archivePrefix = {arXiv},
       eprint = {2007.08753},
 primaryClass = {astro-ph.GA},
       adsurl = {https://ui.adsabs.harvard.edu/abs/2020MNRAS.497.3746C}
}

@ARTICLE{Choudhury21,
       author = {{Choudhury}, Samyaday and {de Grijs}, Richard and {Bekki}, Kenji and {Cioni}, Maria-Rosa L. and {Ivanov}, Valentin D. and {van Loon}, Jacco Th and {Miller}, Amy E. and {Niederhofer}, Florian and {Oliveira}, Joana M. and {Ripepi}, Vincenzo and {Sun}, Ning-Chen and {Subramanian}, Smitha},
        title = "{The VMC survey - XLIV: mapping metallicity trends in the large magellanic cloud using near-infrared passbands}",
      journal = {\mnras},
         year = 2021,
        month = nov,
       volume = {507},
       number = {4},
        pages = {4752-4763},
          doi = {10.1093/mnras/stab2446},
archivePrefix = {arXiv},
       eprint = {2108.10529},
 primaryClass = {astro-ph.GA},
       adsurl = {https://ui.adsabs.harvard.edu/abs/2021MNRAS.507.4752C}
}

@ARTICLE{Cioni11,
       author = {{Cioni}, M. -R.~L. and {Clementini}, G. and {Girardi}, L. and {Guandalini}, R. and {Gullieuszik}, M. and {Miszalski}, B. and {Moretti}, M. -I. and {Ripepi}, V. and {Rubele}, S. and {Bagheri}, G. and {Bekki}, K. and {Cross}, N. and {de Blok}, W.~J.~G. and {de Grijs}, R. and {Emerson}, J.~P. and {Evans}, C.~J. and {Gibson}, B. and {Gonzales-Solares}, E. and {Groenewegen}, M.~A.~T. and {Irwin}, M. and {Ivanov}, V.~D. and {Lewis}, J. and {Marconi}, M. and {Marquette}, J. -B. and {Mastropietro}, C. and {Moore}, B. and {Napiwotzki}, R. and {Naylor}, T. and {Oliveira}, J.~M. and {Read}, M. and {Sutorius}, E. and {van Loon}, J. Th. and {Wilkinson}, M.~I. and {Wood}, P.~R.},
        title = "{The VMC survey. I. Strategy and first data}",
      journal = {\aap},
         year = 2011,
        month = mar,
       volume = {527},
          eid = {A116},
        pages = {A116},
          doi = {10.1051/0004-6361/201016137},
archivePrefix = {arXiv},
       eprint = {1012.5193},
 primaryClass = {astro-ph.CO},
       adsurl = {https://ui.adsabs.harvard.edu/abs/2011A&A...527A.116C}
}

@ARTICLE{Correnti17,
       author = {{Correnti}, Matteo and {Goudfrooij}, Paul and {Bellini}, Andrea and {Kalirai}, Jason S. and {Puzia}, Thomas H.},
        title = "{Dissecting the extended main-sequence turn-off of the young star cluster NGC 1850}",
      journal = {\mnras},
         year = 2017,
        month = may,
       volume = {467},
       number = {3},
        pages = {3628-3641},
          doi = {10.1093/mnras/stx010},
archivePrefix = {arXiv},
       eprint = {1612.08746},
 primaryClass = {astro-ph.SR},
       adsurl = {https://ui.adsabs.harvard.edu/abs/2017MNRAS.467.3628C}
}

@ARTICLE{Cullinane22a,
       author = {{Cullinane}, L.~R. and {Mackey}, A.~D. and {Da Costa}, G.~S. and {Erkal}, D. and {Koposov}, S.~E. and {Belokurov}, V.},
        title = "{The Magellanic Edges Survey - II. Formation of the LMC's northern arm}",
      journal = {\mnras},
         year = 2022,
        month = feb,
       volume = {510},
       number = {1},
        pages = {445-468},
          doi = {10.1093/mnras/stab3350},
archivePrefix = {arXiv},
       eprint = {2106.03274},
 primaryClass = {astro-ph.GA},
       adsurl = {https://ui.adsabs.harvard.edu/abs/2022MNRAS.510..445C}
}

@ARTICLE{Cullinane22b,
       author = {{Cullinane}, L.~R. and {Mackey}, A.~D. and {Da Costa}, G.~S. and {Erkal}, D. and {Koposov}, S.~E. and {Belokurov}, V.},
        title = "{The Magellanic Edges Survey - III. Kinematics of the disturbed LMC outskirts}",
      journal = {\mnras},
         year = 2022,
        month = jun,
       volume = {512},
       number = {4},
        pages = {4798-4818},
          doi = {10.1093/mnras/stac733},
archivePrefix = {arXiv},
       eprint = {2203.05450},
 primaryClass = {astro-ph.GA},
       adsurl = {https://ui.adsabs.harvard.edu/abs/2022MNRAS.512.4798C}
}

@ARTICLE{Erkal19,
       author = {{Erkal}, D. and {Belokurov}, V. and {Laporte}, C.~F.~P. and {Koposov}, S.~E. and {Li}, T.~S. and {Grillmair}, C.~J. and {Kallivayalil}, N. and {Price-Whelan}, A.~M. and {Evans}, N.~W. and {Hawkins}, K. and {Hendel}, D. and {Mateu}, C. and {Navarro}, J.~F. and {del Pino}, A. and {Slater}, C.~T. and {Sohn}, S.~T. and {Orphan Aspen Treasury Collaboration}},
        title = "{The total mass of the Large Magellanic Cloud from its perturbation on the Orphan stream}",
      journal = {\mnras},
         year = 2019,
        month = aug,
       volume = {487},
       number = {2},
        pages = {2685-2700},
          doi = {10.1093/mnras/stz1371},
archivePrefix = {arXiv},
       eprint = {1812.08192},
 primaryClass = {astro-ph.GA},
       adsurl = {https://ui.adsabs.harvard.edu/abs/2019MNRAS.487.2685E}
}

@ARTICLE{Fehrenbach74,
       author = {{Fehrenbach}, Ch. and {Duflot}, M.},
        title = "{Radial velocities from objective-prism plates in the direction of the Large Magellanic Cloud. List of 398 stars, LMC members. List of 1434 galactic stars, in the LMC direction}",
      journal = {\aaps},
         year = 1974,
        month = feb,
       volume = {13},
        pages = {173},
       adsurl = {https://ui.adsabs.harvard.edu/abs/1974A&AS...13..173F}
}

@ARTICLE{Forbes18,
       author = {{Forbes}, Duncan A. and {Bastian}, Nate and {Gieles}, Mark and {Crain}, Robert A. and {Kruijssen}, J.~M. Diederik and {Larsen}, S{\o}ren S. and {Ploeckinger}, Sylvia and {Agertz}, Oscar and {Trenti}, Michele and {Ferguson}, Annette M.~N. and {Pfeffer}, Joel and {Gnedin}, Oleg Y.},
        title = "{Globular cluster formation and evolution in the context of cosmological galaxy assembly: open questions}",
      journal = {Proceedings of the Royal Society of London Series A},
         year = 2018,
        month = feb,
       volume = {474},
       number = {2210},
          eid = {20170616},
        pages = {20170616},
          doi = {10.1098/rspa.2017.0616},
archivePrefix = {arXiv},
       eprint = {1801.05818},
 primaryClass = {astro-ph.GA},
       adsurl = {https://ui.adsabs.harvard.edu/abs/2018RSPSA.47470616F}
}

@ARTICLE{Freedman01,
       author = {{Freedman}, Wendy L. and {Madore}, Barry F. and {Gibson}, Brad K. and {Ferrarese}, Laura and {Kelson}, Daniel D. and {Sakai}, Shoko and {Mould}, Jeremy R. and {Kennicutt}, Jr., Robert C. and {Ford}, Holland C. and {Graham}, John A. and {Huchra}, John P. and {Hughes}, Shaun M.~G. and {Illingworth}, Garth D. and {Macri}, Lucas M. and {Stetson}, Peter B.},
        title = "{Final Results from the Hubble Space Telescope Key Project to Measure the Hubble Constant}",
      journal = {\apj},
         year = 2001,
        month = may,
       volume = {553},
       number = {1},
        pages = {47-72},
          doi = {10.1086/320638},
archivePrefix = {arXiv},
       eprint = {astro-ph/0012376},
 primaryClass = {astro-ph},
       adsurl = {https://ui.adsabs.harvard.edu/abs/2001ApJ...553...47F}
}

@ARTICLE{Freeman83,
       author = {{Freeman}, K.~C. and {Illingworth}, G. and {Oemler}, A., Jr.},
        title = "{The kinematics of globular clusters in the Large Magellanic Cloud.}",
      journal = {\apj},
         year = 1983,
        month = sep,
       volume = {272},
        pages = {488-508},
          doi = {10.1086/161316},
       adsurl = {https://ui.adsabs.harvard.edu/abs/1983ApJ...272..488F}
}

@ARTICLE{Gaia18,
       author = {{Gaia Collaboration} and {Helmi}, A. and {van Leeuwen}, F. and
         {McMillan}, P.~J. and {Massari}, D. and {Antoja}, T. and
         {Robin}, A.~C. and {Lindegren}, L. and {Bastian}, U. and {Arenou}, F. and
         {Babusiaux}, C. and {Biermann}, M. and {Breddels}, M.~A. and
         {Hobbs}, D. and {Jordi}, C. and {Pancino}, E. and {Reyl{\'e}}, C. and
         {Veljanoski}, J. and {Brown}, A.~G.~A. and {Vallenari}, A. and
         {Prusti}, T. and {de Bruijne}, J.~H.~J. and {Bailer-Jones}, C.~A.~L. and
         {Evans}, D.~W. and {Eyer}, L. and {Jansen}, F. and {Klioner}, S.~A. and
         {Lammers}, U. and {Luri}, X. and {Mignard}, F. and {Panem}, C. and
         {Pourbaix}, D. and {Randich}, S. and {Sartoretti}, P. and
         {Siddiqui}, H.~I. and {Soubiran}, C. and {Walton}, N.~A. and
         {Cropper}, M. and {Drimmel}, R. and {Katz}, D. and {Lattanzi}, M.~G. and
         {Bakker}, J. and {Cacciari}, C. and {Casta{\~n}eda}, J. and
         {Chaoul}, L. and {Cheek}, N. and {De Angeli}, F. and {Fabricius}, C. and
         {Guerra}, R. and {Holl}, B. and {Masana}, E. and {Messineo}, R. and
         {Mowlavi}, N. and {Nienartowicz}, K. and {Panuzzo}, P. and
         {Portell}, J. and {Riello}, M. and {Seabroke}, G.~M. and {Tanga}, P. and
         {Th{\'e}venin}, F. and {Gracia-Abril}, G. and {Comoretto}, G. and
         {Garcia-Reinaldos}, M. and {Teyssier}, D. and {Altmann}, M. and
         {Andrae}, R. and {Audard}, M. and {Bellas-Velidis}, I. and
         {Benson}, K. and {Berthier}, J. and {Blomme}, R. and {Burgess}, P. and
         {Busso}, G. and {Carry}, B. and {Cellino}, A. and {Clementini}, G. and
         {Clotet}, M. and {Creevey}, O. and {Davidson}, M. and {De Ridder}, J. and
         {Delchambre}, L. and {Dell'Oro}, A. and {Ducourant}, C. and
         {Fern{\'a}ndez-Hern{\'a}ndez}, J. and {Fouesneau}, M. and
         {Fr{\'e}mat}, Y. and {Galluccio}, L. and {Garc{\'\i}a-Torres}, M. and
         {Gonz{\'a}lez-N{\'u}{\~n}ez}, J. and {Gonz{\'a}lez-Vidal}, J.~J. and
         {Gosset}, E. and {Guy}, L.~P. and {Halbwachs}, J. -L. and
         {Hambly}, N.~C. and {Harrison}, D.~L. and {Hern{\'a}ndez}, J. and
         {Hestroffer}, D. and {Hodgkin}, S.~T. and {Hutton}, A. and
         {Jasniewicz}, G. and {Jean-Antoine-Piccolo}, A. and {Jordan}, S. and
         {Korn}, A.~J. and {Krone-Martins}, A. and {Lanzafame}, A.~C. and
         {Lebzelter}, T. and {L{\"o}ffler}, W. and {Manteiga}, M. and
         {Marrese}, P.~M. and {Mart{\'\i}n-Fleitas}, J.~M. and {Moitinho}, A. and
         {Mora}, A. and {Muinonen}, K. and {Osinde}, J. and {Pauwels}, T. and
         {Petit}, J. -M. and {Recio-Blanco}, A. and {Richards}, P.~J. and
         {Rimoldini}, L. and {Sarro}, L.~M. and {Siopis}, C. and {Smith}, M. and
         {Sozzetti}, A. and {S{\"u}veges}, M. and {Torra}, J. and
         {van Reeven}, W. and {Abbas}, U. and {Abreu Aramburu}, A. and
         {Accart}, S. and {Aerts}, C. and {Altavilla}, G. and
         {{\'A}lvarez}, M.~A. and {Alvarez}, R. and {Alves}, J. and
         {Anderson}, R.~I. and {Andrei}, A.~H. and {Anglada Varela}, E. and
         {Antiche}, E. and {Arcay}, B. and {Astraatmadja}, T.~L. and {Bach}, N. and
         {Baker}, S.~G. and {Balaguer-N{\'u}{\~n}ez}, L. and {Balm}, P. and
         {Barache}, C. and {Barata}, C. and {Barbato}, D. and {Barblan}, F. and
         {Barklem}, P.~S. and {Barrado}, D. and {Barros}, M. and
         {Barstow}, M.~A. and {Bartholom{\'e} Mu{\~n}oz}, S. and
         {Bassilana}, J. -L. and {Becciani}, U. and {Bellazzini}, M. and
         {Berihuete}, A. and {Bertone}, S. and {Bianchi}, L. and
         {Bienaym{\'e}}, O. and {Blanco-Cuaresma}, S. and {Boch}, T. and
         {Boeche}, C. and {Bombrun}, A. and {Borrachero}, R. and {Bossini}, D. and
         {Bouquillon}, S. and {Bourda}, G. and {Bragaglia}, A. and
         {Bramante}, L. and {Bressan}, A. and {Brouillet}, N. and
         {Br{\"u}semeister}, T. and {Brugaletta}, E. and {Bucciarelli}, B. and
         {Burlacu}, A. and {Busonero}, D. and {Butkevich}, A.~G. and
         {Buzzi}, R. and {Caffau}, E. and {Cancelliere}, R. and
         {Cannizzaro}, G. and {Cantat-Gaudin}, T. and {Carballo}, R. and
         {Carlucci}, T. and {Carrasco}, J.~M. and {Casamiquela}, L. and
         {Castellani}, M. and {Castro-Ginard}, A. and {Charlot}, P. and
         {Chemin}, L. and {Chiavassa}, A. and {Cocozza}, G. and {Costigan}, G. and
         {Cowell}, S. and {Crifo}, F. and {Crosta}, M. and {Crowley}, C. and
         {Cuypers}, J. and {Dafonte}, C. and {Damerdji}, Y. and
         {Dapergolas}, A. and {David}, P. and {David}, M. and {de Laverny}, P. and
         {De Luise}, F. and {De March}, R. and {de Martino}, D. and
         {de Souza}, R. and {de Torres}, A. and {Debosscher}, J. and
         {del Pozo}, E. and {Delbo}, M. and {Delgado}, A. and {Delgado}, H.~E. and
         {Di Matteo}, P. and {Diakite}, S. and {Diener}, C. and {Distefano}, E. and
         {Dolding}, C. and {Drazinos}, P. and {Dur{\'a}n}, J. and
         {Edvardsson}, B. and {Enke}, H. and {Eriksson}, K. and {Esquej}, P. and
         {Eynard Bontemps}, G. and {Fabre}, C. and {Fabrizio}, M. and
         {Faigler}, S. and {Falc{\~a}o}, A.~J. and {Farr{\`a}s Casas}, M. and
         {Federici}, L. and {Fedorets}, G. and {Fernique}, P. and
         {Figueras}, F. and {Filippi}, F. and {Findeisen}, K. and {Fonti}, A. and
         {Fraile}, E. and {Fraser}, M. and {Fr{\'e}zouls}, B. and {Gai}, M. and
         {Galleti}, S. and {Garabato}, D. and {Garc{\'\i}a-Sedano}, F. and
         {Garofalo}, A. and {Garralda}, N. and {Gavel}, A. and {Gavras}, P. and
         {Gerssen}, J. and {Geyer}, R. and {Giacobbe}, P. and {Gilmore}, G. and
         {Girona}, S. and {Giuffrida}, G. and {Glass}, F. and {Gomes}, M. and
         {Granvik}, M. and {Gueguen}, A. and {Guerrier}, A. and {Guiraud}, J. and
         {Guti{\'e}rrez-S{\'a}nchez}, R. and {Hofmann}, W. and {Holland}, G. and
         {Huckle}, H.~E. and {Hypki}, A. and {Icardi}, V. and {Jan{\ss}en}, K. and
         {Jevardat de Fombelle}, G. and {Jonker}, P.~G. and
         {Juh{\'a}sz}, {\'A}. L. and {Julbe}, F. and {Karampelas}, A. and
         {Kewley}, A. and {Klar}, J. and {Kochoska}, A. and {Kohley}, R. and
         {Kolenberg}, K. and {Kontizas}, M. and {Kontizas}, E. and
         {Koposov}, S.~E. and {Kordopatis}, G. and {Kostrzewa-Rutkowska}, Z. and
         {Koubsky}, P. and {Lambert}, S. and {Lanza}, A.~F. and {Lasne}, Y. and
         {Lavigne}, J. -B. and {Le Fustec}, Y. and {Le Poncin-Lafitte}, C. and
         {Lebreton}, Y. and {Leccia}, S. and {Leclerc}, N. and
         {Lecoeur-Taibi}, I. and {Lenhardt}, H. and {Leroux}, F. and {Liao}, S. and
         {Licata}, E. and {Lindstr{\o}m}, H.~E.~P. and {Lister}, T.~A. and
         {Livanou}, E. and {Lobel}, A. and {L{\'o}pez}, M. and {Managau}, S. and
         {Mann}, R.~G. and {Mantelet}, G. and {Marchal}, O. and
         {Marchant}, J.~M. and {Marconi}, M. and {Marinoni}, S. and
         {Marschalk{\'o}}, G. and {Marshall}, D.~J. and {Martino}, M. and
         {Marton}, G. and {Mary}, N. and {Matijevi{\v{c}}}, G. and {Mazeh}, T. and
         {Messina}, S. and {Michalik}, D. and {Millar}, N.~R. and {Molina}, D. and
         {Molinaro}, R. and {Moln{\'a}r}, L. and {Montegriffo}, P. and
         {Mor}, R. and {Morbidelli}, R. and {Morel}, T. and {Morris}, D. and
         {Mulone}, A.~F. and {Muraveva}, T. and {Musella}, I. and
         {Nelemans}, G. and {Nicastro}, L. and {Noval}, L. and {O'Mullane}, W. and
         {Ord{\'e}novic}, C. and {Ord{\'o}{\~n}ez-Blanco}, D. and {Osborne}, P. and
         {Pagani}, C. and {Pagano}, I. and {Pailler}, F. and {Palacin}, H. and
         {Palaversa}, L. and {Panahi}, A. and {Pawlak}, M. and
         {Piersimoni}, A.~M. and {Pineau}, F. -X. and {Plachy}, E. and
         {Plum}, G. and {Poggio}, E. and {Poujoulet}, E. and {Pr{\v{s}}a}, A. and
         {Pulone}, L. and {Racero}, E. and {Ragaini}, S. and {Rambaux}, N. and
         {Ramos-Lerate}, M. and {Regibo}, S. and {Riclet}, F. and {Ripepi}, V. and
         {Riva}, A. and {Rivard}, A. and {Rixon}, G. and {Roegiers}, T. and
         {Roelens}, M. and {Romero-G{\'o}mez}, M. and {Rowell}, N. and
         {Royer}, F. and {Ruiz-Dern}, L. and {Sadowski}, G. and
         {Sagrist{\`a} Sell{\'e}s}, T. and {Sahlmann}, J. and {Salgado}, J. and
         {Salguero}, E. and {Sanna}, N. and {Santana-Ros}, T. and {Sarasso}, M. and
         {Savietto}, H. and {Schultheis}, M. and {Sciacca}, E. and {Segol}, M. and
         {Segovia}, J.~C. and {S{\'e}gransan}, D. and {Shih}, I. -C. and
         {Siltala}, L. and {Silva}, A.~F. and {Smart}, R.~L. and {Smith}, K.~W. and
         {Solano}, E. and {Solitro}, F. and {Sordo}, R. and {Soria Nieto}, S. and
         {Souchay}, J. and {Spagna}, A. and {Spoto}, F. and {Stampa}, U. and
         {Steele}, I.~A. and {Steidelm{\"u}ller}, H. and {Stephenson}, C.~A. and
         {Stoev}, H. and {Suess}, F.~F. and {Surdej}, J. and {Szabados}, L. and
         {Szegedi-Elek}, E. and {Tapiador}, D. and {Taris}, F. and {Tauran}, G. and
         {Taylor}, M.~B. and {Teixeira}, R. and {Terrett}, D. and {Teyssand
        ier}, P. and {Thuillot}, W. and {Titarenko}, A. and {Torra Clotet}, F. and
         {Turon}, C. and {Ulla}, A. and {Utrilla}, E. and {Uzzi}, S. and
         {Vaillant}, M. and {Valentini}, G. and {Valette}, V. and
         {van Elteren}, A. and {Van Hemelryck}, E. and {van Leeuwen}, M. and
         {Vaschetto}, M. and {Vecchiato}, A. and {Viala}, Y. and {Vicente}, D. and
         {Vogt}, S. and {von Essen}, C. and {Voss}, H. and {Votruba}, V. and
         {Voutsinas}, S. and {Walmsley}, G. and {Weiler}, M. and {Wertz}, O. and
         {Wevems}, T. and {Wyrzykowski}, {\L}. and {Yoldas}, A. and
         {{\v{Z}}erjal}, M. and {Ziaeepour}, H. and {Zorec}, J. and
         {Zschocke}, S. and {Zucker}, S. and {Zurbach}, C. and {Zwitter}, T.},
        title = "{Gaia Data Release 2. Kinematics of globular clusters and dwarf galaxies around the Milky Way}",
      journal = {\aap},
         year = "2018",
        month = "Aug",
       volume = {616},
          eid = {A12},
        pages = {A12},
          doi = {10.1051/0004-6361/201832698},
archivePrefix = {arXiv},
       eprint = {1804.09381},
 primaryClass = {astro-ph.GA},
       adsurl = {https://ui.adsabs.harvard.edu/abs/2018A&A...616A..12G}
}

@ARTICLE{Gaia23,
       author = {{Gaia Collaboration} and {Vallenari}, A. and {Brown}, A.~G.~A. and {Prusti}, T. and {de Bruijne}, J.~H.~J. and {Arenou}, F. and {Babusiaux}, C. and {Biermann}, M. and {Creevey}, O.~L. and {Ducourant}, C. and {Evans}, D.~W. and {Eyer}, L. and {Guerra}, R. and {Hutton}, A. and {Jordi}, C. and {Klioner}, S.~A. and {Lammers}, U.~L. and {Lindegren}, L. and {Luri}, X. and {Mignard}, F. and {Panem}, C. and {Pourbaix}, D. and {Randich}, S. and {Sartoretti}, P. and {Soubiran}, C. and {Tanga}, P. and {Walton}, N.~A. and {Bailer-Jones}, C.~A.~L. and {Bastian}, U. and {Drimmel}, R. and {Jansen}, F. and {Katz}, D. and {Lattanzi}, M.~G. and {van Leeuwen}, F. and {Bakker}, J. and {Cacciari}, C. and {Casta{\~n}eda}, J. and {De Angeli}, F. and {Fabricius}, C. and {Fouesneau}, M. and {Fr{\'e}mat}, Y. and {Galluccio}, L. and {Guerrier}, A. and {Heiter}, U. and {Masana}, E. and {Messineo}, R. and {Mowlavi}, N. and {Nicolas}, C. and {Nienartowicz}, K. and {Pailler}, F. and {Panuzzo}, P. and {Riclet}, F. and {Roux}, W. and {Seabroke}, G.~M. and {Sordo}, R. and {Th{\'e}venin}, F. and {Gracia-Abril}, G. and {Portell}, J. and {Teyssier}, D. and {Altmann}, M. and {Andrae}, R. and {Audard}, M. and {Bellas-Velidis}, I. and {Benson}, K. and {Berthier}, J. and {Blomme}, R. and {Burgess}, P.~W. and {Busonero}, D. and {Busso}, G. and {C{\'a}novas}, H. and {Carry}, B. and {Cellino}, A. and {Cheek}, N. and {Clementini}, G. and {Damerdji}, Y. and {Davidson}, M. and {de Teodoro}, P. and {Nu{\~n}ez Campos}, M. and {Delchambre}, L. and {Dell'Oro}, A. and {Esquej}, P. and {Fern{\'a}ndez-Hern{\'a}ndez}, J. and {Fraile}, E. and {Garabato}, D. and {Garc{\'\i}a-Lario}, P. and {Gosset}, E. and {Haigron}, R. and {Halbwachs}, J. -L. and {Hambly}, N.~C. and {Harrison}, D.~L. and {Hern{\'a}ndez}, J. and {Hestroffer}, D. and {Hodgkin}, S.~T. and {Holl}, B. and {Jan{\ss}en}, K. and {Jevardat de Fombelle}, G. and {Jordan}, S. and {Krone-Martins}, A. and {Lanzafame}, A.~C. and {L{\"o}ffler}, W. and {Marchal}, O. and {Marrese}, P.~M. and {Moitinho}, A. and {Muinonen}, K. and {Osborne}, P. and {Pancino}, E. and {Pauwels}, T. and {Recio-Blanco}, A. and {Reyl{\'e}}, C. and {Riello}, M. and {Rimoldini}, L. and {Roegiers}, T. and {Rybizki}, J. and {Sarro}, L.~M. and {Siopis}, C. and {Smith}, M. and {Sozzetti}, A. and {Utrilla}, E. and {van Leeuwen}, M. and {Abbas}, U. and {{\'A}brah{\'a}m}, P. and {Abreu Aramburu}, A. and {Aerts}, C. and {Aguado}, J.~J. and {Ajaj}, M. and {Aldea-Montero}, F. and {Altavilla}, G. and {{\'A}lvarez}, M.~A. and {Alves}, J. and {Anders}, F. and {Anderson}, R.~I. and {Anglada Varela}, E. and {Antoja}, T. and {Baines}, D. and {Baker}, S.~G. and {Balaguer-N{\'u}{\~n}ez}, L. and {Balbinot}, E. and {Balog}, Z. and {Barache}, C. and {Barbato}, D. and {Barros}, M. and {Barstow}, M.~A. and {Bartolom{\'e}}, S. and {Bassilana}, J. -L. and {Bauchet}, N. and {Becciani}, U. and {Bellazzini}, M. and {Berihuete}, A. and {Bernet}, M. and {Bertone}, S. and {Bianchi}, L. and {Binnenfeld}, A. and {Blanco-Cuaresma}, S. and {Blazere}, A. and {Boch}, T. and {Bombrun}, A. and {Bossini}, D. and {Bouquillon}, S. and {Bragaglia}, A. and {Bramante}, L. and {Breedt}, E. and {Bressan}, A. and {Brouillet}, N. and {Brugaletta}, E. and {Bucciarelli}, B. and {Burlacu}, A. and {Butkevich}, A.~G. and {Buzzi}, R. and {Caffau}, E. and {Cancelliere}, R. and {Cantat-Gaudin}, T. and {Carballo}, R. and {Carlucci}, T. and {Carnerero}, M.~I. and {Carrasco}, J.~M. and {Casamiquela}, L. and {Castellani}, M. and {Castro-Ginard}, A. and {Chaoul}, L. and {Charlot}, P. and {Chemin}, L. and {Chiaramida}, V. and {Chiavassa}, A. and {Chornay}, N. and {Comoretto}, G. and {Contursi}, G. and {Cooper}, W.~J. and {Cornez}, T. and {Cowell}, S. and {Crifo}, F. and {Cropper}, M. and {Crosta}, M. and {Crowley}, C. and {Dafonte}, C. and {Dapergolas}, A. and {David}, M. and {David}, P. and {de Laverny}, P. and {De Luise}, F. and {De March}, R. and {De Ridder}, J. and {de Souza}, R. and {de Torres}, A. and {del Peloso}, E.~F. and {del Pozo}, E. and {Delbo}, M. and {Delgado}, A. and {Delisle}, J. -B. and {Demouchy}, C. and {Dharmawardena}, T.~E. and {Di Matteo}, P. and {Diakite}, S. and {Diener}, C. and {Distefano}, E. and {Dolding}, C. and {Edvardsson}, B. and {Enke}, H. and {Fabre}, C. and {Fabrizio}, M. and {Faigler}, S. and {Fedorets}, G. and {Fernique}, P. and {Fienga}, A. and {Figueras}, F. and {Fournier}, Y. and {Fouron}, C. and {Fragkoudi}, F. and {Gai}, M. and {Garcia-Gutierrez}, A. and {Garcia-Reinaldos}, M. and {Garc{\'\i}a-Torres}, M. and {Garofalo}, A. and {Gavel}, A. and {Gavras}, P. and {Gerlach}, E. and {Geyer}, R. and {Giacobbe}, P. and {Gilmore}, G. and {Girona}, S. and {Giuffrida}, G. and {Gomel}, R. and {Gomez}, A. and {Gonz{\'a}lez-N{\'u}{\~n}ez}, J. and {Gonz{\'a}lez-Santamar{\'\i}a}, I. and {Gonz{\'a}lez-Vidal}, J.~J. and {Granvik}, M. and {Guillout}, P. and {Guiraud}, J. and {Guti{\'e}rrez-S{\'a}nchez}, R. and {Guy}, L.~P. and {Hatzidimitriou}, D. and {Hauser}, M. and {Haywood}, M. and {Helmer}, A. and {Helmi}, A. and {Sarmiento}, M.~H. and {Hidalgo}, S.~L. and {Hilger}, T. and {H{\l}adczuk}, N. and {Hobbs}, D. and {Holland}, G. and {Huckle}, H.~E. and {Jardine}, K. and {Jasniewicz}, G. and {Jean-Antoine Piccolo}, A. and {Jim{\'e}nez-Arranz}, {\'O}. and {Jorissen}, A. and {Juaristi Campillo}, J. and {Julbe}, F. and {Karbevska}, L. and {Kervella}, P. and {Khanna}, S. and {Kontizas}, M. and {Kordopatis}, G. and {Korn}, A.~J. and {K{\'o}sp{\'a}l}, {\'A}. and {Kostrzewa-Rutkowska}, Z. and {Kruszy{\'n}ska}, K. and {Kun}, M. and {Laizeau}, P. and {Lambert}, S. and {Lanza}, A.~F. and {Lasne}, Y. and {Le Campion}, J. -F. and {Lebreton}, Y. and {Lebzelter}, T. and {Leccia}, S. and {Leclerc}, N. and {Lecoeur-Taibi}, I. and {Liao}, S. and {Licata}, E.~L. and {Lindstr{\o}m}, H.~E.~P. and {Lister}, T.~A. and {Livanou}, E. and {Lobel}, A. and {Lorca}, A. and {Loup}, C. and {Madrero Pardo}, P. and {Magdaleno Romeo}, A. and {Managau}, S. and {Mann}, R.~G. and {Manteiga}, M. and {Marchant}, J.~M. and {Marconi}, M. and {Marcos}, J. and {Marcos Santos}, M.~M.~S. and {Mar{\'\i}n Pina}, D. and {Marinoni}, S. and {Marocco}, F. and {Marshall}, D.~J. and {Martin Polo}, L. and {Mart{\'\i}n-Fleitas}, J.~M. and {Marton}, G. and {Mary}, N. and {Masip}, A. and {Massari}, D. and {Mastrobuono-Battisti}, A. and {Mazeh}, T. and {McMillan}, P.~J. and {Messina}, S. and {Michalik}, D. and {Millar}, N.~R. and {Mints}, A. and {Molina}, D. and {Molinaro}, R. and {Moln{\'a}r}, L. and {Monari}, G. and {Mongui{\'o}}, M. and {Montegriffo}, P. and {Montero}, A. and {Mor}, R. and {Mora}, A. and {Morbidelli}, R. and {Morel}, T. and {Morris}, D. and {Muraveva}, T. and {Murphy}, C.~P. and {Musella}, I. and {Nagy}, Z. and {Noval}, L. and {Oca{\~n}a}, F. and {Ogden}, A. and {Ordenovic}, C. and {Osinde}, J.~O. and {Pagani}, C. and {Pagano}, I. and {Palaversa}, L. and {Palicio}, P.~A. and {Pallas-Quintela}, L. and {Panahi}, A. and {Payne-Wardenaar}, S. and {Pe{\~n}alosa Esteller}, X. and {Penttil{\"a}}, A. and {Pichon}, B. and {Piersimoni}, A.~M. and {Pineau}, F. -X. and {Plachy}, E. and {Plum}, G. and {Poggio}, E. and {Pr{\v{s}}a}, A. and {Pulone}, L. and {Racero}, E. and {Ragaini}, S. and {Rainer}, M. and {Raiteri}, C.~M. and {Rambaux}, N. and {Ramos}, P. and {Ramos-Lerate}, M. and {Re Fiorentin}, P. and {Regibo}, S. and {Richards}, P.~J. and {Rios Diaz}, C. and {Ripepi}, V. and {Riva}, A. and {Rix}, H. -W. and {Rixon}, G. and {Robichon}, N. and {Robin}, A.~C. and {Robin}, C. and {Roelens}, M. and {Rogues}, H.~R.~O. and {Rohrbasser}, L. and {Romero-G{\'o}mez}, M. and {Rowell}, N. and {Royer}, F. and {Ruz Mieres}, D. and {Rybicki}, K.~A. and {Sadowski}, G. and {S{\'a}ez N{\'u}{\~n}ez}, A. and {Sagrist{\`a} Sell{\'e}s}, A. and {Sahlmann}, J. and {Salguero}, E. and {Samaras}, N. and {Sanchez Gimenez}, V. and {Sanna}, N. and {Santove{\~n}a}, R. and {Sarasso}, M. and {Schultheis}, M. and {Sciacca}, E. and {Segol}, M. and {Segovia}, J.~C. and {S{\'e}gransan}, D. and {Semeux}, D. and {Shahaf}, S. and {Siddiqui}, H.~I. and {Siebert}, A. and {Siltala}, L. and {Silvelo}, A. and {Slezak}, E. and {Slezak}, I. and {Smart}, R.~L. and {Snaith}, O.~N. and {Solano}, E. and {Solitro}, F. and {Souami}, D. and {Souchay}, J. and {Spagna}, A. and {Spina}, L. and {Spoto}, F. and {Steele}, I.~A. and {Steidelm{\"u}ller}, H. and {Stephenson}, C.~A. and {S{\"u}veges}, M. and {Surdej}, J. and {Szabados}, L. and {Szegedi-Elek}, E. and {Taris}, F. and {Taylor}, M.~B. and {Teixeira}, R. and {Tolomei}, L. and {Tonello}, N. and {Torra}, F. and {Torra}, J. and {Torralba Elipe}, G. and {Trabucchi}, M. and {Tsounis}, A.~T. and {Turon}, C. and {Ulla}, A. and {Unger}, N. and {Vaillant}, M.~V. and {van Dillen}, E. and {van Reeven}, W. and {Vanel}, O. and {Vecchiato}, A. and {Viala}, Y. and {Vicente}, D. and {Voutsinas}, S. and {Weiler}, M. and {Wevers}, T. and {Wyrzykowski}, {\L}. and {Yoldas}, A. and {Yvard}, P. and {Zhao}, H. and {Zorec}, J. and {Zucker}, S. and {Zwitter}, T.},
        title = "{Gaia Data Release 3. Summary of the content and survey properties}",
      journal = {\aap},
         year = 2023,
        month = jun,
       volume = {674},
          eid = {A1},
        pages = {A1},
          doi = {10.1051/0004-6361/202243940},
archivePrefix = {arXiv},
       eprint = {2208.00211},
 primaryClass = {astro-ph.GA},
       adsurl = {https://ui.adsabs.harvard.edu/abs/2023A&A...674A...1G}
}

@ARTICLE{Graczyk13,
       author = {{Graczyk}, Dariusz and {Pietrzy{\'n}ski}, Grzegorz and
         {Thompson}, Ian B. and {Gieren}, Wolfgang and {Pilecki}, Bogumi{\l} and
         {Konorski}, Piotr and {Udalski}, Andrzej and {Soszy{\'n}ski}, Igor and
         {Villanova}, Sandro and {G{\'o}rski}, Marek and {Suchomska}, Ksenia and
         {Karczmarek}, Paulina and {Kudritzki}, Rolf-Peter and
         {Bresolin}, Fabio and {Gallenne}, Alexandre},
        title = "{The Araucaria Project. The Distance to the Small Magellanic Cloud from Late-type Eclipsing Binaries}",
      journal = {\apj},
         year = "2014",
        month = "Jan",
       volume = {780},
       number = {1},
          eid = {59},
        pages = {59},
          doi = {10.1088/0004-637X/780/1/59},
archivePrefix = {arXiv},
       eprint = {1311.2340},
 primaryClass = {astro-ph.CO},
       adsurl = {https://ui.adsabs.harvard.edu/abs/2014ApJ...780...59G}
}

@ARTICLE{Graczyk20,
       author = {{Graczyk}, Dariusz and {Pietrzy{\'n}ski}, Grzegorz and {Thompson}, Ian B. and {Gieren}, Wolfgang and {Zgirski}, Bart{\l}omiej and {Villanova}, Sandro and {G{\'o}rski}, Marek and {Wielg{\'o}rski}, Piotr and {Karczmarek}, Paulina and {Narloch}, Weronika and {Pilecki}, Bogumi{\l} and {Taormina}, Monica and {Smolec}, Rados{\l}aw and {Suchomska}, Ksenia and {Gallenne}, Alexandre and {Nardetto}, Nicolas and {Storm}, Jesper and {Kudritzki}, Rolf-Peter and {Ka{\l}uszy{\'n}ski}, Miko{\l}aj and {Pych}, Wojciech},
        title = "{A Distance Determination to the Small Magellanic Cloud with an Accuracy of Better than Two Percent Based on Late-type Eclipsing Binary Stars}",
      journal = {\apj},
         year = 2020,
        month = nov,
       volume = {904},
       number = {1},
          eid = {13},
        pages = {13},
          doi = {10.3847/1538-4357/abbb2b},
archivePrefix = {arXiv},
       eprint = {2010.08754},
 primaryClass = {astro-ph.GA},
       adsurl = {https://ui.adsabs.harvard.edu/abs/2020ApJ...904...13G}
}

@ARTICLE{Grady21,
       author = {{Grady}, J. and {Belokurov}, V. and {Evans}, N.~W.},
        title = "{Magellanic Mayhem: Metallicities and Motions}",
      journal = {\apj},
         year = 2021,
        month = mar,
       volume = {909},
       number = {2},
          eid = {150},
        pages = {150},
          doi = {10.3847/1538-4357/abd4e4},
archivePrefix = {arXiv},
       eprint = {2010.05956},
 primaryClass = {astro-ph.GA},
       adsurl = {https://ui.adsabs.harvard.edu/abs/2021ApJ...909..150G}
}

@ARTICLE{Grocholski06,
       author = {{Grocholski}, Aaron J. and {Cole}, Andrew A. and {Sarajedini}, Ata and {Geisler}, Doug and {Smith}, Verne V.},
        title = "{Ca II Triplet Spectroscopy of Large Magellanic Cloud Red Giants. I. Abundances and Velocities for a Sample of Populous Clusters}",
      journal = {\aj},
         year = 2006,
        month = oct,
       volume = {132},
       number = {4},
        pages = {1630-1644},
          doi = {10.1086/507303},
archivePrefix = {arXiv},
       eprint = {astro-ph/0607052},
 primaryClass = {astro-ph},
       adsurl = {https://ui.adsabs.harvard.edu/abs/2006AJ....132.1630G}
}

@ARTICLE{Harris06,
       author = {{Harris}, Jason and {Zaritsky}, Dennis},
        title = "{Spectroscopic Survey of Red Giants in the Small Magellanic Cloud. I. Kinematics}",
      journal = {\aj},
         year = 2006,
        month = may,
       volume = {131},
       number = {5},
        pages = {2514-2524},
          doi = {10.1086/500974},
archivePrefix = {arXiv},
       eprint = {astro-ph/0601025},
 primaryClass = {astro-ph},
       adsurl = {https://ui.adsabs.harvard.edu/abs/2006AJ....131.2514H}
}

@ARTICLE{Harris2020,
  author  = {Harris, Charles R. and Millman, K. Jarrod and
            van der Walt, Stéfan J and Gommers, Ralf and
            Virtanen, Pauli and Cournapeau, David and
            Wieser, Eric and Taylor, Julian and Berg, Sebastian and
            Smith, Nathaniel J. and Kern, Robert and Picus, Matti and
            Hoyer, Stephan and van Kerkwijk, Marten H. and
            Brett, Matthew and Haldane, Allan and
            Fernández del Río, Jaime and Wiebe, Mark and
            Peterson, Pearu and Gérard-Marchant, Pierre and
            Sheppard, Kevin and Reddy, Tyler and Weckesser, Warren and
            Abbasi, Hameer and Gohlke, Christoph and
            Oliphant, Travis E.},
  title   = {Array programming with {NumPy}},
  journal = {Nature},
  year    = {2020},
  volume  = {585},
  pages   = {357–362},
  doi     = {10.1038/s41586-020-2649-2}
}

@ARTICLE{Hernquist90,
       author = {{Hernquist}, Lars},
        title = "{An Analytical Model for Spherical Galaxies and Bulges}",
      journal = {\apj},
         year = 1990,
        month = jun,
       volume = {356},
        pages = {359},
          doi = {10.1086/168845},
       adsurl = {https://ui.adsabs.harvard.edu/abs/1990ApJ...356..359H}
}

@ARTICLE{Hidalgo18,
       author = {{Hidalgo}, Sebastian L. and {Pietrinferni}, Adriano and {Cassisi}, Santi and {Salaris}, Maurizio and {Mucciarelli}, Alessio and {Savino}, Alessandro and {Aparicio}, Antonio and {Silva Aguirre}, Victor and {Verma}, Kuldeep},
        title = "{The Updated BaSTI Stellar Evolution Models and Isochrones. I. Solar-scaled Calculations}",
      journal = {\apj},
         year = 2018,
        month = apr,
       volume = {856},
       number = {2},
          eid = {125},
        pages = {125},
          doi = {10.3847/1538-4357/aab158},
archivePrefix = {arXiv},
       eprint = {1802.07319},
 primaryClass = {astro-ph.GA},
       adsurl = {https://ui.adsabs.harvard.edu/abs/2018ApJ...856..125H}
}

@ARTICLE{Horta21,
       author = {{Horta}, Danny and {Hughes}, Meghan E. and {Pfeffer}, Joel L. and {Bastian}, Nate and {Kruijssen}, J.~M. Diederik and {Reina-Campos}, Marta and {Crain}, Rob A.},
        title = "{Linking globular cluster formation at low and high redshift through the age-metallicity relation in E-MOSAICS}",
      journal = {\mnras},
         year = 2021,
        month = jan,
       volume = {500},
       number = {4},
        pages = {4768-4778},
          doi = {10.1093/mnras/staa3522},
archivePrefix = {arXiv},
       eprint = {2010.10522},
 primaryClass = {astro-ph.GA},
       adsurl = {https://ui.adsabs.harvard.edu/abs/2021MNRAS.500.4768H}
}

@Article{Hunter07,
  Author    = {Hunter, J. D.},
  Title     = {Matplotlib: A 2D graphics environment},
  Journal   = {Computing in Science \& Engineering},
  Volume    = {9},
  Number    = {3},
  Pages     = {90--95},
  publisher = {IEEE COMPUTER SOC},
  doi       = {10.1109/MCSE.2007.55},
  year      = 2007
}

@ARTICLE{Jimenez23,
       author = {{Jim{\'e}nez-Arranz}, {\'O}. and {Romero-G{\'o}mez}, M. and {Luri}, X. and {McMillan}, P.~J. and {Antoja}, T. and {Chemin}, L. and {Roca-F{\`a}brega}, S. and {Masana}, E. and {Muros}, A.},
        title = "{Kinematic analysis of the Large Magellanic Cloud using Gaia DR3}",
      journal = {\aap},
         year = 2023,
        month = jan,
       volume = {669},
          eid = {A91},
        pages = {A91},
          doi = {10.1051/0004-6361/202244601},
archivePrefix = {arXiv},
       eprint = {2210.01728},
 primaryClass = {astro-ph.GA},
       adsurl = {https://ui.adsabs.harvard.edu/abs/2023A&A...669A..91J}
}

@ARTICLE{Jimenez25,
       author = {{Jim{\'e}nez-Arranz}, {\'O}. and {Roca-F{\`a}brega}, S.},
        title = "{Tidal interaction can stop galactic bars: On the LMC non-rotating bar}",
      journal = {\aap},
         year = 2025,
        month = jun,
       volume = {698},
          eid = {L7},
        pages = {L7},
          doi = {10.1051/0004-6361/202555019},
archivePrefix = {arXiv},
       eprint = {2504.01870},
 primaryClass = {astro-ph.GA},
       adsurl = {https://ui.adsabs.harvard.edu/abs/2025A&A...698L...7J}
}

@ARTICLE{Kacharov24,
       author = {{Kacharov}, Nikolay and {Tahmasebzadeh}, Behzad and {Cioni}, Maria-Rosa L. and {van de Ven}, Glenn and {Zhu}, Ling and {Khoperskov}, Sergey},
        title = "{Equilibrium dynamical models in the inner region of the Large Magellanic Cloud based on Gaia DR3 kinematics}",
      journal = {\aap},
         year = 2024,
        month = dec,
       volume = {692},
          eid = {A40},
        pages = {A40},
          doi = {10.1051/0004-6361/202451578},
archivePrefix = {arXiv},
       eprint = {2410.05374},
 primaryClass = {astro-ph.GA},
       adsurl = {https://ui.adsabs.harvard.edu/abs/2024A&A...692A..40K}
}

@ARTICLE{Kallivayalil13,
   author = {{Kallivayalil}, N. and {van der Marel}, R.~P. and {Besla}, G. and 
	{Anderson}, J. and {Alcock}, C.},
    title = "{Third-epoch Magellanic Cloud Proper Motions. I. Hubble Space Telescope/WFC3 Data and Orbit Implications}",
  journal = {\apj},
archivePrefix = "arXiv",
   eprint = {1301.0832},
     year = 2013,
    month = feb,
   volume = 764,
      eid = {161},
    pages = {161},
      doi = {10.1088/0004-637X/764/2/161},
   adsurl = {http://esoads.eso.org/abs/2013ApJ...764..161K}
}

@ARTICLE{Kamann23,
       author = {{Kamann}, S. and {Saracino}, S. and {Bastian}, N. and {Gossage}, S. and {Usher}, C. and {Baade}, D. and {Cabrera-Ziri}, I. and {de Mink}, S.~E. and {Ekstrom}, S. and {Georgy}, C. and {Hilker}, M. and {Larsen}, S.~S. and {Mackey}, D. and {Niederhofer}, F. and {Platais}, I. and {Yong}, D.},
        title = "{The effects of stellar rotation along the main sequence of the 100-Myr-old massive cluster NGC 1850}",
      journal = {\mnras},
         year = 2023,
        month = jan,
       volume = {518},
       number = {1},
        pages = {1505-1521},
          doi = {10.1093/mnras/stac3170},
archivePrefix = {arXiv},
       eprint = {2211.00693},
 primaryClass = {astro-ph.SR},
       adsurl = {https://ui.adsabs.harvard.edu/abs/2023MNRAS.518.1505K}
}

@INCOLLECTION{Kluyver16,
       author = {{Kluyver}, Thomas and {Ragan-Kelley}, Benjain and {P{\'e}rez}, Fernando and {Granger}, Brian and {Bussonnier}, Matthias and {Frederic}, Jonathan and {Kelley}, Kyle and {Hamrick}, Jessica and {Grout}, Jason and {Corlay}, Sylvain and {Ivanov}, Paul and {Avila}, Dami{\'a}n and {Abdalla}, Safia and {Willing}, Carol and {Jupyter Development Team}},
        title = "{Jupyter Notebooks{\textemdash}a publishing format for reproducible computational workflows}",
    booktitle = {IOS Press},
         year = 2016,
        pages = {87-90},
          doi = {10.3233/978-1-61499-649-1-87},
       adsurl = {https://ui.adsabs.harvard.edu/abs/2016ppap.book...87K}
}

@ARTICLE{Libralato18b,
       author = {{Libralato}, Mattia and {Bellini}, Andrea and {Bedin}, Luigi R. and {Moreno D.}, Edmundo and {Fern{\'a}ndez-Trincado}, Jos{\'e} G. and {Pichardo}, Barbara and {van der Marel}, Roeland P. and {Anderson}, Jay and {Apai}, D{\'a}niel and {Burgasser}, Adam J. and {Fabiola Marino}, Anna and {Milone}, Antonino P. and {Rees}, Jon M. and {Watkins}, Laura L.},
        title = "{The HST Large Programme on {\ensuremath{\omega}} Centauri. III. Absolute Proper Motion}",
      journal = {\apj},
         year = 2018,
        month = feb,
       volume = {854},
       number = {1},
          eid = {45},
        pages = {45},
          doi = {10.3847/1538-4357/aaa59e},
archivePrefix = {arXiv},
       eprint = {1801.01502},
 primaryClass = {astro-ph.GA},
       adsurl = {https://ui.adsabs.harvard.edu/abs/2018ApJ...854...45L}
}

@ARTICLE{Libralato22,
       author = {{Libralato}, Mattia and {Bellini}, Andrea and {Vesperini}, Enrico and {Piotto}, Giampaolo and {Milone}, Antonino P. and {van der Marel}, Roeland P. and {Anderson}, Jay and {Aparicio}, Antonio and {Barbuy}, Beatriz and {Bedin}, Luigi R. and {Borsato}, Luca and {Cassisi}, Santi and {Dalessandro}, Emanuele and {Ferraro}, Francesco R. and {King}, Ivan R. and {Lanzoni}, Barbara and {Nardiello}, Domenico and {Ortolani}, Sergio and {Sarajedini}, Ata and {Sohn}, Sangmo Tony},
        title = "{The Hubble Space Telescope UV Legacy Survey of Galactic Globular Clusters. XXIII. Proper-motion Catalogs and Internal Kinematics}",
      journal = {\apj},
         year = 2022,
        month = aug,
       volume = {934},
       number = {2},
          eid = {150},
        pages = {150},
          doi = {10.3847/1538-4357/ac7727},
archivePrefix = {arXiv},
       eprint = {2206.09924},
 primaryClass = {astro-ph.GA},
       adsurl = {https://ui.adsabs.harvard.edu/abs/2022ApJ...934..150L}
}

@ARTICLE{Marino18,
       author = {{Marino}, A.~F. and {Przybilla}, N. and {Milone}, A.~P. and {Da Costa}, G. and {D'Antona}, F. and {Dotter}, A. and {Dupree}, A.},
        title = "{Different Stellar Rotations in the Two Main Sequences of the Young Globular Cluster NGC 1818: The First Direct Spectroscopic Evidence}",
      journal = {\aj},
         year = 2018,
        month = sep,
       volume = {156},
       number = {3},
          eid = {116},
        pages = {116},
          doi = {10.3847/1538-3881/aad3cd},
archivePrefix = {arXiv},
       eprint = {1807.04493},
 primaryClass = {astro-ph.SR},
       adsurl = {https://ui.adsabs.harvard.edu/abs/2018AJ....156..116M}
}

@ARTICLE{Massana22,
       author = {{Massana}, P. and {Ruiz-Lara}, T. and {No{\"e}l}, N.~E.~D. and {Gallart}, C. and {Nidever}, D.~L. and {Choi}, Y. and {Sakowska}, J.~D. and {Besla}, G. and {Olsen}, K.~A.~G. and {Monelli}, M. and {Dorta}, A. and {Stringfellow}, G.~S. and {Cassisi}, S. and {Bernard}, E.~J. and {Zaritsky}, D. and {Cioni}, M. -R.~L. and {Monachesi}, A. and {van der Marel}, R.~P. and {de Boer}, T.~J.~L. and {Walker}, A.~R.},
        title = "{The synchronized dance of the magellanic clouds' star formation history}",
      journal = {\mnras},
         year = 2022,
        month = jun,
       volume = {513},
       number = {1},
        pages = {L40-L45},
          doi = {10.1093/mnrasl/slac030},
archivePrefix = {arXiv},
       eprint = {2203.09523},
 primaryClass = {astro-ph.GA},
       adsurl = {https://ui.adsabs.harvard.edu/abs/2022MNRAS.513L..40M}
}

@ARTICLE{Massari21,
       author = {{Massari}, Davide and {Raso}, Silvia and {Libralato}, Mattia and {Bellini}, Andrea},
        title = "{Kinematic complexity around NGC 419: resolving the proper motion of the cluster, the Small Magellanic Cloud, and the Magellanic bridge}",
      journal = {\mnras},
         year = 2021,
        month = jan,
       volume = {500},
       number = {2},
        pages = {2012-2019},
          doi = {10.1093/mnras/staa3497},
archivePrefix = {arXiv},
       eprint = {2011.03288},
 primaryClass = {astro-ph.GA},
       adsurl = {https://ui.adsabs.harvard.edu/abs/2021MNRAS.500.2012M}
}

@ARTICLE{Massari23,
       author = {{Massari}, Davide and {Aguado-Agelet}, Fernando and {Monelli}, Matteo and {Cassisi}, Santi and {Pancino}, Elena and {Saracino}, Sara and {Gallart}, Carme and {Ruiz-Lara}, Tom{\'a}s and {Fern{\'a}ndez-Alvar}, Emma and {Surot}, Francisco and {Stokholm}, Amalie and {Salaris}, Maurizio and {Miglio}, Andrea and {Ceccarelli}, Edoardo},
        title = "{Cluster Ages to Reconstruct the Milky Way Assembly (CARMA). I. The final word on the origin of NGC 6388 and NGC 6441}",
      journal = {\aap},
         year = 2023,
        month = dec,
       volume = {680},
          eid = {A20},
        pages = {A20},
          doi = {10.1051/0004-6361/202347289},
archivePrefix = {arXiv},
       eprint = {2310.01495},
 primaryClass = {astro-ph.GA},
       adsurl = {https://ui.adsabs.harvard.edu/abs/2023A&A...680A..20M}
}

@ARTICLE{Mazzi21,
       author = {{Mazzi}, Alessandro and {Girardi}, L{\'e}o and {Zaggia}, Simone and {Pastorelli}, Giada and {Rubele}, Stefano and {Bressan}, Alessandro and {Cioni}, Maria-Rosa L. and {Clementini}, Gisella and {Cusano}, Felice and {Rocha}, Jo{\~a}o Pedro and {Gullieuszik}, Marco and {Kerber}, Leandro and {Marigo}, Paola and {Ripepi}, Vincenzo and {Bekki}, Kenji and {Bell}, Cameron P.~M. and {de Grijs}, Richard and {Groenewegen}, Martin A.~T. and {Ivanov}, Valentin D. and {Oliveira}, Joana M. and {Sun}, Ning-Chen and {van Loon}, Jacco Th},
        title = "{The VMC survey - XLIII. The spatially resolved star formation history across the Large Magellanic Cloud}",
      journal = {\mnras},
         year = 2021,
        month = nov,
       volume = {508},
       number = {1},
        pages = {245-266},
          doi = {10.1093/mnras/stab2399},
archivePrefix = {arXiv},
       eprint = {2108.07225},
 primaryClass = {astro-ph.GA},
       adsurl = {https://ui.adsabs.harvard.edu/abs/2021MNRAS.508..245M}
}

@ARTICLE{Milone18,
       author = {{Milone}, A.~P. and {Marino}, A.~F. and {Di Criscienzo}, M. and {D'Antona}, F. and {Bedin}, L.~R. and {Da Costa}, G. and {Piotto}, G. and {Tailo}, M. and {Dotter}, A. and {Angeloni}, R. and {Anderson}, J. and {Jerjen}, H. and {Li}, C. and {Dupree}, A. and {Granata}, V. and {Lagioia}, E.~P. and {Mackey}, A.~D. and {Nardiello}, D. and {Vesperini}, E.},
        title = "{Multiple stellar populations in Magellanic Cloud clusters - VI. A survey of multiple sequences and Be stars in young clusters}",
      journal = {\mnras},
         year = 2018,
        month = jun,
       volume = {477},
       number = {2},
        pages = {2640-2663},
          doi = {10.1093/mnras/sty661},
archivePrefix = {arXiv},
       eprint = {1802.10538},
 primaryClass = {astro-ph.SR},
       adsurl = {https://ui.adsabs.harvard.edu/abs/2018MNRAS.477.2640M}
}

@ARTICLE{Milone23a,
       author = {{Milone}, A.~P. and {Cordoni}, G. and {Marino}, A.~F. and {D'Antona}, F. and {Bellini}, A. and {Di Criscienzo}, M. and {Dondoglio}, E. and {Lagioia}, E.~P. and {Langer}, N. and {Legnardi}, M.~V. and {Libralato}, M. and {Baumgardt}, H. and {Bettinelli}, M. and {Cavecchi}, Y. and {de Grijs}, R. and {Deng}, L. and {Hastings}, B. and {Li}, C. and {Mohandasan}, A. and {Renzini}, A. and {Vesperini}, E. and {Wang}, C. and {Ziliotto}, T. and {Carlos}, M. and {Costa}, G. and {Dell'Agli}, F. and {Di Stefano}, S. and {Jang}, S. and {Martorano}, M. and {Simioni}, M. and {Tailo}, M. and {Ventura}, P.},
        title = "{Hubble Space Telescope survey of Magellanic Cloud star clusters. Photometry and astrometry of 113 clusters and early results}",
      journal = {\aap},
         year = 2023,
        month = apr,
       volume = {672},
          eid = {A161},
        pages = {A161},
          doi = {10.1051/0004-6361/202244798},
archivePrefix = {arXiv},
       eprint = {2212.07978},
 primaryClass = {astro-ph.SR},
       adsurl = {https://ui.adsabs.harvard.edu/abs/2023A&A...672A.161M}
}

@ARTICLE{Milone23b,
       author = {{Milone}, A.~P. and {Cordoni}, G. and {Marino}, A.~F. and {Muratore}, F. and {D'Antona}, F. and {Di Criscienzo}, M. and {Dondoglio}, E. and {Lagioia}, E.~P. and {Legnardi}, M.~V. and {Mohandasan}, A. and {Ziliotto}, T. and {Dell'Agli}, F. and {Tailo}, M. and {Ventura}, P.},
        title = "{Hubble Space Telescope survey of Magellanic Cloud star clusters: UV-dim stars in young clusters}",
      journal = {\mnras},
         year = 2023,
        month = oct,
       volume = {524},
       number = {4},
        pages = {6149-6158},
          doi = {10.1093/mnras/stad2242},
archivePrefix = {arXiv},
       eprint = {2307.10020},
 primaryClass = {astro-ph.SR},
       adsurl = {https://ui.adsabs.harvard.edu/abs/2023MNRAS.524.6149M}
}

@ARTICLE{Mucciarelli06,
       author = {{Mucciarelli}, Alessio and {Origlia}, Livia and {Ferraro}, Francesco R. and {Maraston}, Claudia and {Testa}, Vincenzo},
        title = "{Red Giant Stars in the Large Magellanic Cloud Clusters}",
      journal = {\apj},
         year = 2006,
        month = aug,
       volume = {646},
       number = {2},
        pages = {939-950},
          doi = {10.1086/504969},
archivePrefix = {arXiv},
       eprint = {astro-ph/0604139},
 primaryClass = {astro-ph},
       adsurl = {https://ui.adsabs.harvard.edu/abs/2006ApJ...646..939M}
}

@ARTICLE{Mucciarelli08,
       author = {{Mucciarelli}, Alessio and {Carretta}, Eugenio and {Origlia}, Livia and {Ferraro}, Francesco R.},
        title = "{The Chemical Composition of Red Giant Stars in Four Intermediate-Age Clusters of the Large Magellanic Cloud}",
      journal = {\aj},
         year = 2008,
        month = jul,
       volume = {136},
       number = {1},
        pages = {375-388},
          doi = {10.1088/0004-6256/136/1/375},
archivePrefix = {arXiv},
       eprint = {0804.4061},
 primaryClass = {astro-ph},
       adsurl = {https://ui.adsabs.harvard.edu/abs/2008AJ....136..375M}
}

@ARTICLE{Mucciarelli11,
       author = {{Mucciarelli}, A. and {Cristallo}, S. and {Brocato}, E. and {Pasquini}, L. and {Straniero}, O. and {Caffau}, E. and {Raimondo}, G. and {Kaufer}, A. and {Musella}, I. and {Ripepi}, V. and {Romaniello}, M. and {Walker}, A.~R.},
        title = "{NGC 1866: a milestone for understanding the chemical evolution of stellar populations in the Large Magellanic Cloud}",
      journal = {\mnras},
         year = 2011,
        month = may,
       volume = {413},
       number = {2},
        pages = {837-851},
          doi = {10.1111/j.1365-2966.2010.18167.x},
archivePrefix = {arXiv},
       eprint = {1012.1476},
 primaryClass = {astro-ph.SR},
       adsurl = {https://ui.adsabs.harvard.edu/abs/2011MNRAS.413..837M}
}

@ARTICLE{Mucciarelli14,
       author = {{Mucciarelli}, A. and {Dalessandro}, E. and {Ferraro}, F.~R. and {Origlia}, L. and {Lanzoni}, B.},
        title = "{No Evidence of Chemical Anomalies in the Bimodal Turnoff Cluster NGC 1806 in the Large Magellanic Cloud}",
      journal = {\apjl},
         year = 2014,
        month = sep,
       volume = {793},
       number = {1},
          eid = {L6},
        pages = {L6},
          doi = {10.1088/2041-8205/793/1/L6},
archivePrefix = {arXiv},
       eprint = {1409.0259},
 primaryClass = {astro-ph.SR},
       adsurl = {https://ui.adsabs.harvard.edu/abs/2014ApJ...793L...6M}
}

@ARTICLE{Mucciarelli21,
       author = {{Mucciarelli}, A. and {Massari}, D. and {Minelli}, A. and {Romano}, D. and {Bellazzini}, M. and {Ferraro}, F.~R. and {Matteucci}, F. and {Origlia}, L.},
        title = "{A relic from a past merger event in the Large Magellanic Cloud}",
      journal = {Nature Astronomy},
         year = 2021,
        month = dec,
       volume = {5},
        pages = {1247-1254},
          doi = {10.1038/s41550-021-01493-y},
archivePrefix = {arXiv},
       eprint = {2110.10561},
 primaryClass = {astro-ph.GA},
       adsurl = {https://ui.adsabs.harvard.edu/abs/2021NatAs...5.1247M}
}

@ARTICLE{Narloch22,
       author = {{Narloch}, W. and {Pietrzy{\'n}ski}, G. and {Gieren}, W. and {Piatti}, A.~E. and {Karczmarek}, P. and {G{\'o}rski}, M. and {Graczyk}, D. and {Smolec}, R. and {Hajdu}, G. and {Suchomska}, K. and {Zgirski}, B. and {Wielg{\'o}rski}, P. and {Pilecki}, B. and {Taormina}, M. and {Ka{\l}uszy{\'n}ski}, M. and {Pych}, W. and {Rojas Garc{\'\i}a}, G. and {Lewis}, M.~O.},
        title = "{Metallicities and ages for star clusters and their surrounding fields in the Large Magellanic Cloud}",
      journal = {\aap},
         year = 2022,
        month = oct,
       volume = {666},
          eid = {A80},
        pages = {A80},
          doi = {10.1051/0004-6361/202243378},
archivePrefix = {arXiv},
       eprint = {2207.13153},
 primaryClass = {astro-ph.GA},
       adsurl = {https://ui.adsabs.harvard.edu/abs/2022A&A...666A..80N}
}

@ARTICLE{Navarrete23,
       author = {{Navarrete}, Camila and {Aguado}, David S. and {Belokurov}, Vasily and {Erkal}, Denis and {Deason}, Alis and {Cullinane}, Lara and {Carballo-Bello}, Julio},
        title = "{The 3D kinematics of stellar substructures in the periphery of the Large Magellanic Cloud}",
      journal = {\mnras},
         year = 2023,
        month = aug,
       volume = {523},
       number = {3},
        pages = {4720-4738},
          doi = {10.1093/mnras/stad1698},
archivePrefix = {arXiv},
       eprint = {2302.04579},
 primaryClass = {astro-ph.GA},
       adsurl = {https://ui.adsabs.harvard.edu/abs/2023MNRAS.523.4720N}
}

@ARTICLE{Navarro97,
       author = {{Navarro}, Julio F. and {Frenk}, Carlos S. and {White}, Simon D.~M.},
        title = "{A Universal Density Profile from Hierarchical Clustering}",
      journal = {\apj},
         year = 1997,
        month = dec,
       volume = {490},
       number = {2},
        pages = {493-508},
          doi = {10.1086/304888},
archivePrefix = {arXiv},
       eprint = {astro-ph/9611107},
 primaryClass = {astro-ph},
       adsurl = {https://ui.adsabs.harvard.edu/abs/1997ApJ...490..493N}
}

@ARTICLE{Niederhofer22,
       author = {{Niederhofer}, Florian and {Cioni}, Maria-Rosa L. and {Schmidt}, Thomas and {Bekki}, Kenji and {de Grijs}, Richard and {Ivanov}, Valentin D. and {Oliveira}, Joana M. and {Ripepi}, Vincenzo and {Subramanian}, Smitha and {van Loon}, Jacco Th},
        title = "{The VMC survey - XLVI. Stellar proper motions in the centre of the Large Magellanic Cloud}",
      journal = {\mnras},
         year = 2022,
        month = jun,
       volume = {512},
       number = {4},
        pages = {5423-5439},
          doi = {10.1093/mnras/stac712},
archivePrefix = {arXiv},
       eprint = {2203.14369},
 primaryClass = {astro-ph.GA},
       adsurl = {https://ui.adsabs.harvard.edu/abs/2022MNRAS.512.5423N}
}

@ARTICLE{Niederhofer24,
       author = {{Niederhofer}, F. and {Bellini}, A. and {Kozhurina-Platais}, V. and {Libralato}, M. and {H{\"a}berle}, M. and {Kacharov}, N. and {Kamann}, S. and {Bastian}, N. and {Cabrera-Ziri}, I. and {Cioni}, M. -R.~L. and {Dresbach}, F. and {Martocchia}, S. and {Massari}, D. and {Saracino}, S.},
        title = "{Hubble Space Telescope proper motions of Large Magellanic Cloud star clusters: I. Catalogues and results for NGC 1850}",
      journal = {\aap},
         year = 2024,
        month = sep,
       volume = {689},
          eid = {A162},
        pages = {A162},
          doi = {10.1051/0004-6361/202450255},
archivePrefix = {arXiv},
       eprint = {2406.17347},
 primaryClass = {astro-ph.GA},
       adsurl = {https://ui.adsabs.harvard.edu/abs/2024A&A...689A.162N}
}

@ARTICLE{Niederhofer25,
       author = {{Niederhofer}, F. and {Massari}, D. and {Aguado-Agelet}, F. and {Cassisi}, S. and {Bellini}, A. and {Kozhurina-Platais}, V. and {Libralato}, M. and {Kacharov}, N. and {Mucciarelli}, A. and {Monelli}, M. and {Bastian}, N. and {Cabrera-Ziri}, I. and {Ceccarelli}, E. and {Cioni}, M. -R.~L. and {Dresbach}, F. and {H{\"a}berle}, M. and {Martocchia}, S. and {Saracino}, S.},
        title = "{Cluster Ages to Reconstruct the Milky Way Assembly (CARMA) IV. Chrono-dynamics of seven old star clusters in the Large Magellanic Cloud and the peculiar origin of NGC 1841}",
      journal = {arXiv e-prints},
         year = 2025,
        month = sep,
          eid = {arXiv:2509.10144},
        pages = {arXiv:2509.10144},
archivePrefix = {arXiv},
       eprint = {2509.10144},
 primaryClass = {astro-ph.GA},
       adsurl = {https://ui.adsabs.harvard.edu/abs/2025arXiv250910144N}
}

@ARTICLE{Olsen11,
       author = {{Olsen}, Knut A.~G. and {Zaritsky}, Dennis and {Blum}, Robert D. and
         {Boyer}, Martha L. and {Gordon}, Karl D.},
        title = "{A Population of Accreted Small Magellanic Cloud Stars in the Large Magellanic Cloud}",
      journal = {\apj},
         year = "2011",
        month = "Aug",
       volume = {737},
       number = {1},
          eid = {29},
        pages = {29},
          doi = {10.1088/0004-637X/737/1/29},
archivePrefix = {arXiv},
       eprint = {1106.0044},
 primaryClass = {astro-ph.GA},
       adsurl = {https://ui.adsabs.harvard.edu/abs/2011ApJ...737...29O}
}

@ARTICLE{Omkumar25,
       author = {{Omkumar}, Abinaya O. and {Cioni}, Maria-Rosa L. and {Subramanian}, Smitha and {de Bruijne}, Jos and {Nair}, Aparna and {Dias}, Bruno},
        title = "{Str{\"o}mgren photometric metallicity map of the Magellanic Cloud stars using Gaia DR3{\textendash}XP spectra}",
      journal = {\aap},
         year = 2025,
        month = aug,
       volume = {700},
          eid = {A74},
        pages = {A74},
          doi = {10.1051/0004-6361/202452510},
archivePrefix = {arXiv},
       eprint = {2506.10749},
 primaryClass = {astro-ph.GA},
       adsurl = {https://ui.adsabs.harvard.edu/abs/2025A&A...700A..74O}
}

@ARTICLE{Pagel98,
       author = {{Pagel}, B.~E.~J. and {Tautvaisiene}, G.},
        title = "{Chemical evolution of the Magellanic Clouds: analytical models}",
      journal = {\mnras},
         year = 1998,
        month = sep,
       volume = {299},
       number = {2},
        pages = {535-544},
          doi = {10.1046/j.1365-8711.1998.01792.x},
archivePrefix = {arXiv},
       eprint = {astro-ph/9801221},
 primaryClass = {astro-ph},
       adsurl = {https://ui.adsabs.harvard.edu/abs/1998MNRAS.299..535P}
}

@ARTICLE{Perez07,  author={F. {Perez} and B. E. {Granger}},  journal={Computing in Science   Engineering},  title={IPython: A System for Interactive Scientific Computing},   year={2007},  volume={9},  number={3},  pages={21-29},}

@ARTICLE{Piatti19,
       author = {{Piatti}, Andr{\'e}s E. and {Alfaro}, Emilio J. and {Cantat-Gaudin}, Tristan},
        title = "{Two kinematically distinct old globular cluster populations in the Large Magellanic Cloud}",
      journal = {\mnras},
         year = 2019,
        month = mar,
       volume = {484},
       number = {1},
        pages = {L19-L23},
          doi = {10.1093/mnrasl/sly240},
archivePrefix = {arXiv},
       eprint = {1812.10709},
 primaryClass = {astro-ph.GA},
       adsurl = {https://ui.adsabs.harvard.edu/abs/2019MNRAS.484L..19P}
}

@ARTICLE{Piatti19b,
       author = {{Piatti}, Andr{\'e}s E. and {Pietrzy{\'n}ski}, Grzegorz and {Narloch}, Weronika and {G{\'o}rski}, Marek and {Graczyk}, Dariusz},
        title = "{Metallicity estimates of young clusters in the Magellanic Clouds from Str{\"o}mgren photometry of supergiant stars}",
      journal = {\mnras},
         year = 2019,
        month = mar,
       volume = {483},
       number = {4},
        pages = {4766-4773},
          doi = {10.1093/mnras/sty3473},
archivePrefix = {arXiv},
       eprint = {1812.07911},
 primaryClass = {astro-ph.GA},
       adsurl = {https://ui.adsabs.harvard.edu/abs/2019MNRAS.483.4766P}
}

@ARTICLE{Pietrinferni21,
       author = {{Pietrinferni}, Adriano and {Hidalgo}, Sebastian and {Cassisi}, Santi and {Salaris}, Maurizio and {Savino}, Alessandro and {Mucciarelli}, Alessio and {Verma}, Kuldeep and {Silva Aguirre}, Victor and {Aparicio}, Antonio and {Ferguson}, Jason W.},
        title = "{Updated BaSTI Stellar Evolution Models and Isochrones. II. {\ensuremath{\alpha}}-enhanced Calculations}",
      journal = {\apj},
         year = 2021,
        month = feb,
       volume = {908},
       number = {1},
          eid = {102},
        pages = {102},
          doi = {10.3847/1538-4357/abd4d5},
archivePrefix = {arXiv},
       eprint = {2012.10085},
 primaryClass = {astro-ph.SR},
       adsurl = {https://ui.adsabs.harvard.edu/abs/2021ApJ...908..102P}
}

@ARTICLE{Pietrzynski19,
       author = {{Pietrzy{\'n}ski}, G. and {Graczyk}, D. and {Gallenne}, A. and
         {Gieren}, W. and {Thompson}, I.~B. and {Pilecki}, B. and
         {Karczmarek}, P. and {G{\'o}rski}, M. and {Suchomska}, K. and
         {Taormina}, M. and {Zgirski}, B. and {Wielg{\'o}rski}, P. and
         {Ko{\l}aczkowski}, Z. and {Konorski}, P. and {Villanova}, S. and
         {Nardetto}, N. and {Kervella}, P. and {Bresolin}, F. and
         {Kudritzki}, R.~P. and {Storm}, J. and {Smolec}, R. and {Narloch}, W.},
        title = "{A distance to the Large Magellanic Cloud that is precise to one per cent}",
      journal = {\nat},
         year = "2019",
        month = "Mar",
       volume = {567},
       number = {7747},
        pages = {200-203},
          doi = {10.1038/s41586-019-0999-4},
archivePrefix = {arXiv},
       eprint = {1903.08096},
 primaryClass = {astro-ph.GA},
       adsurl = {https://ui.adsabs.harvard.edu/abs/2019Natur.567..200P}
}

@ARTICLE{Rathore25,
       author = {{Rathore}, Himansh and {Besla}, Gurtina and {Daniel}, Kathryne J. and {Beraldo e Silva}, Leandro},
        title = "{Response of the LMC's Bar to a Recent SMC Collision and Implications for the SMC's Dark Matter Profile}",
      journal = {\apj},
         year = 2025,
        month = jul,
       volume = {988},
       number = {1},
          eid = {79},
        pages = {79},
          doi = {10.3847/1538-4357/ade0ae},
archivePrefix = {arXiv},
       eprint = {2504.16163},
 primaryClass = {astro-ph.GA},
       adsurl = {https://ui.adsabs.harvard.edu/abs/2025ApJ...988...79R}
}

@ARTICLE{Sabbi16,
       author = {{Sabbi}, E. and {Lennon}, D.~J. and {Anderson}, J. and {Cignoni}, M. and {van der Marel}, R.~P. and {Zaritsky}, D. and {De Marchi}, G. and {Panagia}, N. and {Gouliermis}, D.~A. and {Grebel}, E.~K. and {Gallagher}, J.~S., III and {Smith}, L.~J. and {Sana}, H. and {Aloisi}, A. and {Tosi}, M. and {Evans}, C.~J. and {Arab}, H. and {Boyer}, M. and {de Mink}, S.~E. and {Gordon}, K. and {Koekemoer}, A.~M. and {Larsen}, S.~S. and {Ryon}, J.~E. and {Zeidler}, P.},
        title = "{Hubble Tarantula Treasury Project. III. Photometric Catalog and Resulting Constraints on the Progression of Star Formation in the 30 Doradus Region}",
      journal = {\apjs},
         year = 2016,
        month = jan,
       volume = {222},
       number = {1},
          eid = {11},
        pages = {11},
          doi = {10.3847/0067-0049/222/1/11},
archivePrefix = {arXiv},
       eprint = {1511.06021},
 primaryClass = {astro-ph.GA},
       adsurl = {https://ui.adsabs.harvard.edu/abs/2016ApJS..222...11S}
}

@ARTICLE{Sakari17,
       author = {{Sakari}, Charli M. and {McWilliam}, Andrew and {Wallerstein}, George},
        title = "{Chemical Abundances of Two Stars in the Large Magellanic Cloud Globular Cluster NGC 1718}",
      journal = {\mnras},
         year = 2017,
        month = may,
       volume = {467},
       number = {1},
        pages = {1112-1125},
          doi = {10.1093/mnras/stx074},
archivePrefix = {arXiv},
       eprint = {1701.03802},
 primaryClass = {astro-ph.GA},
       adsurl = {https://ui.adsabs.harvard.edu/abs/2017MNRAS.467.1112S}
}

@ARTICLE{Salaris93,
       author = {{Salaris}, Maurizio and {Chieffi}, Alessandro and {Straniero}, Oscar},
        title = "{The alpha -enhanced Isochrones and Their Impact on the FITS to the Galactic Globular Cluster System}",
      journal = {\apj},
         year = 1993,
        month = sep,
       volume = {414},
        pages = {580},
          doi = {10.1086/173105},
       adsurl = {https://ui.adsabs.harvard.edu/abs/1993ApJ...414..580S}
}

@ARTICLE{Sharma10,
       author = {{Sharma}, Saurabh and {Borissova}, J. and {Kurtev}, R. and {Ivanov}, V.~D. and {Geisler}, D.},
        title = "{Toward the General Red Giant Branch Slope-Metallicity-Age Calibration. I. Metallicities, Ages, and Kinematics for Eight Large Magellanic Cloud Clusters}",
      journal = {\aj},
         year = 2010,
        month = mar,
       volume = {139},
       number = {3},
        pages = {878-897},
          doi = {10.1088/0004-6256/139/3/878},
archivePrefix = {arXiv},
       eprint = {0912.5208},
 primaryClass = {astro-ph.GA},
       adsurl = {https://ui.adsabs.harvard.edu/abs/2010AJ....139..878S}
}

@ARTICLE{Shipp21,
       author = {{Shipp}, Nora and {Erkal}, Denis and {Drlica-Wagner}, Alex and {Li}, Ting S. and {Pace}, Andrew B. and {Koposov}, Sergey E. and {Cullinane}, Lara R. and {Da Costa}, Gary S. and {Ji}, Alexander P. and {Kuehn}, Kyler and {Lewis}, Geraint F. and {Mackey}, Dougal and {Simpson}, Jeffrey D. and {Wan}, Zhen and {Zucker}, Daniel B. and {Bland-Hawthorn}, Joss and {Ferguson}, Peter S. and {Lilleengen}, Sophia and {Lilleengen}, Sophia},
        title = "{Measuring the Mass of the Large Magellanic Cloud with Stellar Streams Observed by S $^{5}$}",
      journal = {\apj},
         year = 2021,
        month = dec,
       volume = {923},
       number = {2},
          eid = {149},
        pages = {149},
          doi = {10.3847/1538-4357/ac2e93},
archivePrefix = {arXiv},
       eprint = {2107.13004},
 primaryClass = {astro-ph.GA},
       adsurl = {https://ui.adsabs.harvard.edu/abs/2021ApJ...923..149S}
}

@ARTICLE{Song21,
       author = {{Song}, Ying-Yi and {Mateo}, Mario and {Bailey}, John I. and {Walker}, Matthew G. and {Roederer}, Ian U. and {Olszewski}, Edward W. and {Reiter}, Megan and {Kremin}, Anthony},
        title = "{Dynamical masses and mass-to-light ratios of resolved massive star clusters - II. Results for 26 star clusters in the Magellanic Clouds}",
      journal = {\mnras},
         year = 2021,
        month = jul,
       volume = {504},
       number = {3},
        pages = {4160-4191},
          doi = {10.1093/mnras/stab1065},
archivePrefix = {arXiv},
       eprint = {2104.06882},
 primaryClass = {astro-ph.GA},
       adsurl = {https://ui.adsabs.harvard.edu/abs/2021MNRAS.504.4160S}
}

@ARTICLE{Subramanian12,
       author = {{Subramanian}, Smitha and {Subramaniam}, Annapurni},
        title = "{The Three-dimensional Structure of the Small Magellanic Cloud}",
      journal = {\apj},
         year = 2012,
        month = jan,
       volume = {744},
       number = {2},
          eid = {128},
        pages = {128},
          doi = {10.1088/0004-637X/744/2/128},
archivePrefix = {arXiv},
       eprint = {1109.3980},
 primaryClass = {astro-ph.CO},
       adsurl = {https://ui.adsabs.harvard.edu/abs/2012ApJ...744..128S}
}

@ARTICLE{Usher19,
       author = {{Usher}, Christopher and {Beckwith}, Thomas and {Bellstedt}, Sabine and {Alabi}, Adebusola and {Chevalier}, Leonie and {Pastorello}, Nicola and {Cerulo}, Pierluigi and {Dalgleish}, Hannah S. and {Fraser-McKelvie}, Amelia and {Kamann}, Sebastian and {Penny}, Samantha and {Foster}, Caroline and {McDermid}, Richard and {Schiavon}, Ricardo P. and {Villaume}, Alexa},
        title = "{The WAGGS project - II. The reliability of the calcium triplet as a metallicity indicator in integrated stellar light}",
      journal = {\mnras},
         year = 2019,
        month = jan,
       volume = {482},
       number = {1},
        pages = {1275-1303},
          doi = {10.1093/mnras/sty2611},
archivePrefix = {arXiv},
       eprint = {1809.07650},
 primaryClass = {astro-ph.GA},
       adsurl = {https://ui.adsabs.harvard.edu/abs/2019MNRAS.482.1275U}
}

@ARTICLE{vanderMarel01b,
       author = {{van der Marel}, Roeland P. and {Cioni}, Maria-Rosa L.},
        title = "{Magellanic Cloud Structure from Near-Infrared Surveys. I. The Viewing Angles of the Large Magellanic Cloud}",
      journal = {\aj},
         year = 2001,
        month = oct,
       volume = {122},
       number = {4},
        pages = {1807-1826},
          doi = {10.1086/323099},
archivePrefix = {arXiv},
       eprint = {astro-ph/0105339},
 primaryClass = {astro-ph},
       adsurl = {https://ui.adsabs.harvard.edu/abs/2001AJ....122.1807V}
}

@ARTICLE{vanderMarel02,
       author = {{van der Marel}, Roeland P. and {Alves}, David R. and {Hardy}, Eduardo and {Suntzeff}, Nicholas B.},
        title = "{New Understanding of Large Magellanic Cloud Structure, Dynamics, and Orbit from Carbon Star Kinematics}",
      journal = {\aj},
         year = 2002,
        month = nov,
       volume = {124},
       number = {5},
        pages = {2639-2663},
          doi = {10.1086/343775},
archivePrefix = {arXiv},
       eprint = {astro-ph/0205161},
 primaryClass = {astro-ph},
       adsurl = {https://ui.adsabs.harvard.edu/abs/2002AJ....124.2639V}
}

@ARTICLE{vanderMarel14,
   author = {{van der Marel}, R.~P. and {Kallivayalil}, N.},
    title = "{Third-epoch Magellanic Cloud Proper Motions. II. The Large Magellanic Cloud Rotation Field in Three Dimensions}",
  journal = {\apj},
archivePrefix = "arXiv",
   eprint = {1305.4641},
     year = 2014,
    month = feb,
   volume = 781,
      eid = {121},
    pages = {121},
      doi = {10.1088/0004-637X/781/2/121},
   adsurl = {http://adsabs.harvard.edu/abs/2014ApJ...781..121V}
}

@ARTICLE{Vasiliev19,
       author = {{Vasiliev}, Eugene},
        title = "{AGAMA: action-based galaxy modelling architecture}",
      journal = {\mnras},
         year = 2019,
        month = jan,
       volume = {482},
       number = {2},
        pages = {1525-1544},
          doi = {10.1093/mnras/sty2672},
archivePrefix = {arXiv},
       eprint = {1802.08239},
 primaryClass = {astro-ph.GA},
       adsurl = {https://ui.adsabs.harvard.edu/abs/2019MNRAS.482.1525V}
}

@ARTICLE{Vasiliev21b,
       author = {{Vasiliev}, Eugene and {Belokurov}, Vasily and {Erkal}, Denis},
        title = "{Tango for three: Sagittarius, LMC, and the Milky Way}",
      journal = {\mnras},
         year = 2021,
        month = feb,
       volume = {501},
       number = {2},
        pages = {2279-2304},
          doi = {10.1093/mnras/staa3673},
archivePrefix = {arXiv},
       eprint = {2009.10726},
 primaryClass = {astro-ph.GA},
       adsurl = {https://ui.adsabs.harvard.edu/abs/2021MNRAS.501.2279V}
}

@ARTICLE{Vijayasree25,
       author = {{Vijayasree}, Sreepriya and {Niederhofer}, Florian and {Cioni}, Maria-Rosa L. and {Cullinane}, Lara and {Bekki}, Kenji and {van Loon}, Jacco Th. and {Kacharov}, Nikolay and {de Grijs}, Richard and {Ivanov}, Valentin D. and {Oliveira}, Joana M. and {Dresbach}, Francesca and {Groenewegen}, Martin A.~T. and {Erkal}, Denis},
        title = "{The VMC survey: LIV. The internal kinematics of the Large Magellanic Cloud with new VISTA observations}",
      journal = {\aap},
         year = 2025,
        month = aug,
       volume = {700},
          eid = {A279},
        pages = {A279},
          doi = {10.1051/0004-6361/202453145},
archivePrefix = {arXiv},
       eprint = {2503.13039},
 primaryClass = {astro-ph.GA},
       adsurl = {https://ui.adsabs.harvard.edu/abs/2025A&A...700A.279V}
}

@ARTICLE{Virtanen20,
       author = {{Virtanen}, Pauli and {Gommers}, Ralf and {Oliphant},
         Travis E. and {Haberland}, Matt and {Reddy}, Tyler and
         {Cournapeau}, David and {Burovski}, Evgeni and {Peterson}, Pearu
         and {Weckesser}, Warren and {Bright}, Jonathan and {van der Walt},
         St{\'e}fan J.  and {Brett}, Matthew and {Wilson}, Joshua and
         {Jarrod Millman}, K.  and {Mayorov}, Nikolay and {Nelson}, Andrew
         R.~J. and {Jones}, Eric and {Kern}, Robert and {Larson}, Eric and
         {Carey}, CJ and {Polat}, {\.I}lhan and {Feng}, Yu and {Moore},
         Eric W. and {Vand erPlas}, Jake and {Laxalde}, Denis and
         {Perktold}, Josef and {Cimrman}, Robert and {Henriksen}, Ian and
         {Quintero}, E.~A. and {Harris}, Charles R and {Archibald}, Anne M.
         and {Ribeiro}, Ant{\^o}nio H. and {Pedregosa}, Fabian and
         {van Mulbregt}, Paul and {Contributors}, SciPy 1. 0},
        title = "{SciPy 1.0: Fundamental Algorithms for Scientific
                  Computing in Python}",
      journal = {Nature Methods},
      year = "2020",
      volume={17},
      pages={261--272},
      adsurl = {https://rdcu.be/b08Wh},
      doi = {https://doi.org/10.1038/s41592-019-0686-2},
}

@ARTICLE{Watkins24,
       author = {{Watkins}, Laura L. and {van der Marel}, Roeland P. and {Bennet}, Paul},
        title = "{The Mass of the Large Magellanic Cloud from the Three-dimensional Kinematics of Its Globular Clusters}",
      journal = {\apj},
         year = 2024,
        month = mar,
       volume = {963},
       number = {2},
          eid = {84},
        pages = {84},
          doi = {10.3847/1538-4357/ad1f58},
archivePrefix = {arXiv},
       eprint = {2401.14458},
 primaryClass = {astro-ph.GA},
       adsurl = {https://ui.adsabs.harvard.edu/abs/2024ApJ...963...84W}
}

@ARTICLE{Yang18,
       author = {{Yang}, Yujiao and {Li}, Chengyuan and {Deng}, Licai and {de Grijs}, Richard and {Milone}, Antonino P.},
        title = "{New Insights into the Formation of the Blue Main Sequence in NGC 1850}",
      journal = {\apj},
         year = 2018,
        month = jun,
       volume = {859},
       number = {2},
          eid = {98},
        pages = {98},
          doi = {10.3847/1538-4357/aabe26},
archivePrefix = {arXiv},
       eprint = {1804.03948},
 primaryClass = {astro-ph.SR},
       adsurl = {https://ui.adsabs.harvard.edu/abs/2018ApJ...859...98Y}
}

@ARTICLE{Zivick18,
       author = {{Zivick}, Paul and {Kallivayalil}, Nitya and {van der Marel}, Roeland
         P. and {Besla}, Gurtina and {Linden}, Sean T. and
         {Koz{\l}owski}, Szymon and {Fritz}, Tobias K. and {Kochanek}, C.~S. and
         {Anderson}, J. and {Sohn}, Sangmo Tony and {Geha}, Marla C. and
         {Alcock}, Charles R.},
        title = "{The Proper Motion Field of the Small Magellanic Cloud: Kinematic Evidence for Its Tidal Disruption}",
      journal = {\apj},
         year = "2018",
        month = "Sep",
       volume = {864},
       number = {1},
          eid = {55},
        pages = {55},
          doi = {10.3847/1538-4357/aad4b0},
archivePrefix = {arXiv},
       eprint = {1804.04110},
 primaryClass = {astro-ph.GA},
       adsurl = {https://ui.adsabs.harvard.edu/abs/2018ApJ...864...55Z}
}

%--------------------------------------------------------------------

\begin{appendix}

\section{List of observations\label{app:obs}}
Tables \ref{tab:ngc1751obs} and \ref{tab:ngc1818obs} list the observations used for creating the astro-photometric catalogues of NGC~1751 and NGC~1818, respectively. The tables give the programme ID number, the PI, the epoch of observations, the used instrument, camera and filter combination, as well as the number of exposures along with the respective exposure times.

\begin{table*}
\centering
\caption{Observations of NGC~1751\label{tab:ngc1751obs}}
\begin{tabular}{@{}l@{ }c@{ }c@{ }c@{ }c@{ }c@{ }c@{ }}
\hline\hline
\noalign{\smallskip}
Programme ID & ~~~~~~~~PI~~~~~~~~ & ~~~~~~~Epoch~~~~~~~ & ~~~~~Instrument/Camera~~~~~ & ~~~~~Filter~~~~~ & ~~~~~Exposures~~~~~\\
& & (yyyy/mm) & & & (N\,$\times$\,t$_{\mathrm{exp}}$) \\
\noalign{\smallskip}
\hline
\noalign{\smallskip}
GO-9891 &  G. Gilmore & 2003/10 & ACS/WFC & F555W & 1\,$\times$\,300\,s \\
 \noalign{\smallskip}
&   &  & & F814W & 1\,$\times$\,200\,s \\
 \noalign{\smallskip}
\hline
\noalign{\smallskip}
GO-10595 & P. Goudfrooij & 2006/10 & ACS/WFC & F435W & 1\,$\times$\,90\,s\\
&  &  &  & & 2\,$\times$\,340\,s\\
 \noalign{\smallskip}
&  &  &  & F555W & 1\,$\times$\,40\,s\\
&  &  &  & & 2\,$\times$\,340\,s\\
 \noalign{\smallskip}
 &  &  &  & F814W & 1\,$\times$\,8\,s\\
&  &  &  &  & 2\,$\times$\,340\,s\\
\noalign{\smallskip}
\hline
\noalign{\smallskip}
GO-12257 & L. Girardi & 2011/10 & WFC3/UVIS & F336W & 1\,$\times$\,1190\,s\\
&  &  &  &  & 2\,$\times$\,1200\,s\\
\noalign{\smallskip}
\hline
\end{tabular}

\end{table*}

%--------------------------------------------------------------------

\begin{table*}
\centering
\caption{Observations of NGC~1818\label{tab:ngc1818obs}}
\begin{tabular}{@{}l@{ }c@{ }c@{ }c@{ }c@{ }c@{ }c@{ }}
\hline\hline
\noalign{\smallskip}
Programme ID & ~~~~~~~~PI~~~~~~~~ & ~~~~~~~Epoch~~~~~~~ & ~~~~~Instrument/Camera~~~~~ & ~~~~~Filter~~~~~ & ~~~~~Exposures~~~~~\\
& & (yyyy/mm) & & & (N\,$\times$\,t$_{\mathrm{exp}}$) \\
\noalign{\smallskip}
\hline
\noalign{\smallskip}
GO-12116 &  J. Dalcanton & 2013/03 & WFC3/UVIS & F475W & 1\,$\times$\,100\,s \\
\noalign{\smallskip}
\hline
\noalign{\smallskip}
GO-13727 & J. Kalirai & 2015/10 & WFC3/UVIS & F336W & 1\,$\times$\,10\,s\\
&  &  &  & & 1\,$\times$\,100\,s\\
&  &  &  & & 1\,$\times$\,790\,s\\
&  &  &  & & 3\,$\times$\,947\,s\\
\noalign{\smallskip}
\hline
\noalign{\smallskip}
GO-14710 & A. Milone & 2017/02 & WFC3/UVIS & F814W & 1\,$\times$\,90\,s\\
&  &  &  &  & 1\,$\times$\,666\,s\\
\noalign{\smallskip}
\hline
\noalign{\smallskip}
GO-15945 & G. Cordoni & 2020/06 & WFC3/UVIS & F606W & 1\,$\times$\,724\,s\\
&  &  &  &  & 2\,$\times$\,772\,s\\
\noalign{\smallskip}
&  &  &  & F814W & 2\,$\times$\,795\,s\\
&  &  &  &  & 1\,$\times$\,810\,s\\
\noalign{\smallskip}
\hline
\end{tabular}

\end{table*}

%--------------------------------------------------------------------
\section{PM-based selection of cluster stars \label{app:pm_sel}}

\begin{figure*}

\includegraphics[width=1\textwidth]{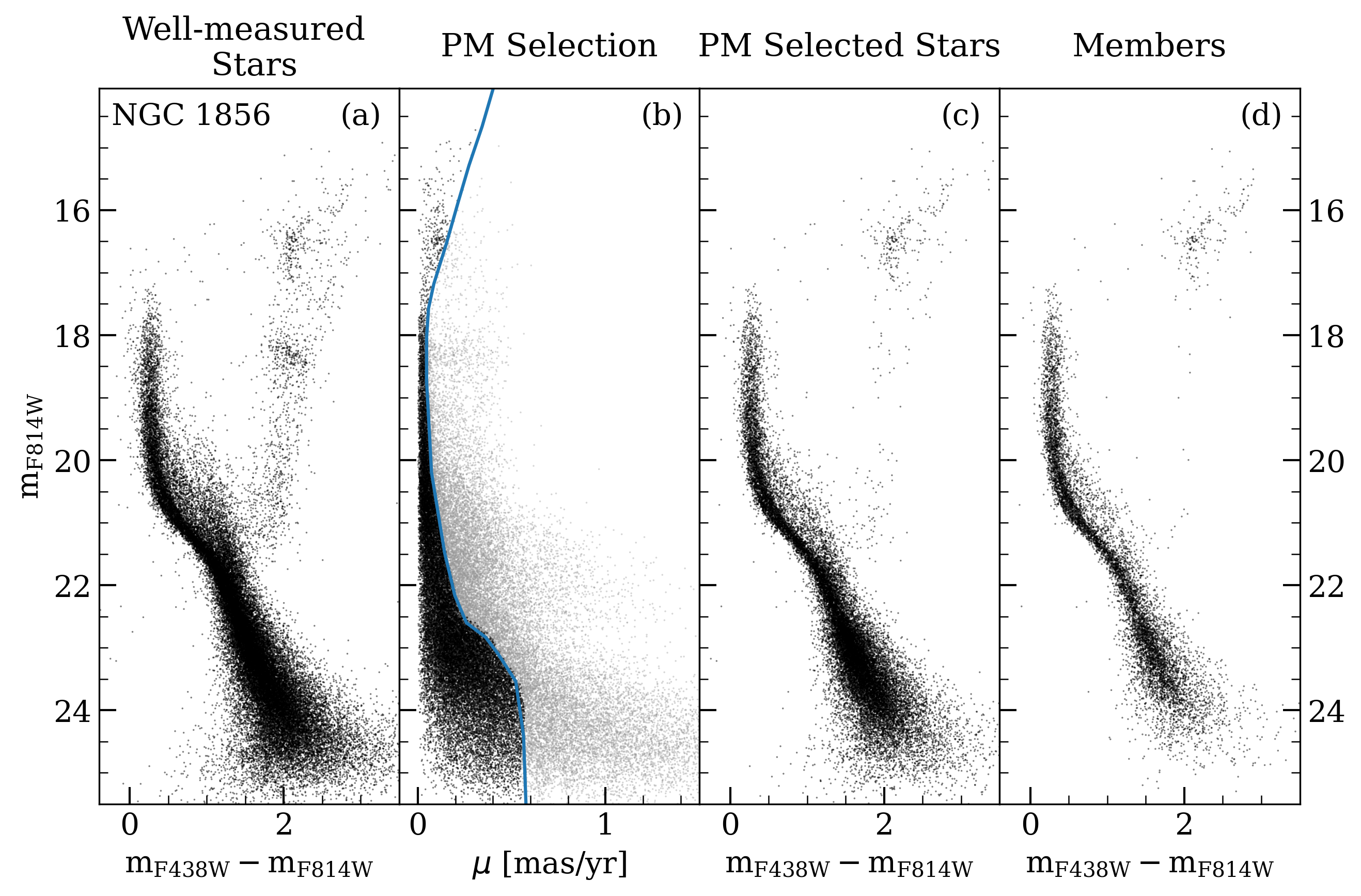} 
\caption{Selection of cluster member stars using the example of NGC~1856. (a) The m$_{\rm F814W}$ vs m$_{\rm F438W}$--m$_{\rm F814W}$ CMD of well-measured stars in the field of the cluster for which PMs have been determined (b) 1-D relative PMs as a function of the m$_{\rm F814W}$ magnitude. The blue line (drawn by hand) follows our selection of likely cluster members (black dots) based on their motions. The dependence on the magnitude is to account for the varying PM uncertainties as a function of magnitude. (c) CMD of PM-selected cluster members. (d) CMD of PM-selected cluster stars, including only stars within one effective radius from
the cluster centre.
\label{fig:ngc1856_pm_sel}
}
\end{figure*}

Figure~\ref{fig:ngc1856_pm_sel} illustrates the procedure described in Section~\ref{sec:selection} to select likely cluster-member stars based on their measured relative PMs. 

%--------------------------------------------------------------------

\section{Differential reddening correction\label{app:dr}}
\begin{figure*}
\begin{tabular}{cc}
\includegraphics[width=1\columnwidth]{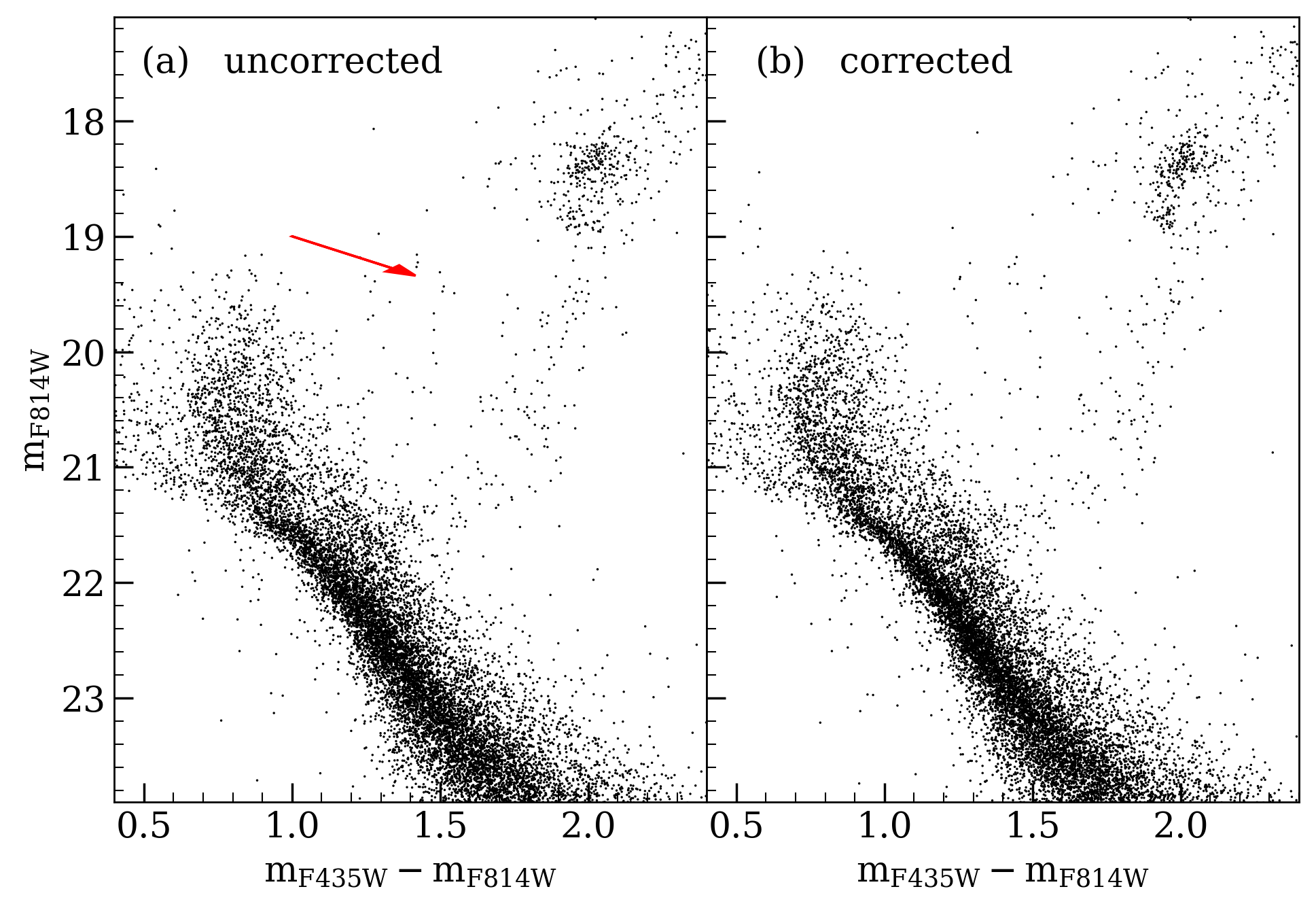} &
\includegraphics[width=0.95\columnwidth]{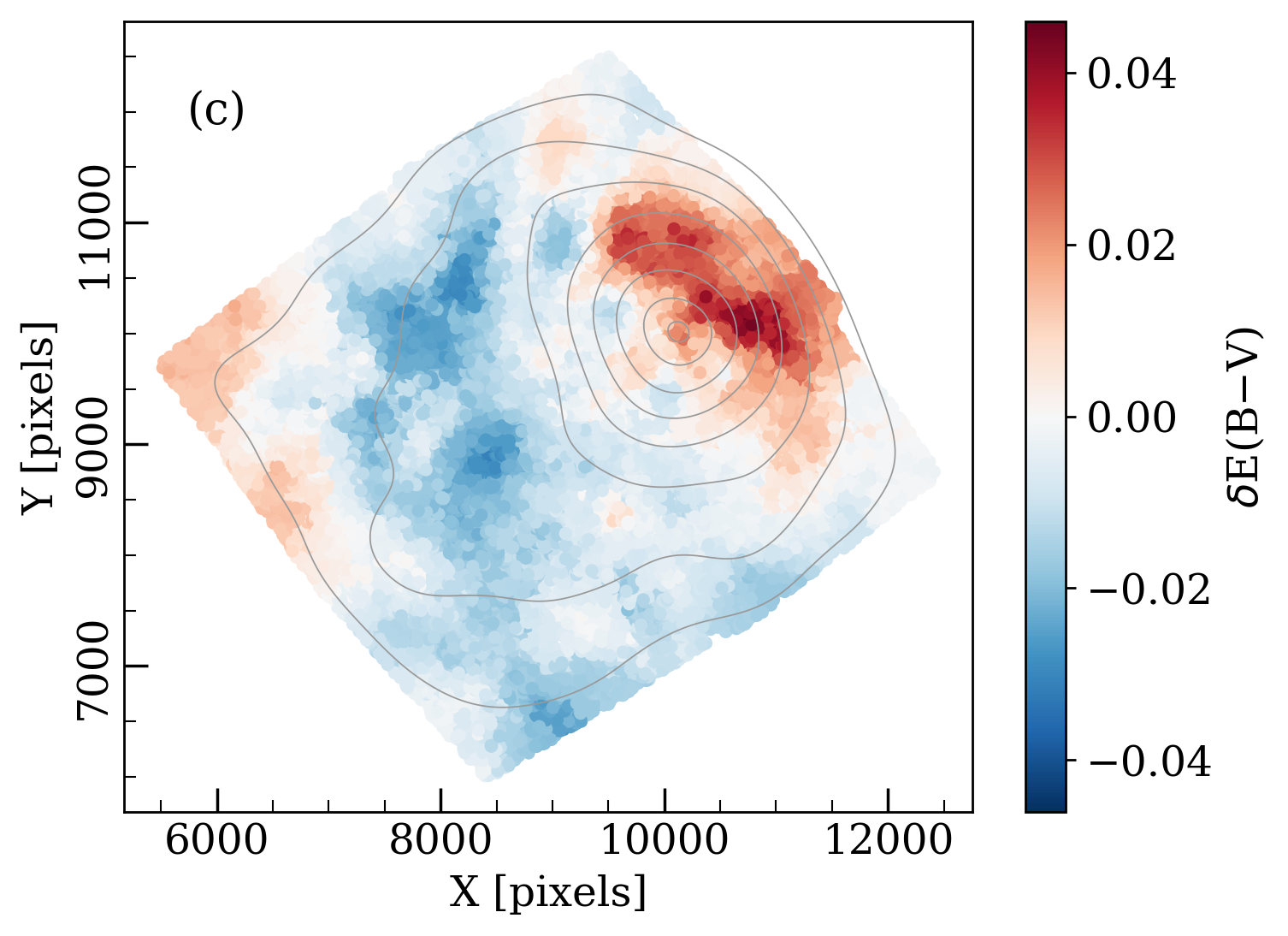} \\
\end{tabular}
\caption{Illustration of the differential reddening correction towards NGC~1751. Panel (a) shows the original, uncorrected $m_{\rm F814W}$ vs. $m_{\rm F435W}-m_{\rm F814W}$ CMD of well-measured stars. Also shown as a red arrow is the direction of the reddening vector. Panel (b) shows the same CMD after the corrections for the effects of differential reddening have been applied. Panel (c) presents the resulting reddening map. The grey contours follow the density distribution of the reference stars used for the correction. 
\label{fig:ngc1751_dr}
}
\end{figure*}

Figure~\ref{fig:ngc1751_dr} illustrates the effect of our differential reddening corrections using the example of NGC~1751.

%--------------------------------------------------------------------

\section{Isochrone fitting results\label{app:iso_fit_ngc1978}}
\begin{figure*}
     \centering
     \begin{subfigure}[b]{\columnwidth}
         \centering
         \includegraphics[width=0.75\textwidth]{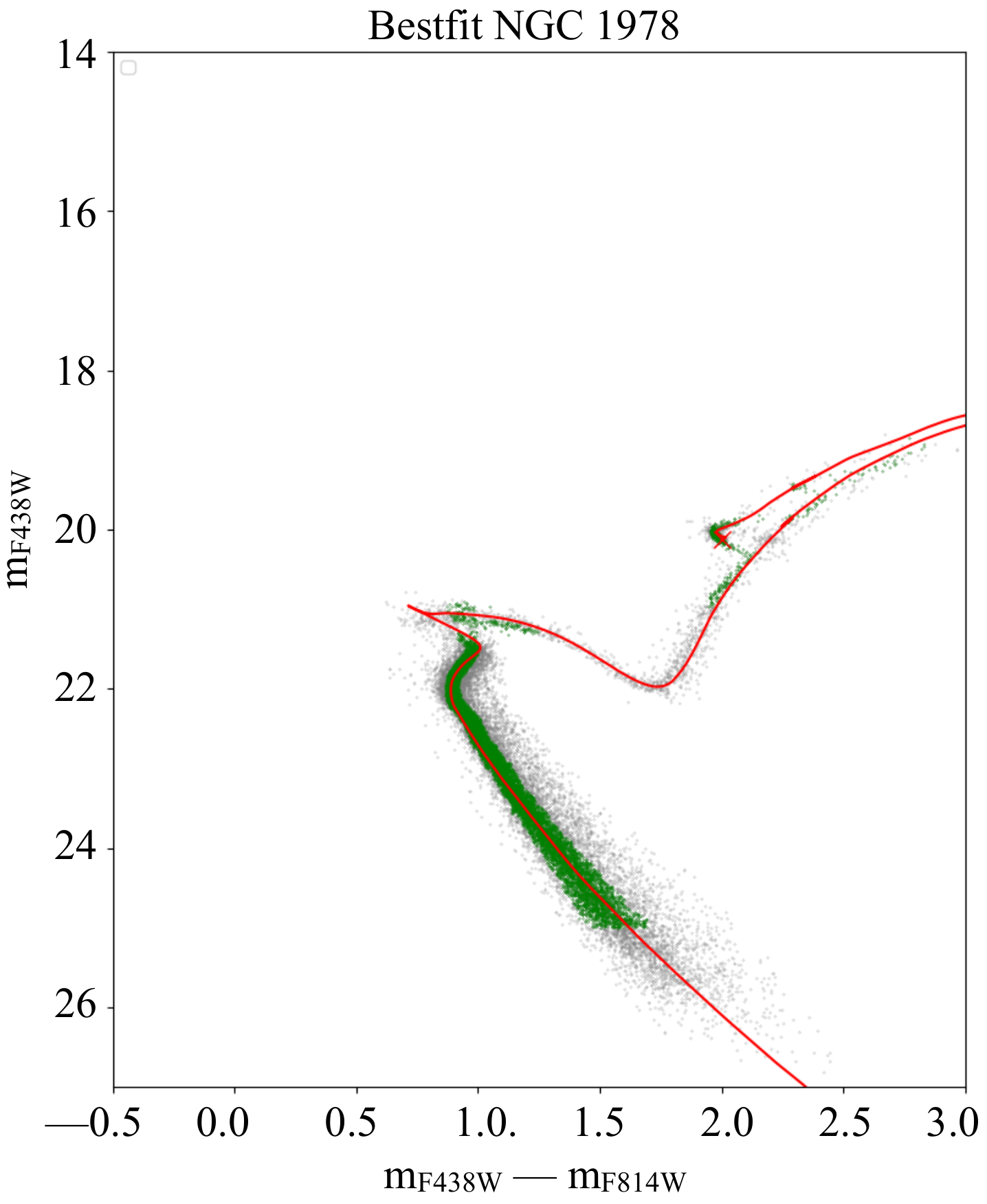}
         \caption{}
     \end{subfigure}
     \hfill
     \begin{subfigure}[b]{\columnwidth}
         \centering
         \includegraphics[width=0.75\textwidth]{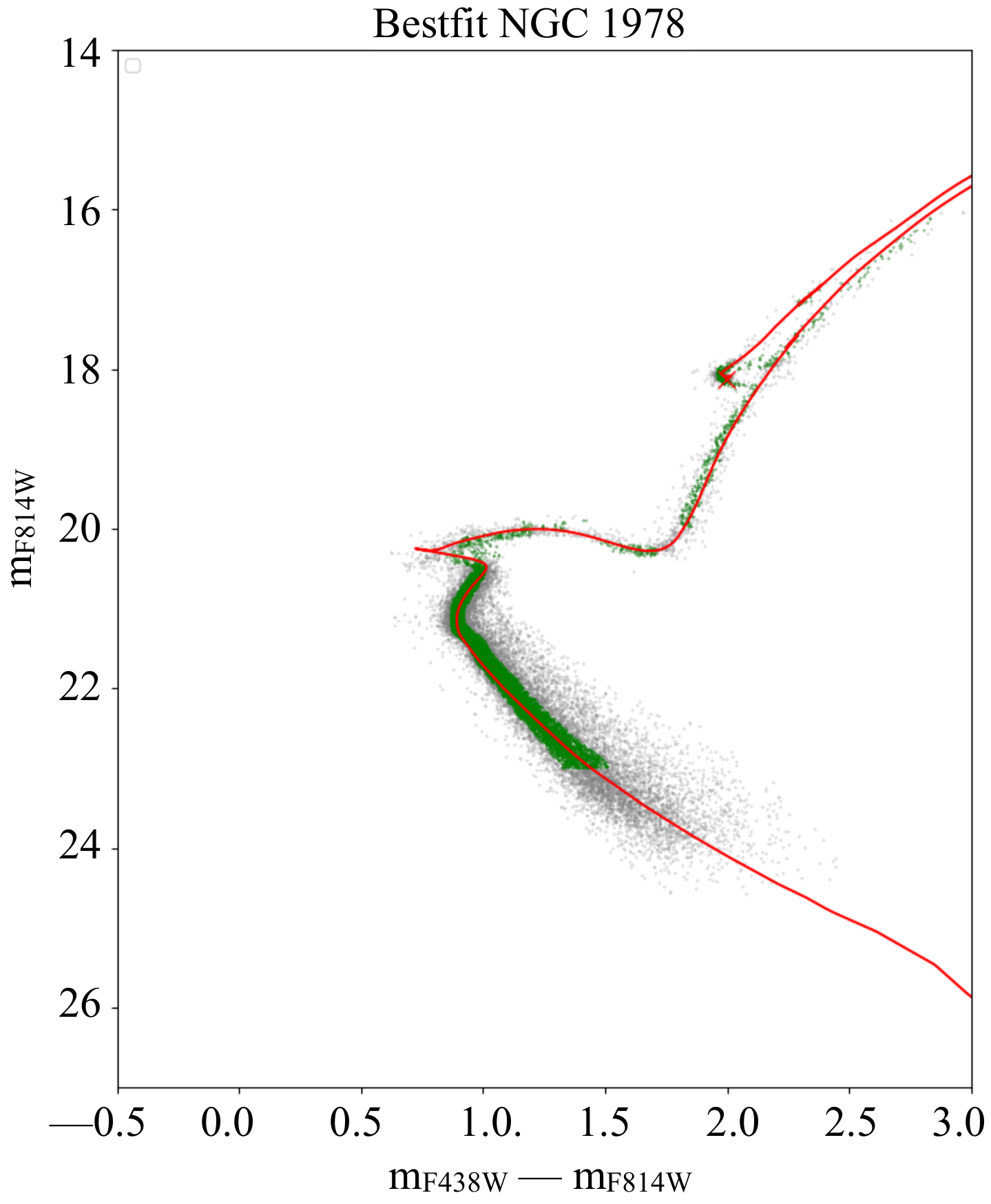}
         \caption{ }
     \end{subfigure}

     \begin{subfigure}[b]{\columnwidth}
         \centering
         \includegraphics[width=0.75\textwidth]{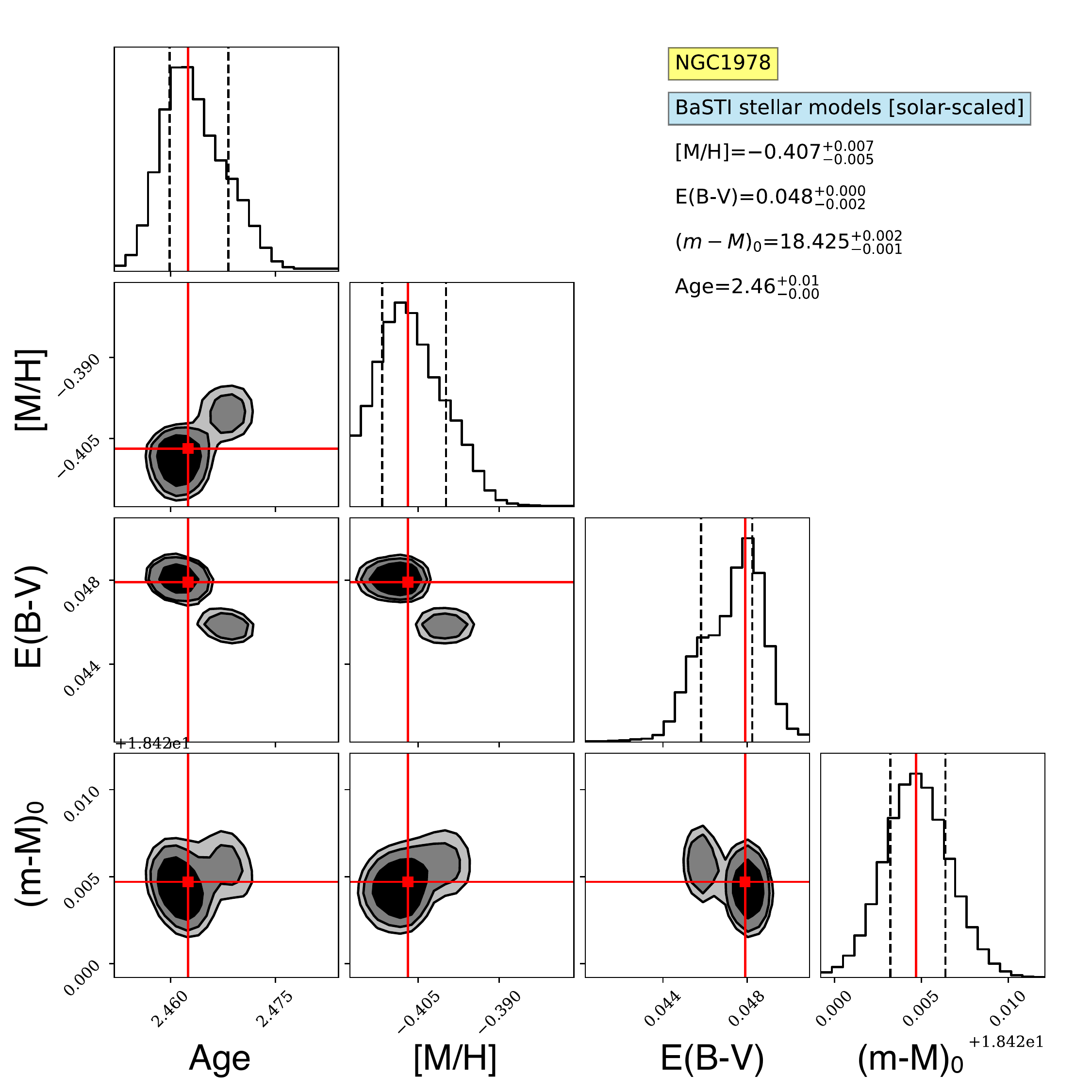}
         \caption{}
     \end{subfigure}
     \hfill
     \begin{subfigure}[b]{\columnwidth}
         \centering
         \includegraphics[width=0.75\textwidth]{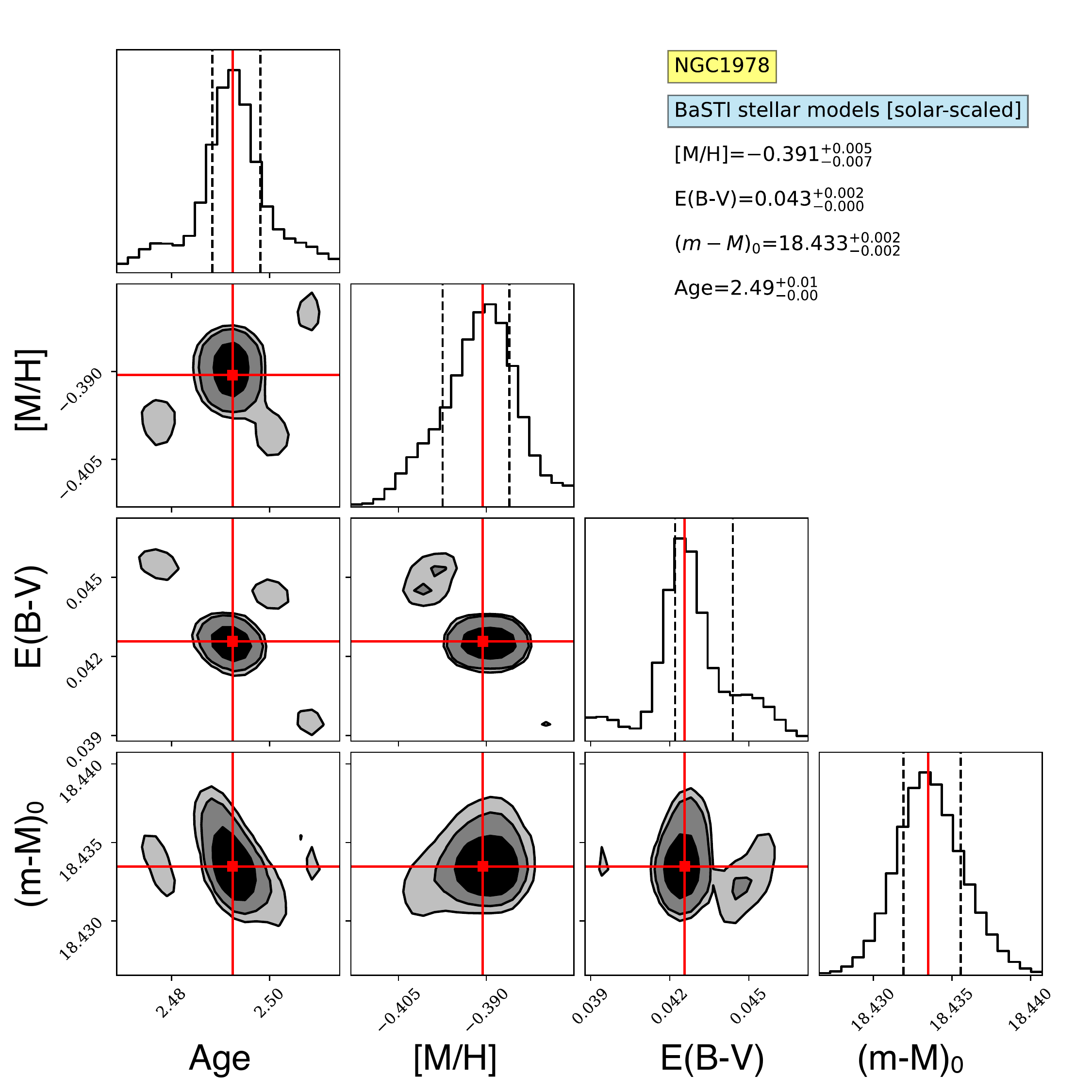}
         \caption{}
     \end{subfigure}
        \caption{Isochrone fitting results using the example of NGC~1978. (a) Best-fit isochrone model in the $m_{\rm F438W}$ vs. $m_{\rm F438W}-m_{\rm F814W}$ CMD. (b) Best-fit isochrone model in the $m_{\rm F814W}$ vs. $m_{\rm F438W}-m_{\rm F814W}$ CMD. (c): Corner plot of the posterior probability distributions of pairwise model parameters for the $m_{\rm F438W}$ vs. $m_{\rm F438W}-m_{\rm F814W}$ CMD. The best-fit parameters are quoted in the labels. 
        (d): as per (c) but for the $m_{\rm F814W}$ vs. $m_{\rm F438W}-m_{\rm F814W}$ CMD. 
        }
        \label{fig:ngc1978_isofit}
\end{figure*}

Figure~\ref{fig:ngc1978_isofit} shows the results of the isochrone fitting for the cluster NGC~1978 as an example. The upper two panels show two different CMDs, with the best-fitting isochrone models overplotted as a red solid line. Stars actually used for the fit are highlighted in green. The lower two panels show the corner plots of the posterior probability distribution.  Figure~\ref{fig:cmds} presents for the remaining 18 clusters one representative CMD with the best-fitting isochrone.

\begin{figure*}
\begin{tabular}{ccc}
\includegraphics[width=0.60\columnwidth]{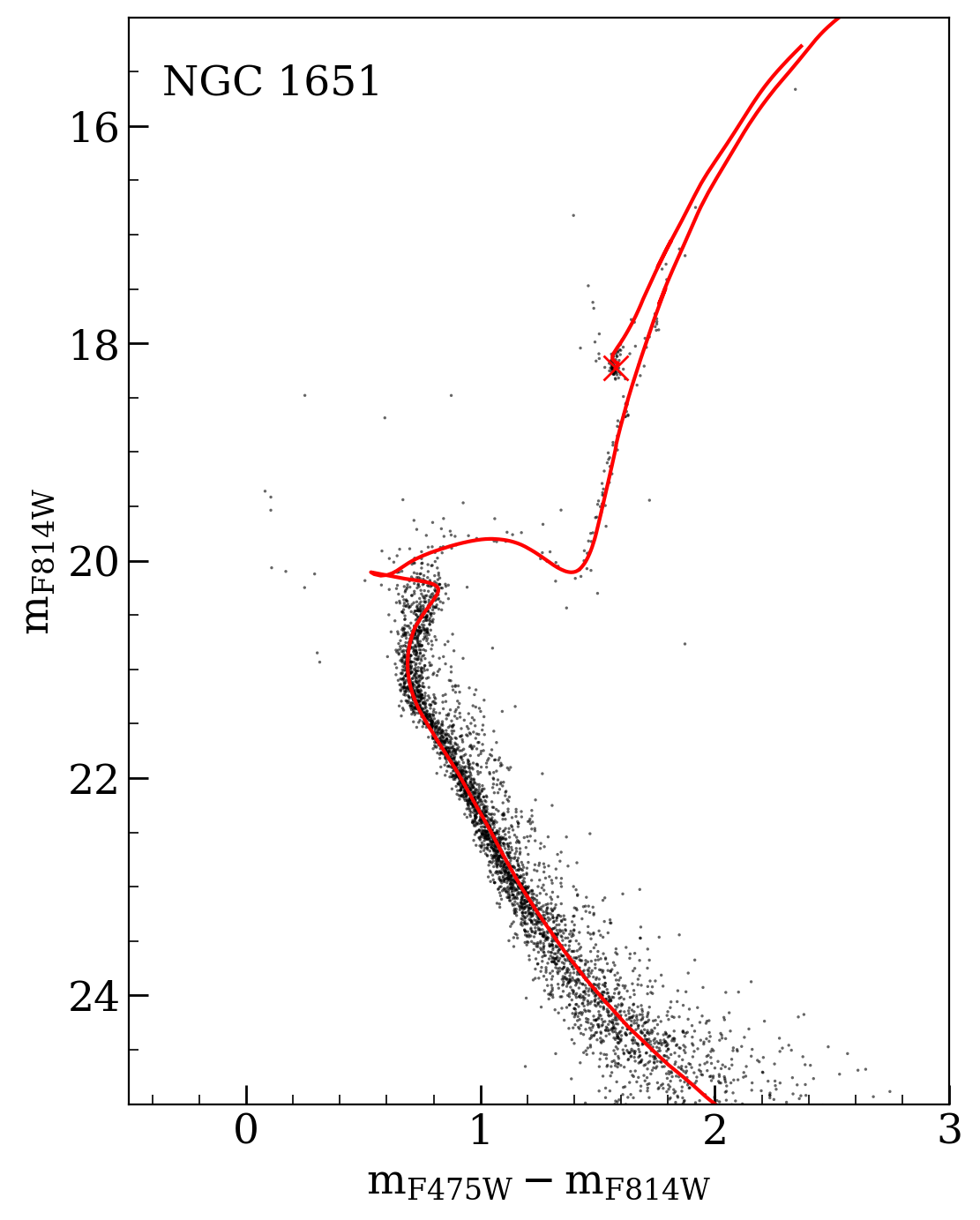} &
\includegraphics[width=0.60\columnwidth]{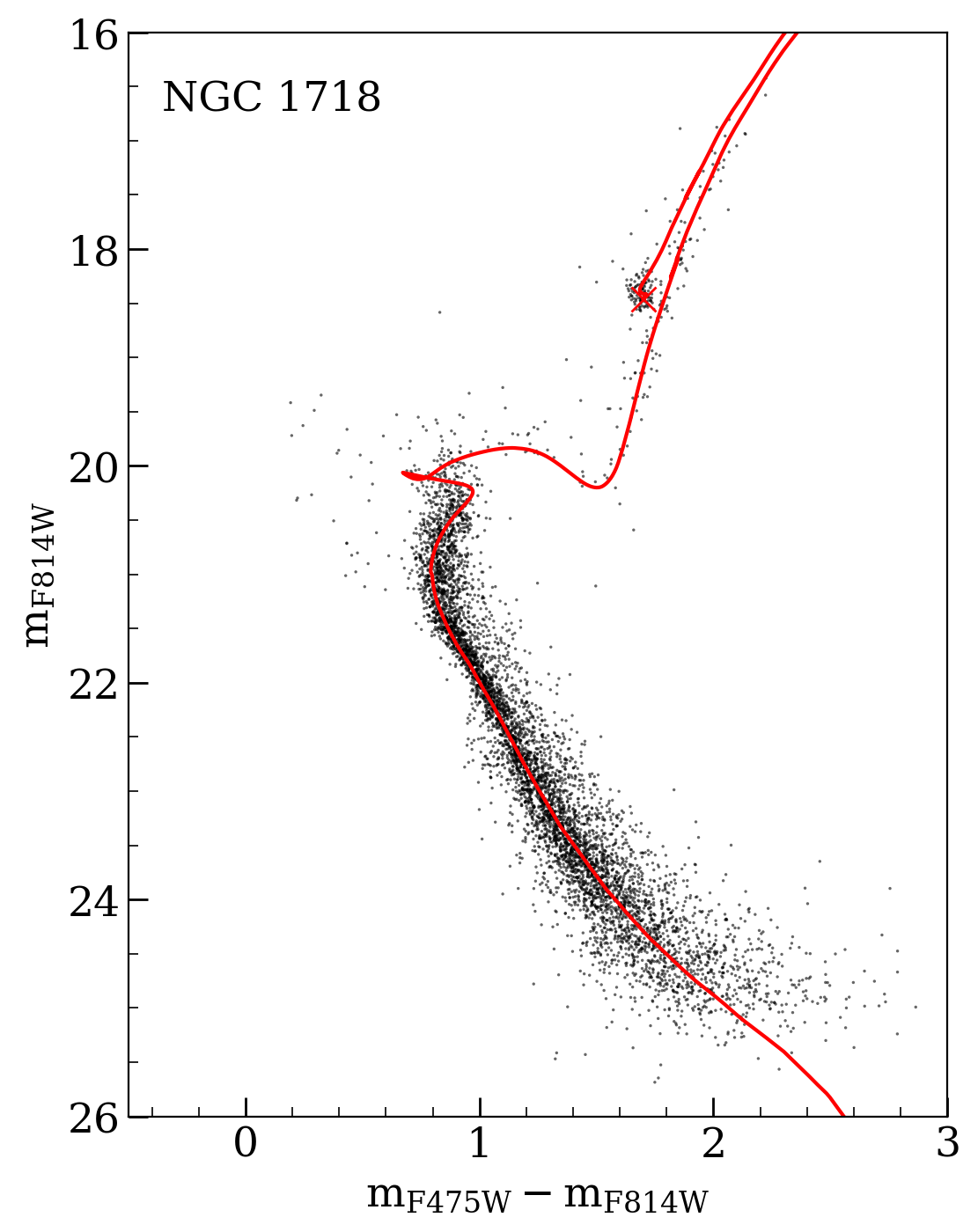} &
\includegraphics[width=0.60\columnwidth]{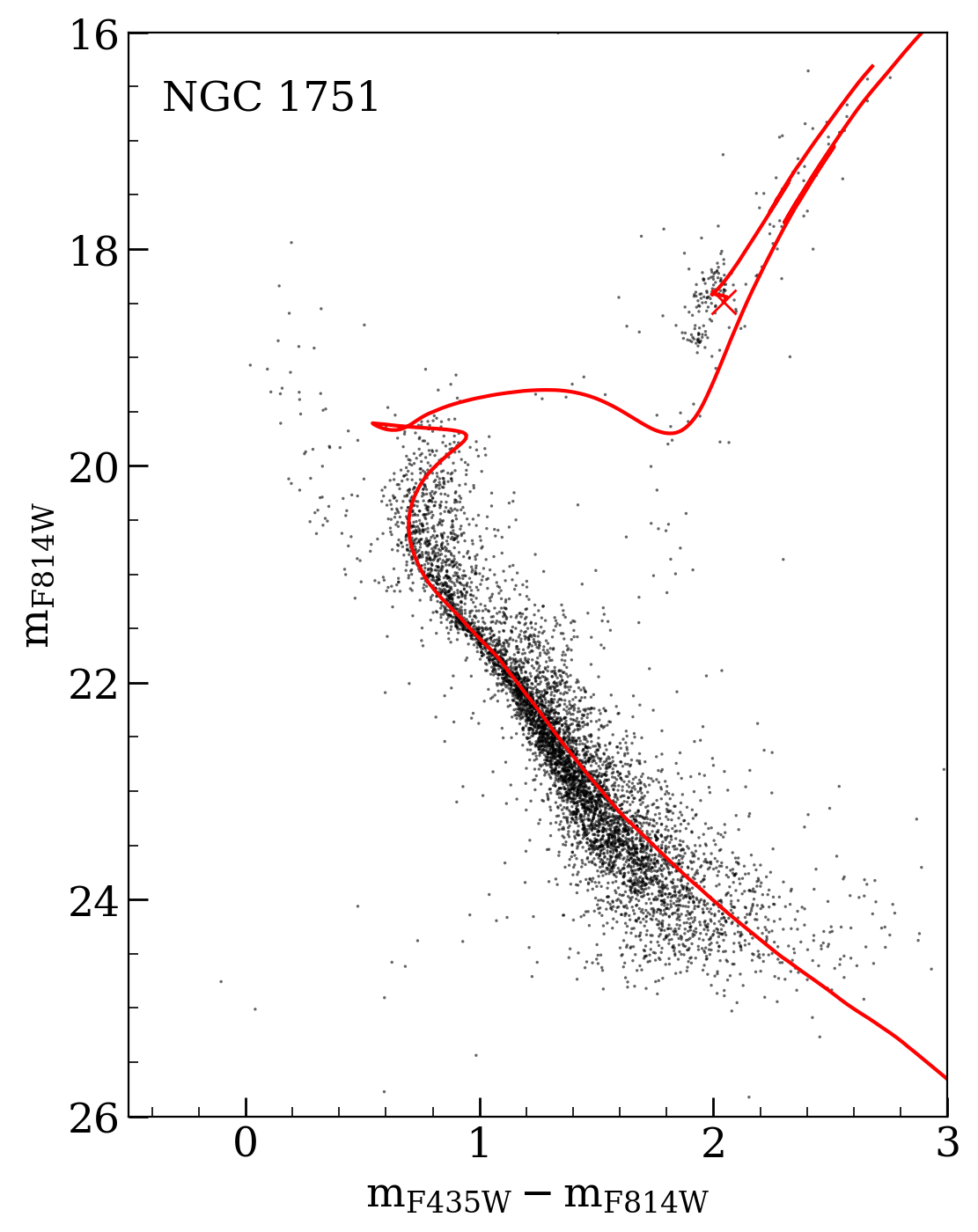} \\
\includegraphics[width=0.60\columnwidth]{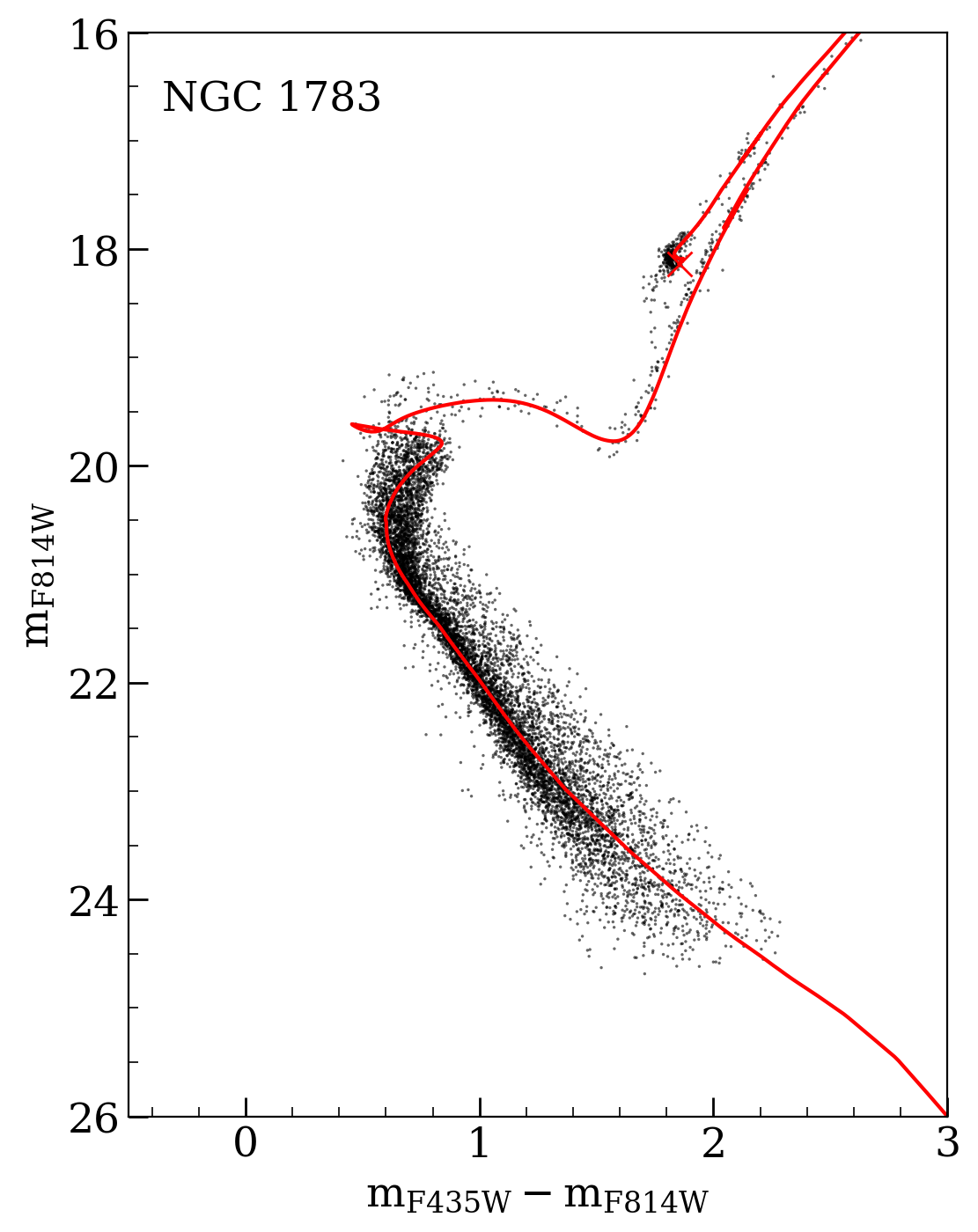} &
\includegraphics[width=0.60\columnwidth]{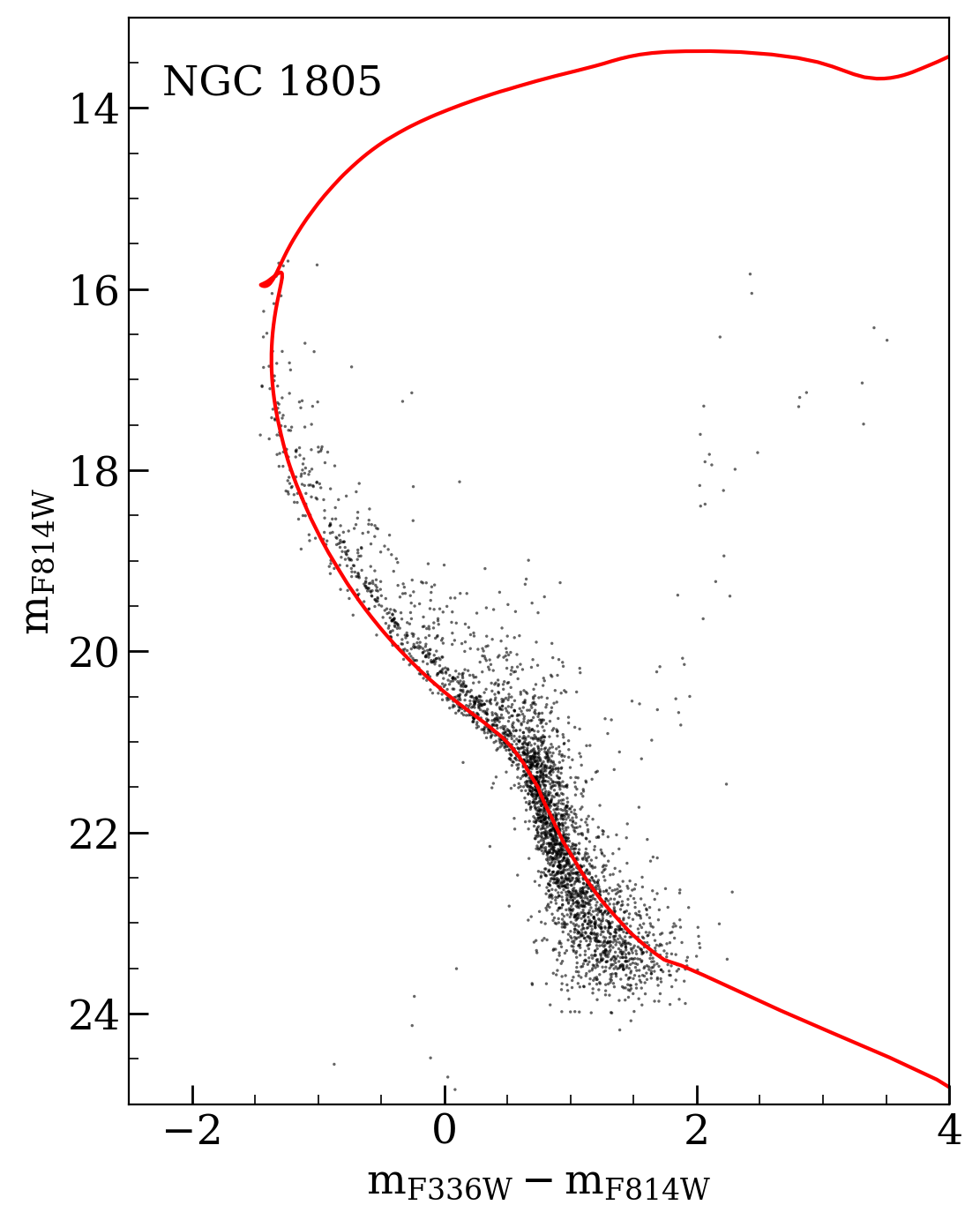} &
\includegraphics[width=0.60\columnwidth]{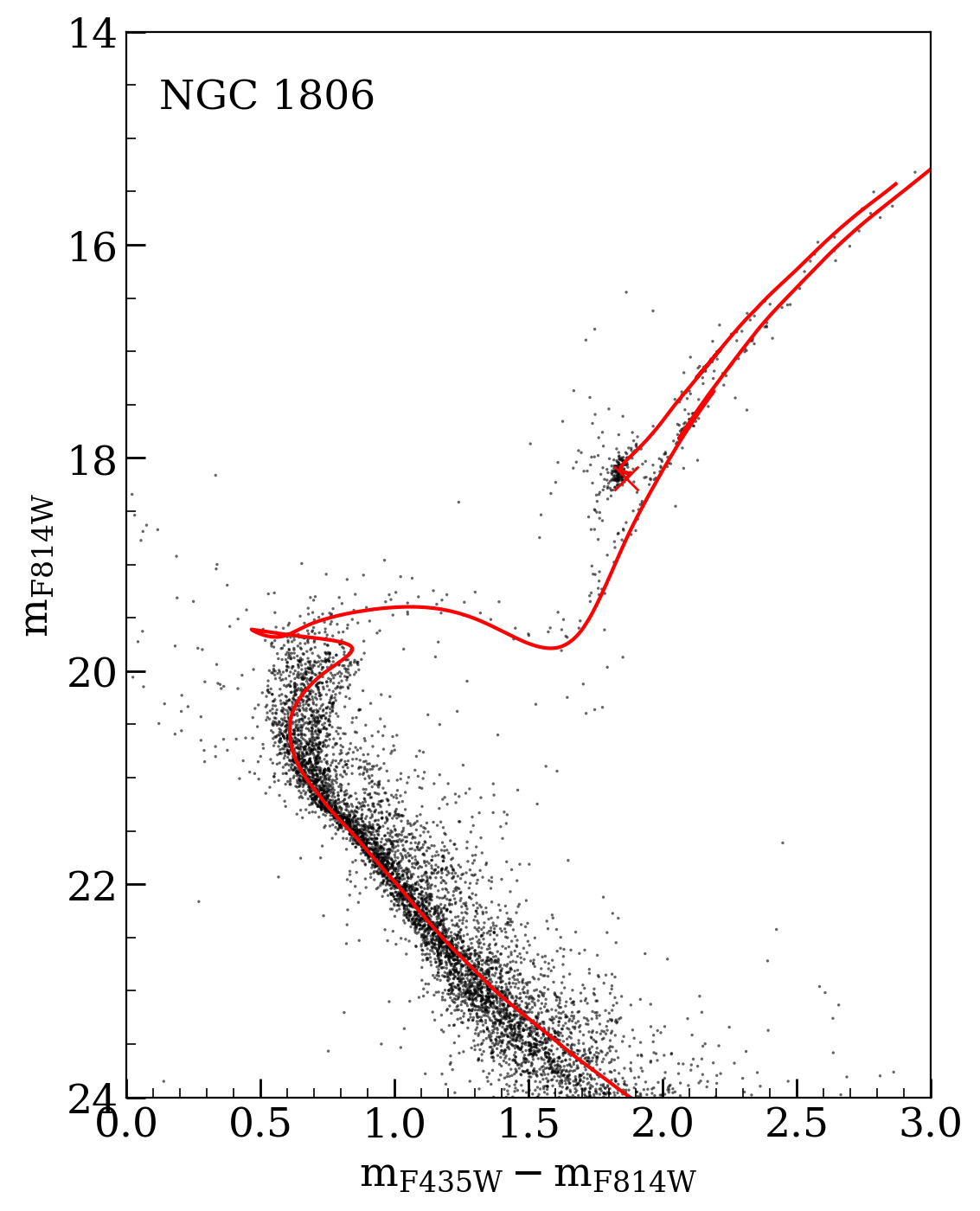} \\
\includegraphics[width=0.60\columnwidth]{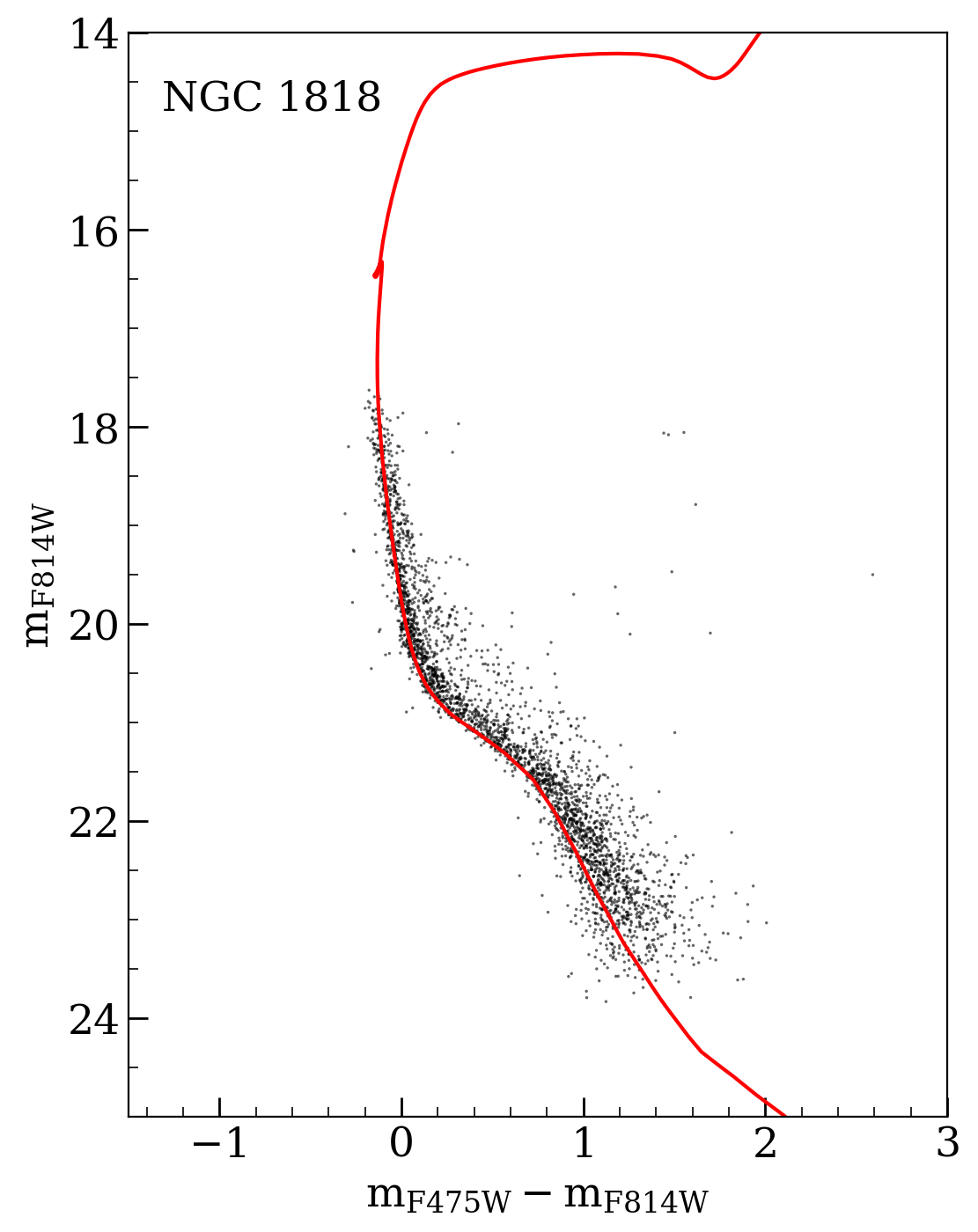} &
\includegraphics[width=0.60\columnwidth]{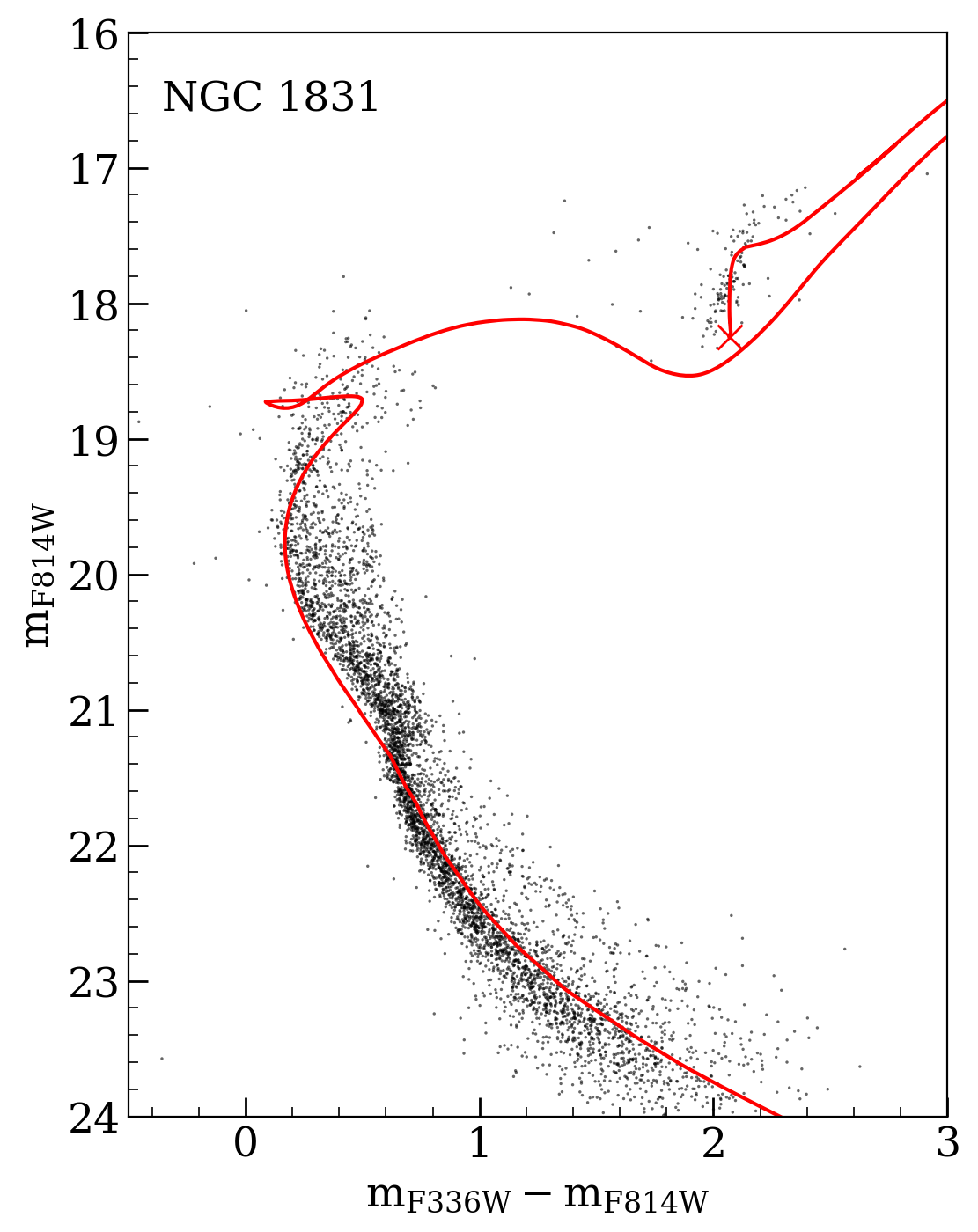} &
\includegraphics[width=0.60\columnwidth]{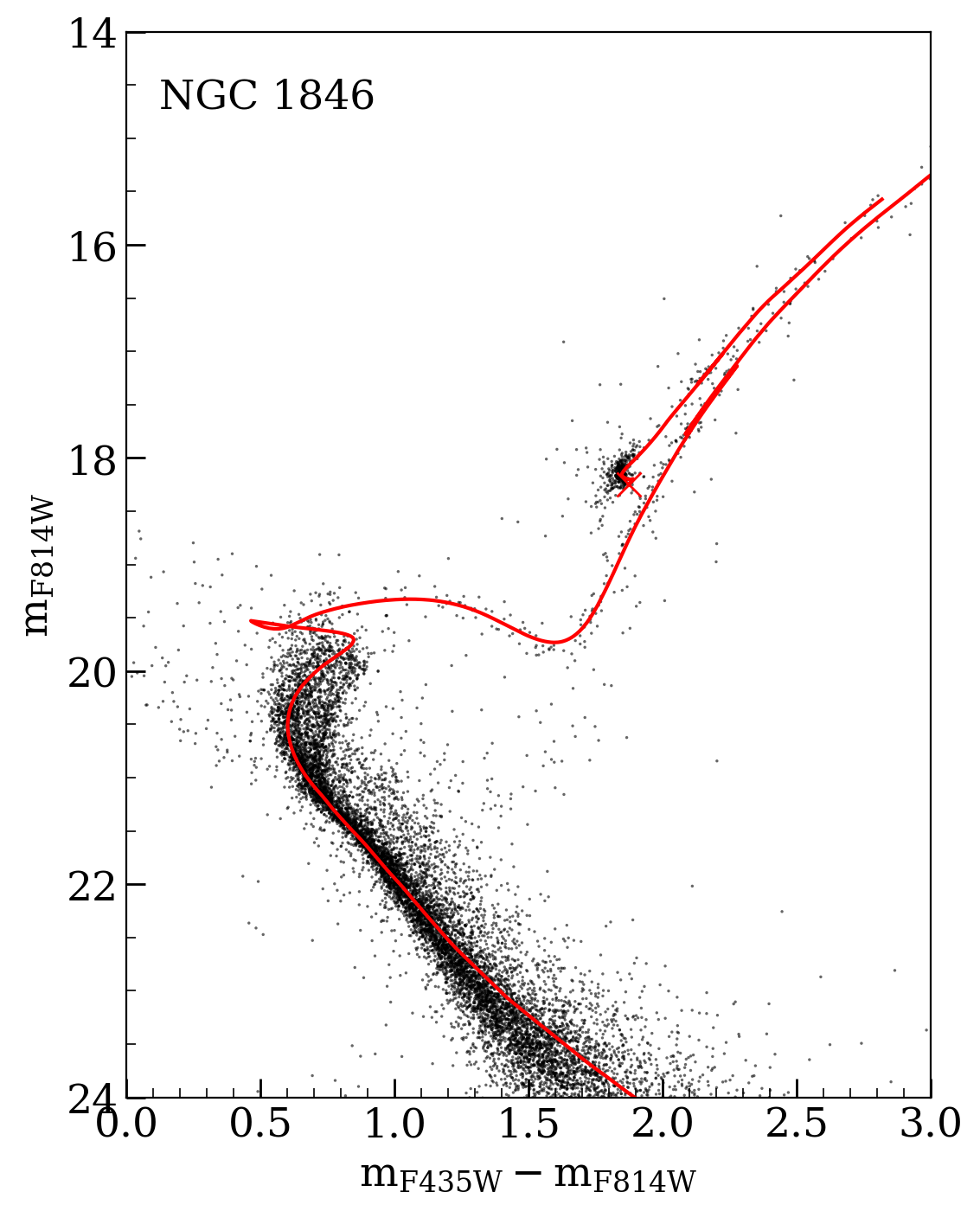} \\

\end{tabular}
\caption{Representative CMDs for the clusters studied in this work. The best-fitting isochrone model for each cluster is shown as a red solid line. The names of the clusters are given in the top left corner within each panel. 
\label{fig:cmds}
}
\end{figure*}

\begin{figure*}
\ContinuedFloat
\begin{tabular}{ccc}
\includegraphics[width=0.60\columnwidth]{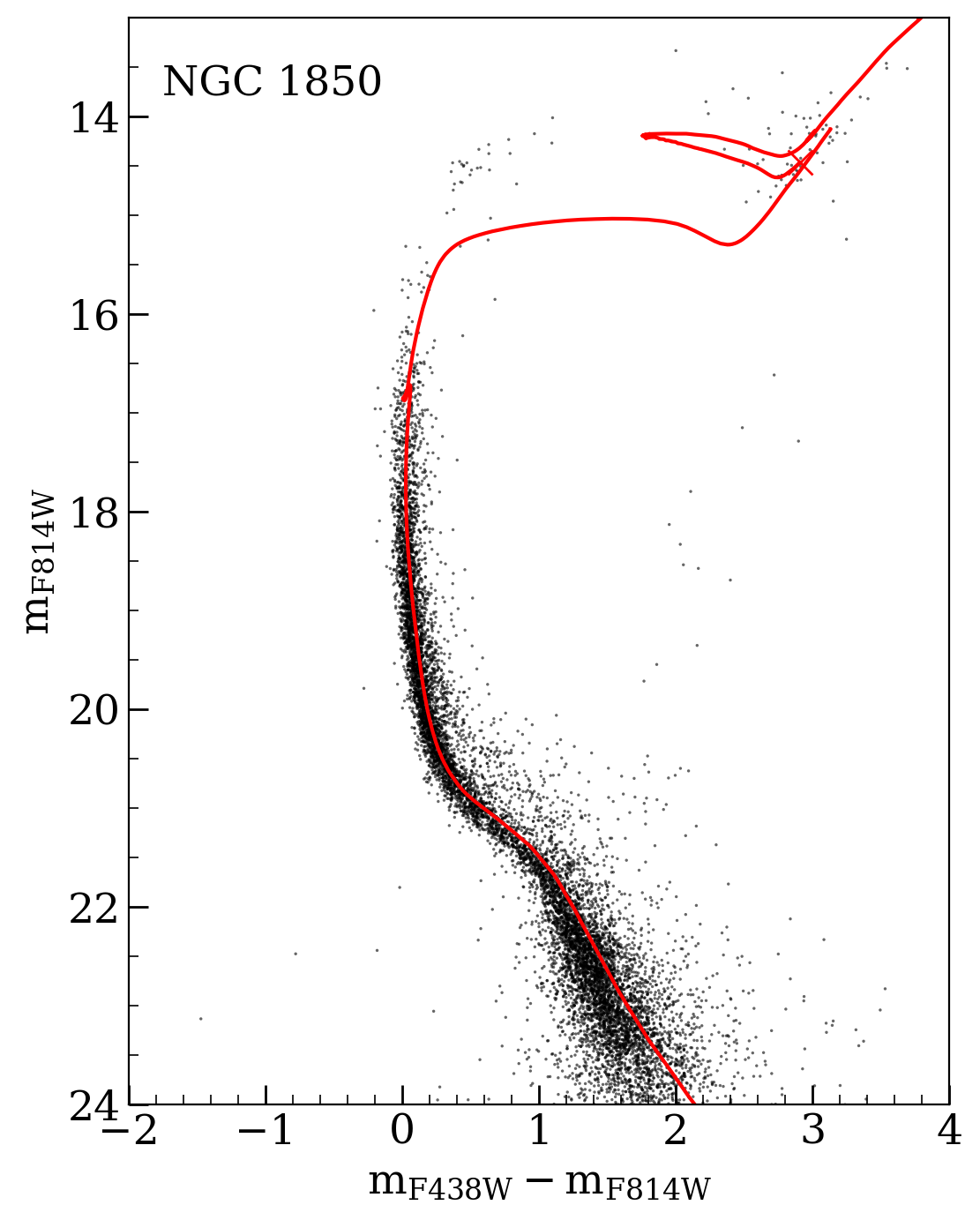} &
\includegraphics[width=0.60\columnwidth]{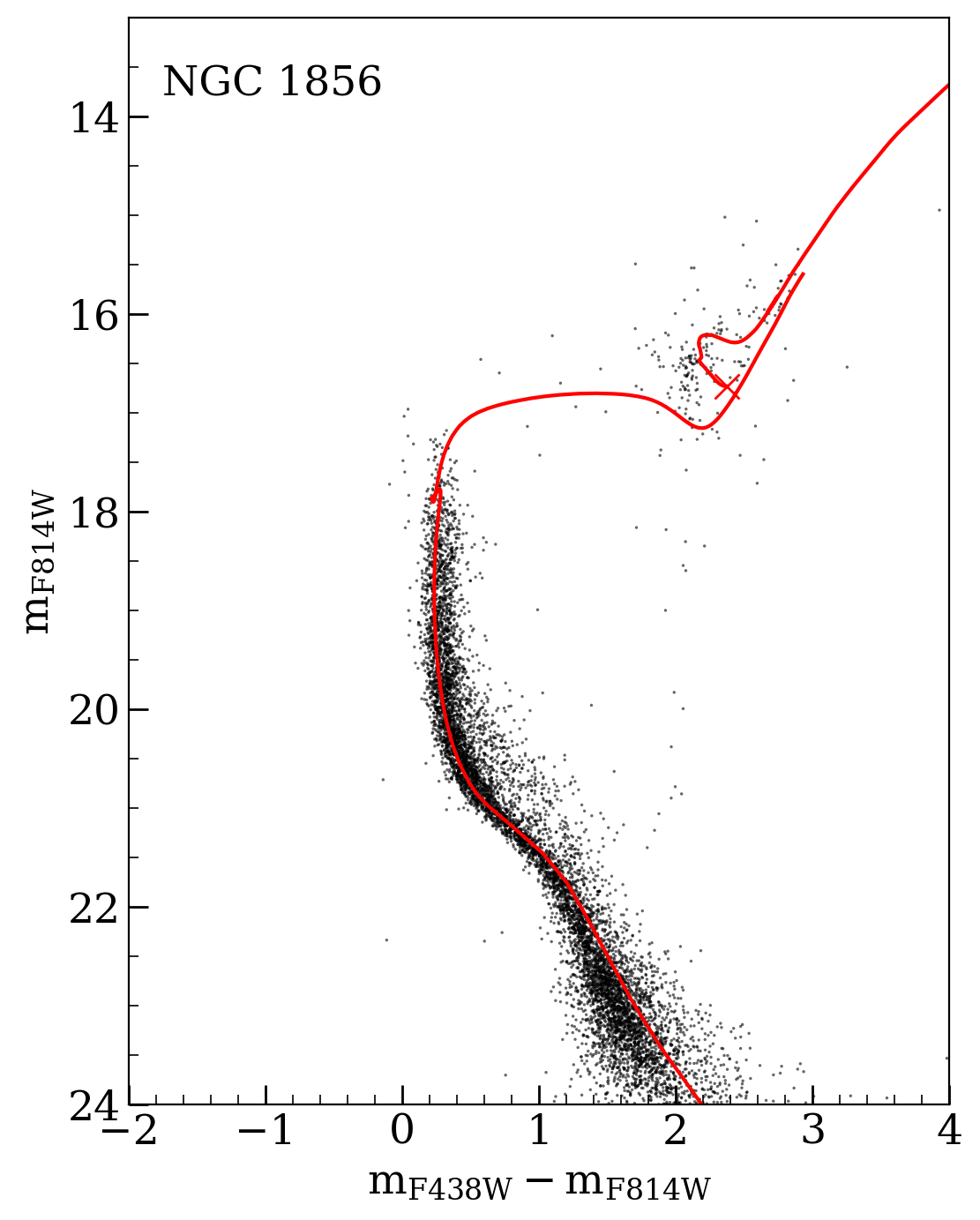} &
\includegraphics[width=0.60\columnwidth]{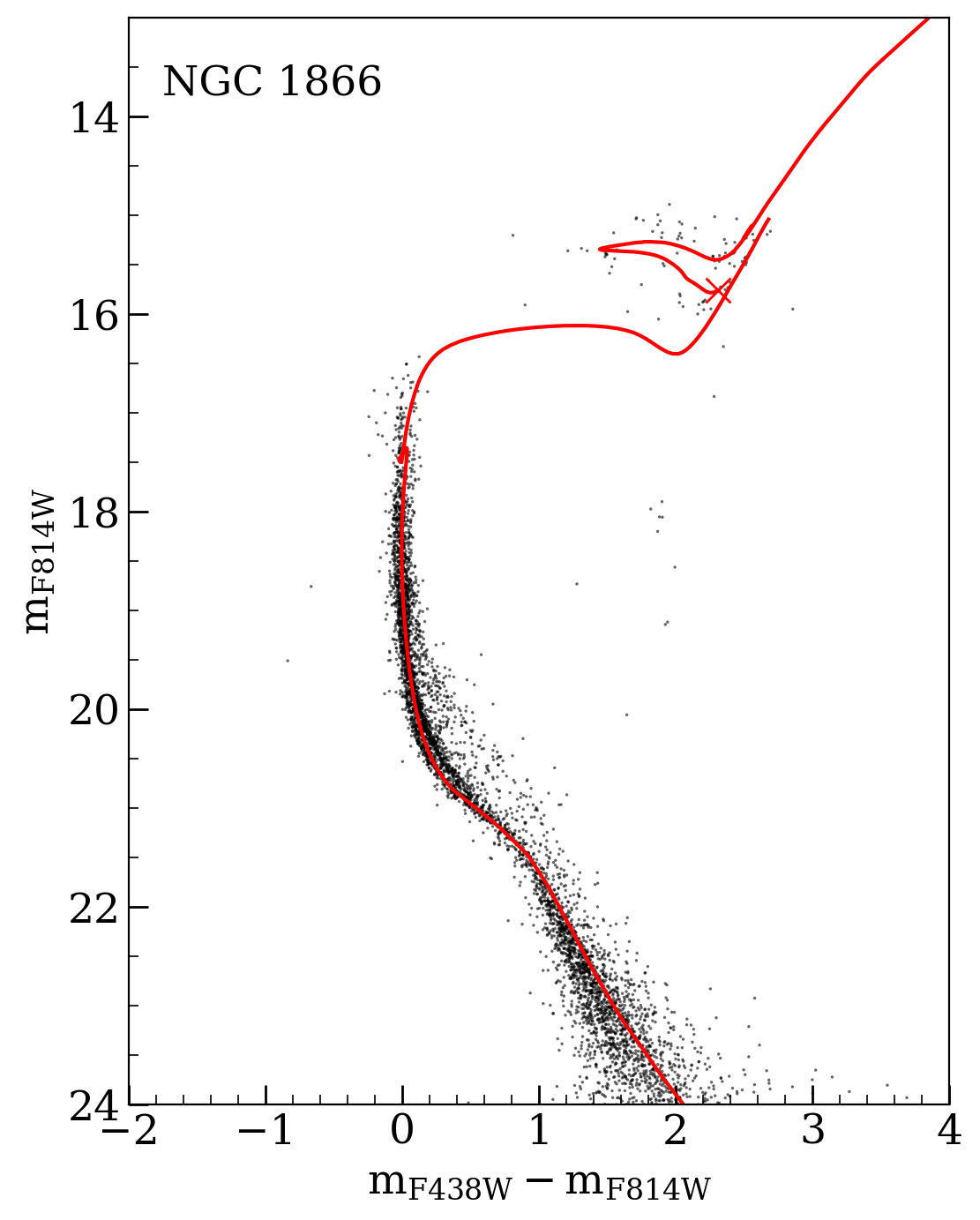} \\
\includegraphics[width=0.60\columnwidth]{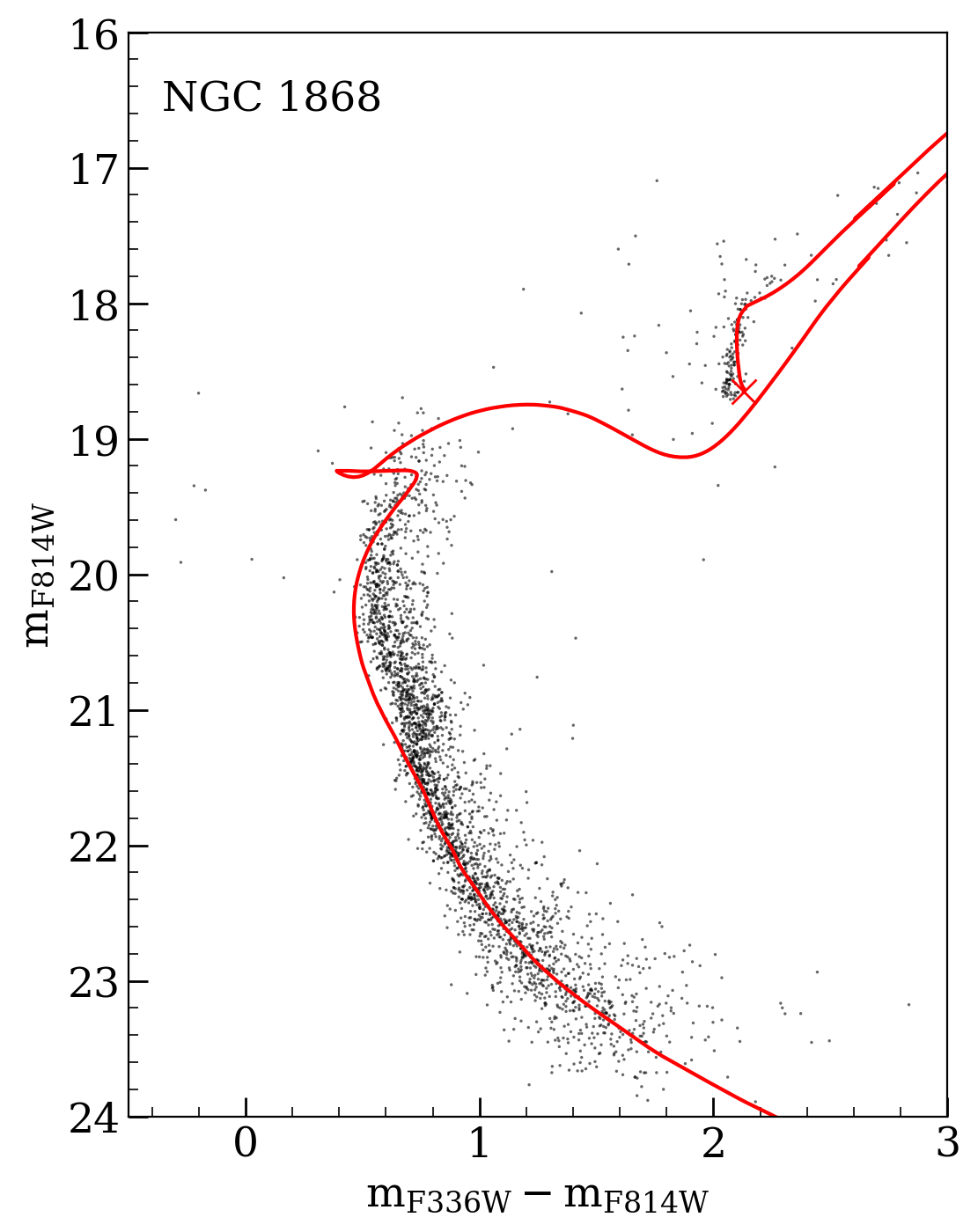} &
\includegraphics[width=0.60\columnwidth]{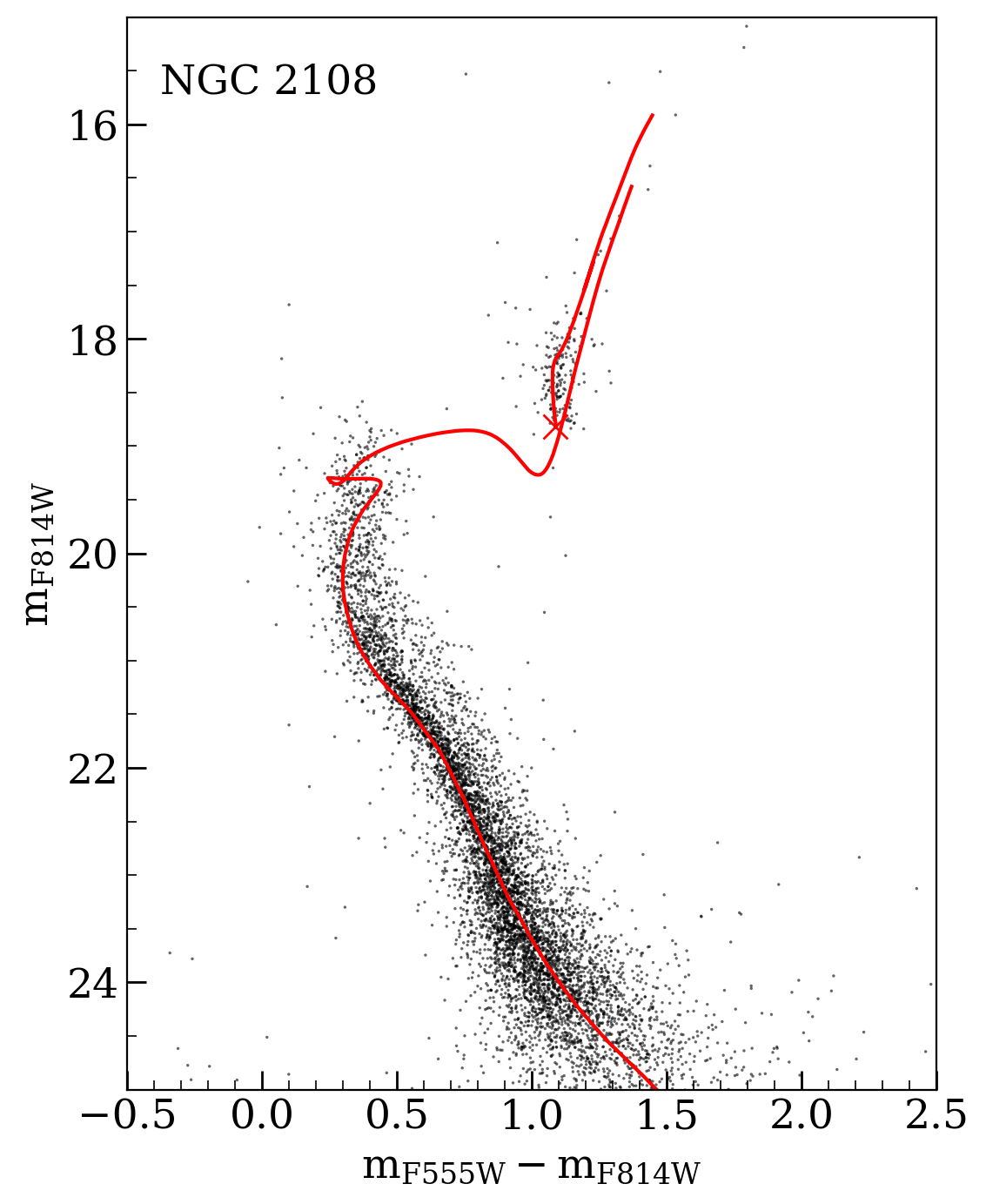} &
\includegraphics[width=0.60\columnwidth]{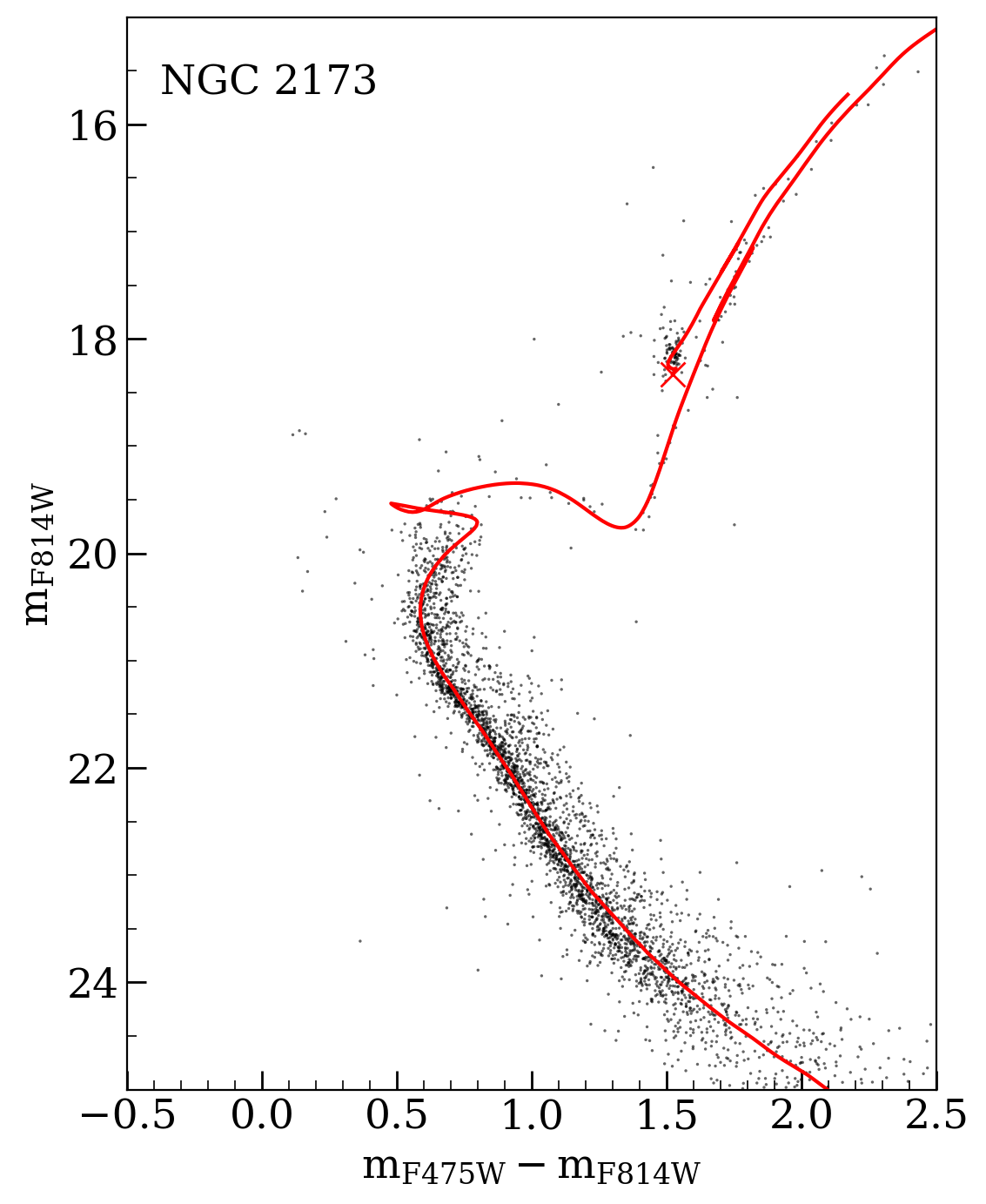} \\
\includegraphics[width=0.60\columnwidth]{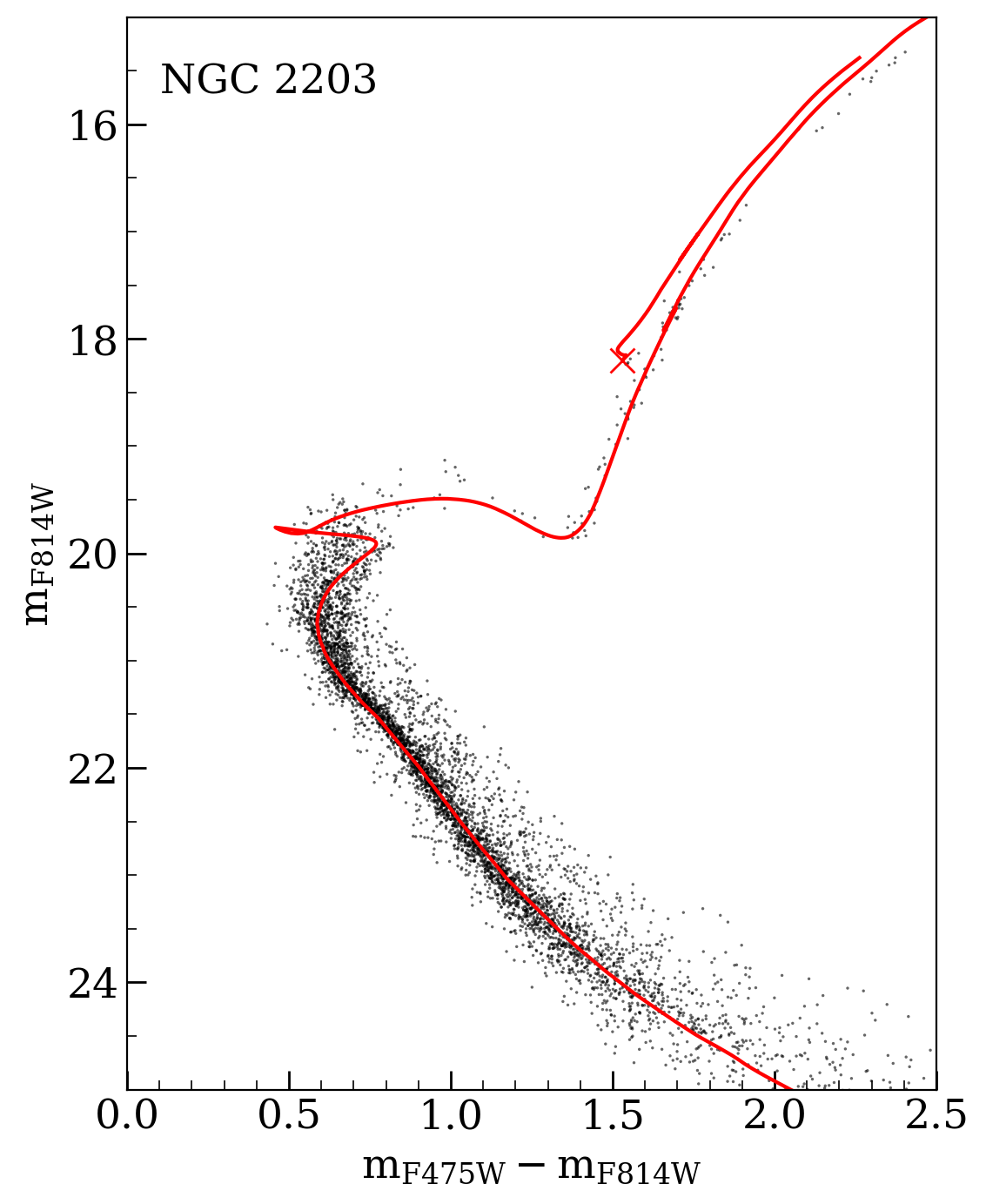} &
\includegraphics[width=0.60\columnwidth]{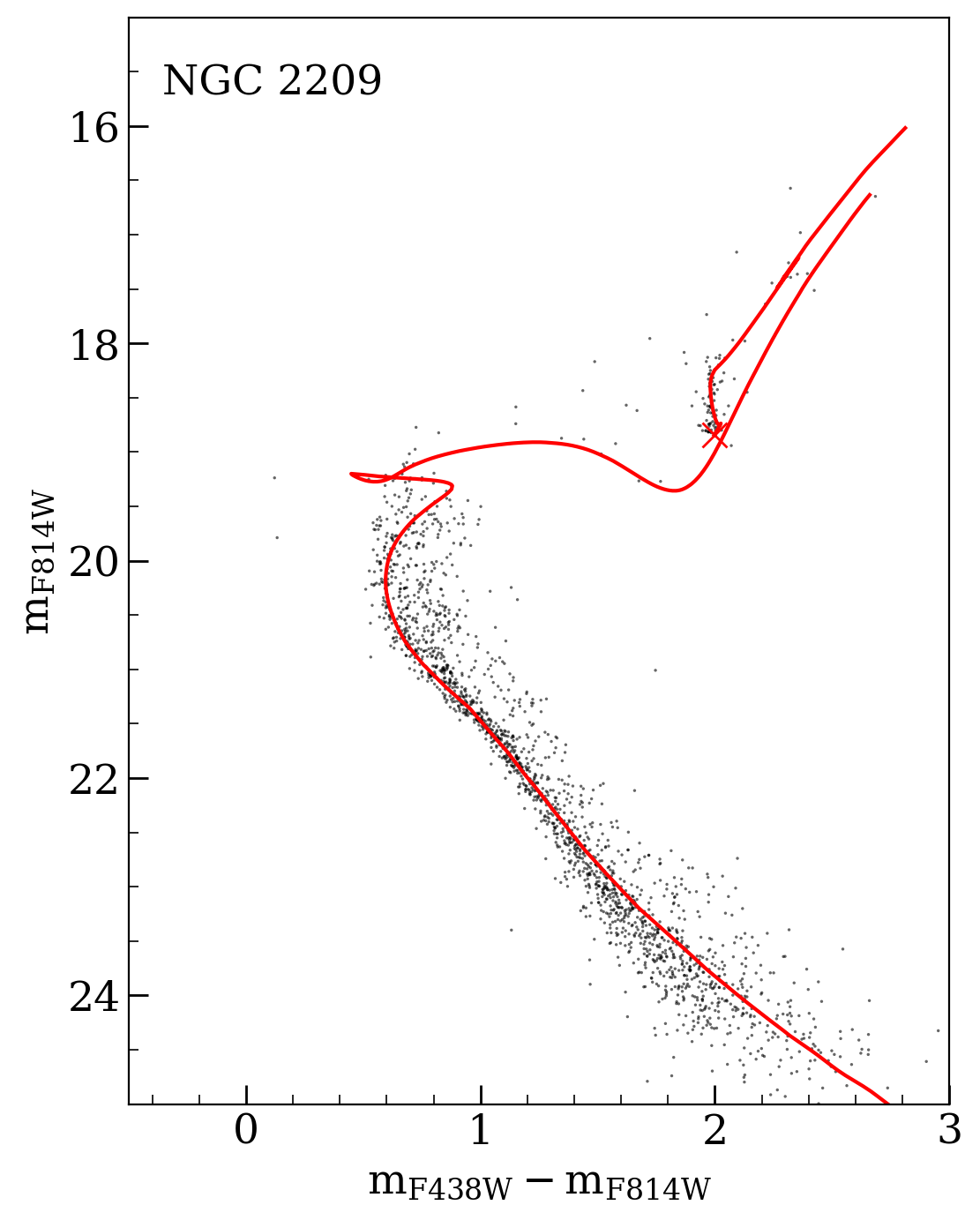} &
\includegraphics[width=0.60\columnwidth]{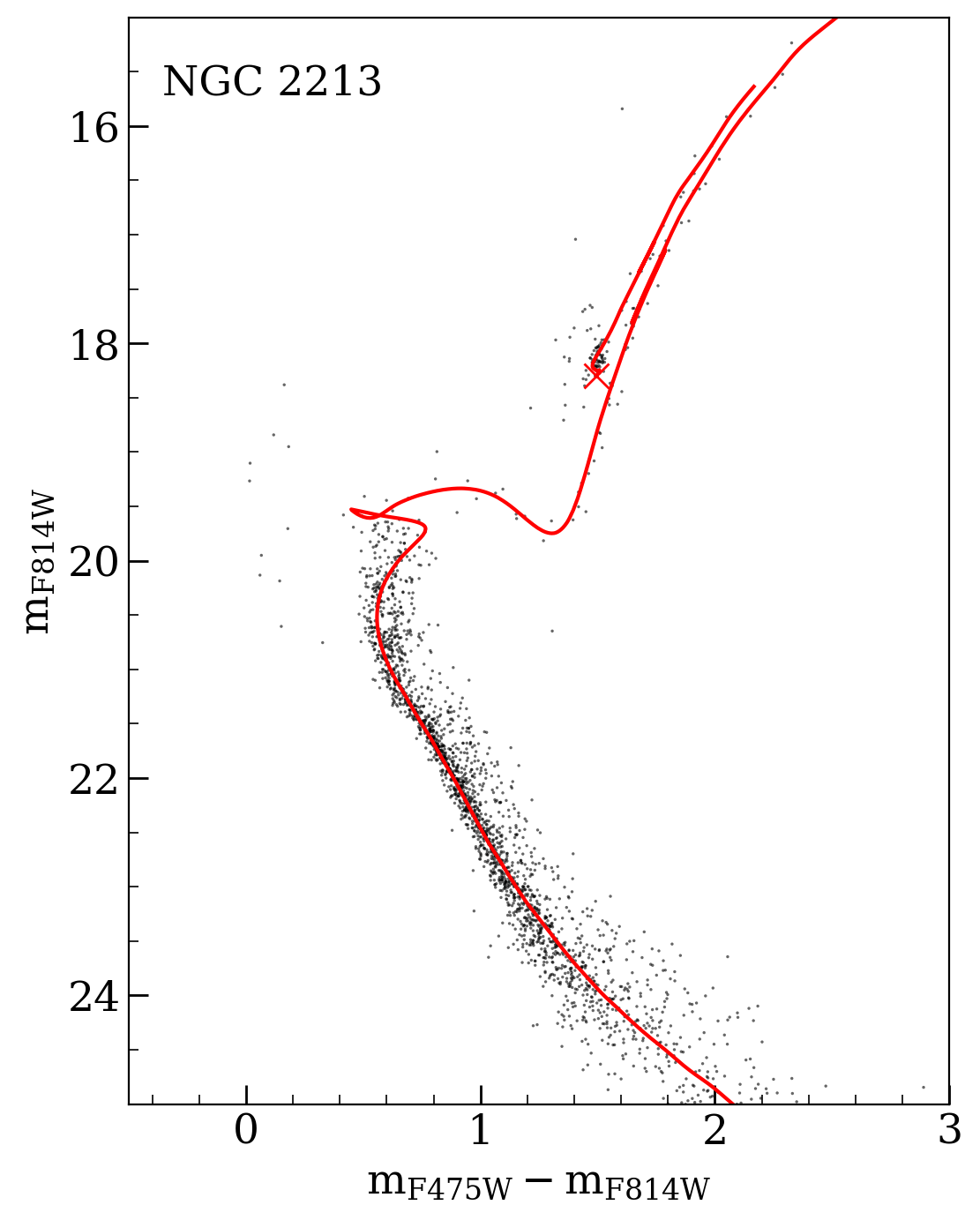} \\

\end{tabular}
\caption{Continued.
}
\end{figure*}

%--------------------------------------------------------------------

\section{Orbit parameters of young clusters}

Table~\ref{tab:orbit_params} list the orbit parameters of the young star clusters in our sample, reporting for each cluster: the inclination angle, the maximum height above the LMC disc plane (Z$_{\mathrm{max}}$), the apocentric distance (r$_{\mathrm{apo}}$), the eccentricity and the birth radius (R$_0$).

\begin{table*}
\centering
\caption{Orbital parameters of the young clusters \label{tab:orbit_params}}
\begin{tabular} {l r@{\,}c@{\,}l  r@{\,}c@{\,}l r@{\,}c@{\,}l r@{\,}c@{\,}l r@{\,}c@{\,}l}
\hline\hline
\noalign{\smallskip}
Cluster ID
      &  \multicolumn{3}{c}{Inclination} 
        & \multicolumn{3}{c}{Z$_{\mathrm{max}}$} 
            & \multicolumn{3}{c}{r$_{\mathrm{apo}}$}
                    & \multicolumn{3}{c}{Eccentricity} 
                    & \multicolumn{3}{c}{R$_0$} 
                    \\
&   \multicolumn{3}{c}{[deg]}
      & \multicolumn{3}{c}{[kpc]}
      & \multicolumn{3}{c}{[kpc]}
      & \multicolumn{3}{c}{}
      & \multicolumn{3}{c}{[kpc]}
      \\
\noalign{\smallskip}
\hline
\noalign{\smallskip}
NGC 1805  &   22 & $\pm$ & 5  &   1.42 & $\pm$ &  0.37  &    4.46 & $\pm$ &   0.32  &     0.08 &  $\pm$ &   0.03  & 4.35 & $\pm$ & 0.31\\ 
NGC 1818  &    9 & $\pm$ & 5  &   0.34 & $\pm$ &  0.27  &    4.85 & $\pm$ &   0.55  &     0.25 &  $\pm$ &   0.05  & 4.85 & $\pm$ & 0.55\\ 
NGC 1831  &   15 & $\pm$ & 13 &   1.15 & $\pm$ &  1.12  &    4.42 & $\pm$ &   0.93  &     0.27 &  $\pm$ &   0.12  & 3.78 & $\pm$ & 0.96\\ 
NGC 1850  &   75 & $\pm$ & 6  &   1.48 & $\pm$ &  0.18  &    1.91 & $\pm$ &   0.15  &     0.32 &  $\pm$ &   0.07  & 1.63 & $\pm$ & 0.13\\ 
NGC 1856  &      & --    &    &   1.16 & $\pm$ &  0.74  &    1.36 & $\pm$ &   0.59  &     0.76 &  $\pm$ &   0.24  & 0.77 & $\pm$ & 0.54\\ 
NGC 1866  &   19 & $\pm$ & 10 &   1.37 & $\pm$ &  0.82  &    5.68 & $\pm$ &   0.84  &     0.16 &  $\pm$ &   0.06  & 3.98 & $\pm$ & 0.53\\ 
\noalign{\smallskip}
\hline
\end{tabular}
\tablefoot{Due to the orbital shape of NGC~1856, no inclination angle can be given.}

\end{table*}

%--------------------------------------------------------------------

\section{Orbits of young clusters assuming positions in the disc plane\label{app:orbits}}

\begin{figure}

\includegraphics[width=1\columnwidth]{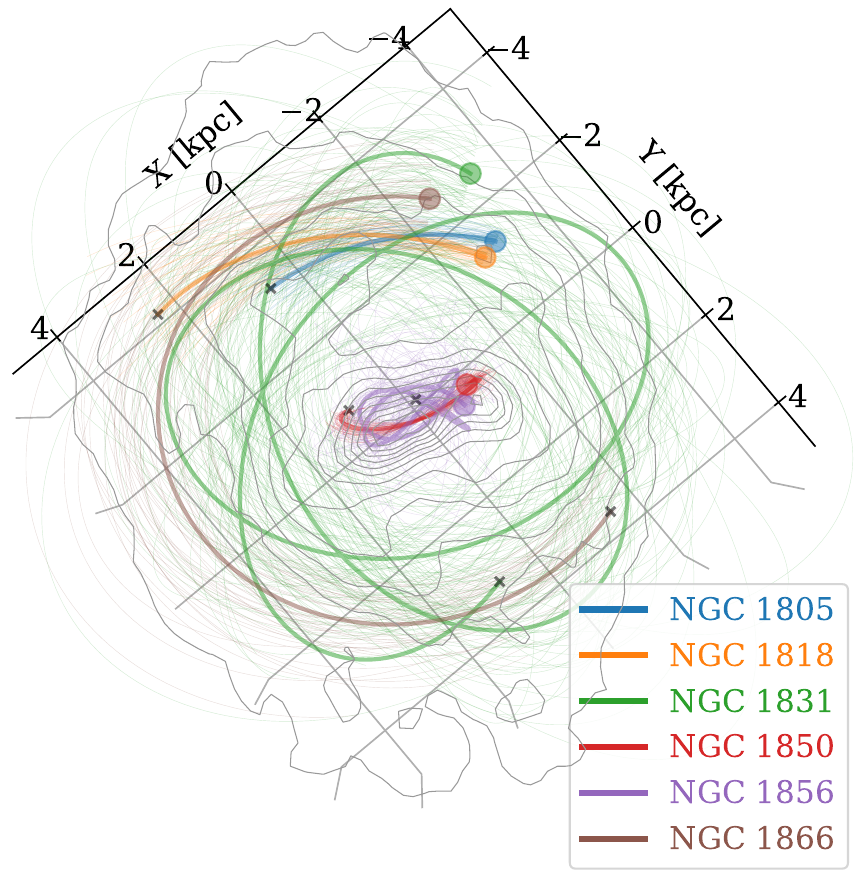} 
\caption{Similar to Fig.~\ref{fig:young_clusters_orbits} but now assuming all clusters are currently located within the disc plane of the LMC. The plot is oriented such that the LMC disc is seen directly from above, and North is to the top and East to the right. 
\label{fig:young_cluster_orbits_disc}
}
\end{figure}

Fig.~\ref{fig:young_cluster_orbits_disc} shows the reconstructed orbits of the six young ($<$1 Gyr) clusters within the LMC, now assuming that all clusters are currently located within the disc plane of the LMC.  

\section{Model parameters for the LMC and SMC}

Table~\ref{tab:model_params} lists the model parameters for the present-day positions and velocities of the LMC and SMC used as initial conditions for the dynamical models.

\begin{table*}
\centering
\caption{Model parameters for the present-day positions and velocities of the LMC and SMC. \label{tab:model_params}}
\begin{tabular} {l c c c }
\hline\hline
\noalign{\smallskip}
Parameter
      &  Value 
          & Unit
           &  Reference
      \\
\noalign{\smallskip}
\hline
\noalign{\smallskip}
LMC $\alpha_0$                            &           79.88           & degrees        & \citet{vanderMarel14} \\
LMC $\delta_0$                            &           $-$69.59        & degrees        & \citet{vanderMarel14} \\
LMC V$_{\mathrm{LOS},~0}$                 & 261.1     $\pm$     2.2   & km\,s$^{-1}$   & \citet{vanderMarel14} \\
LMC $\mu_{\alpha} \rm cos \delta_{~0}$    & $-$1.895  $\pm$     0.024 & mas\,yr$^{-1}$ & \citet{vanderMarel14} \\
LMC $\mu_{\delta,~0}$                     & 0.287     $\pm$     0.054 & mas\,yr$^{-1}$ & \citet{vanderMarel14} \\
LMC D$_0$                                 & 50.1      $\pm$     2.5   & kpc            & \citet{Freedman01}    \\
SMC $\alpha_0$                            &           13.38           & degrees        & \citet{Subramanian12} \\
SMC $\delta_0$                            &           $-$73.0         & degrees        & \citet{Subramanian12} \\
SMC V$_{\mathrm{LOS},~0}$                 & 145.6     $\pm$     0.6   & km\,s$^{-1}$   & \citet{Harris06}      \\
SMC $\mu_{\alpha} \rm cos \delta_{~0}$    & 0.772     $\pm$     0.063 & mas\,yr$^{-1}$ & \citet{Kallivayalil13} \\
SMC $\mu_{\delta,~0}$                     & $-$1.117  $\pm$     0.061 & mas\,yr$^{-1}$ & \citet{Kallivayalil13} \\
SMC D$_0$                                 & 62.1      $\pm$     1.9   & kpc            & \citet{Graczyk13}    \\

\noalign{\smallskip}
\hline
\end{tabular}
\end{table*}

\end{appendix}

\end{document}